\documentclass{jfm_ase_dec2010}

\DeclareGraphicsExtensions{.pdf,.eps,.jpg,.jpeg,.png}

\graphicspath{{./Graphics/}}

\usepackage{amsmath}
\usepackage{amssymb}
\usepackage[colorlinks=true, urlcolor=black,linkcolor=blue, citecolor=blue]{hyperref}
\usepackage{mathrsfs}
\usepackage{amsbsy}
\usepackage{graphicx}
\usepackage{graphics}
\usepackage{nomencl}
\usepackage{enumerate}
\usepackage{fancyhdr}
\usepackage{subfig}
\usepackage{makeidx}
\usepackage{index}
\usepackage{natbib}

\title{Cross-linked polymers in strain: Structure and anisotropic stress}
\crest{\includegraphics[width=4cm]{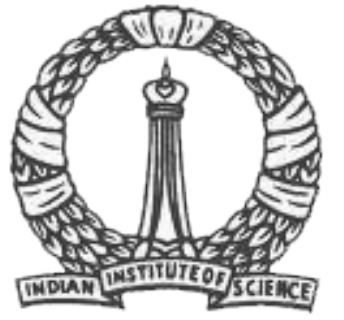}}
\titlegraphic{\includegraphics[width=0.7\textwidth]{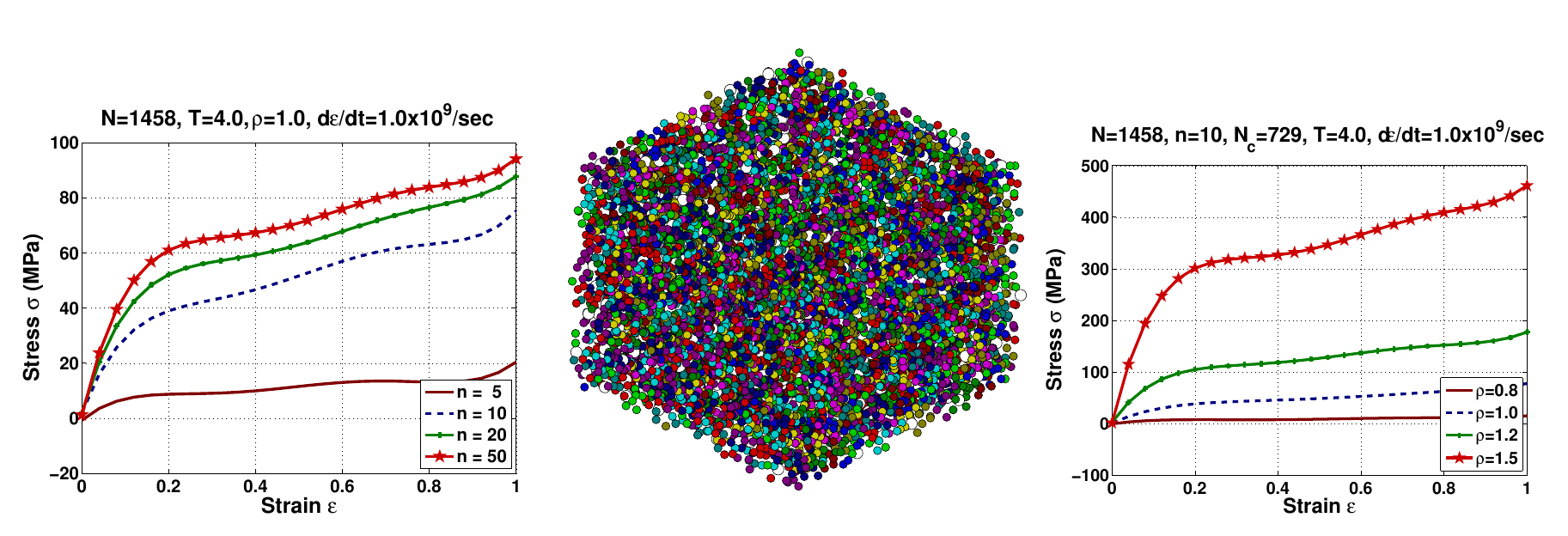}}
\author{Prashant Kumar Srivastava}
\degree{Graduate Student}
\advisor{Kartik Venkatraman}
\advdesignation{Associate Professor}
\reportnumber{ASELAB-TR-2011-MAY-04}
\date{\today}
\collegeordept{\href{http://www.aero.iisc.ernet.in}{Department of Aerospace Engineering}}
\university{\href{http://www.iisc.ernet.in}{Indian Institute of Science}}
\lab{Aeroservoelasticity Laboratory}
\address{Bangalore-560012, India}

\makenomenclature

\begin{document}

\maketitle

\begin{abstract}

Molecular dynamic simulation enables one to correlate the evolution of the micro-structure with anisotropic stress when a material is subject to strain. The anisotropic stress due to a constant strain-rate load in a cross-linked polymer is primarily dependent on the mean-square bond length and mean-square bond angle. Excluded volume interactions due to chain stacking and spatial distribution also has a bearing on the stress response. The bond length distribution along the chain is not uniform. Rather, the bond lengths at the end of the chains are larger and uniformly decrease towards the middle of the chain from both ends. The effect is due to the presence of cross-linkers. As with linear polymers, at high density values, changes in mean-square bond length dominates over changes in radius of gyration and end-to-end length. That is, bond deformations dominate over changes in size and shape. A large change in the mean-square bond length reflects in a jump in the stress response. Short-chain polymers more or less behave like rigid molecules. Temperature has a peculiar effect on the response in the sense that even though bond lengths increase with temperature, stress response decreases with increasing temperature. This is due to the dominance of excluded volume effects which result in lower stresses at higher temperatures. At low strain rates, some relaxation in the bond stretch is observed from $\epsilon=0.2$ to $\epsilon=0.5$. At high strain rates, internal deformation of the chains dominate over their uncoiling leading to a rise in the stress levels. 
 
\end{abstract}

\tableofcontents
\newpage

\addcontentsline{toc}{section}{List of figures}
\listoffigures
\newpage

\addcontentsline{toc}{section}{List of tables}
\listoftables
\newpage

\addcontentsline{toc}{section}{Nomenclature}
\printnomenclature

\newcommand{\vecr}{\mathbf{r}}
\newcommand{\bvecr}{\bar{\mathbf{r}}}
\newcommand{\barr}{\bar{r}}
\newcommand{\force}{\mathbf{f}}
\newcommand{\bforce}{\bar{\mathbf{f}}}
\newcommand{\stress}{\boldsymbol{\tau}}
\newcommand{\strain}{\boldsymbol{\epsilon}}
\newcommand{\PiolaT}{\mathbf{T}}
\newcommand{\PiolaS}{\mathbf{S}}
\newcommand{\Hamiltonian}{\mathscr{H}}
\newcommand{\Lagrangian}{\mathscr{L}}
\newcommand{\Defgrad}{\mathbf{F}}
\newcommand{\Kinpar}{\mathscr{A}}
\newcommand{\Force}{\mathscr{F}}
\newcommand{\partition}{\mathscr{Z}}
\newcommand{\partitionQ}{\mathscr{Q}}
\newcommand{\Cellscale}{\mathbf{H}}
\newcommand{\Mom}{\mathbf{p}}
\newcommand{\Pos}{\mathbf{q}}
\newcommand{\refconf}{\mathbf{X}}
\newcommand{\curconf}{\mathbf{x}}
\newcommand{\trace}{\textrm{Tr}}
\newcommand{\boltz}{k_B\mathcal{T}}
\newcommand{\etab}{\boldsymbol{\eta}}
\newcommand{\zetab}{\boldsymbol{\zeta}}
\newcommand{\Jb}{\mathbf{J}}
\newcommand{\scoor}{\mathbf{s}}
\newcommand{\dhhi}{\dot{\mathbf{H}}\mathbf{H}^{-1}}
\newcommand{\bond}{\mathbf{b}}
\newcommand{\Helmholtz}{\mathcal{A}}
\newcommand{\entropy}{\mathcal{S}}
\newcommand{\temperature}{\mathcal{T}}
\newcommand{\energy}{\mathcal{E}}

\pagenumbering{arabic}
\setcounter{section}{0}

\section{Introduction}

Polymers are finding an increasing number of applications keeping with the wide variety of properties, in varying environmental conditions, they exhibit. In part, the behavior of these materials can also be attributed to the set of complicated inherent constraints these materials possess. For instance, covalent bonds present in the polymers cannot cross each other. Further, because of the cross-linking between chains, they are not free to move with respect to each other. The monomers in the chain are also constrained to move because of their connectivity with other monomers in the neighborhood. This kind of molecular structure provides them the capability to undergo nonlinear and elastic deformation. Unlike metals, these materials can undergo high strains. This is possible because the major contribution to the strain comes from the unfolding of the chains rather than the actual deformation of the bonds. 

Uniaxial extension of the polymer may result in the alignment of the chains in the direction of the applied loading. This could be a significant factor in the evolution of the microstructure.  An increase in the amount of the alignment may result in a mechanically stronger material, but extension beyond a particular limit may result in local breakage of the bonds. To understand the effect of the molecular structure on the macroscopic properties of such systems we need an accurate model of the interaction between the monomers. An accurate model not only helps in establishing the correlation between the molecular structure and its macroscopic behavior, but also could be useful in studying the morphological evolution and in altering the properties of the system according to the need. 

Maxwell and Kelvin-Voigt models \citep{Rajagopal,Thien} that use simple mechanical elements to describe the behavior of viscoelastic materials, are very useful to predict the response of cross-linked polymers to mechanical loading. But these models fail to correlate the mechanical behavior of such materials with their internal structure. More detailed descriptions of the behavior of polymeric materials is given using continuum mechanics models \citep{Kankanala}. Here the material constitutive relations are described in terms of a free energy function. The free energy function is usually obtained by fitting experimental data. These models too suffer from non-uniqueness as the choice of the free energy function is arbitrary. These methods also face the same difficulty of not being able to correlate the properties of the system to their specific structural origin. Few statistical mechanics models \citep{Doi} such as freely jointed chain models, phantom chain models, tube models, and reptation models, to name a few, are also capable of describing the behavior, but with a limited range and validity. Further, many of them are based on certain assumptions which may turn out to be physically non-admissible. For instance, phantom chain models \citep{Weiner} assume that the bonds between the two beads are virtual and hence there is no restriction on the bonds crossing each other. 

One of the very early attempts using statistical mechanical models to study the behavior of polymer chains was made by \cite{James}. Here, they computed the probabilities associated with a given element in a given network position and the probabilities of a given relative separation between two elements. A theory of elasticity based on a statistical mechanics approach for a polymer network was developed by \cite{Flory, Erman}. They have shown that as opposed to the phantom networks, restriction on the fluctuations of the network junctions imposed by neighboring chains leads to an increment in the stress. In this work they assume that dimensions of the domain are transformed linearly, or affinely, by the macroscopic strain. Furthermore, they show that this assumption leads to the less restrictive constraint in the principle direction of extension. This leads to an increase in the stretch which in turn diminishes the relative enhancement in the stress. Many molecular dynamics studies have been performed on the kinetics of the end cross-linking in a polymer network \citep{Grest1992, Grest1993}. In these studies, the authors have shown that for stoichiometric number of cross-linkers, the number of free ends and the number of unsaturated cross-linkers decay with time as a power law $t^{-1/2}$. However, for non-stoichiometric numbers of cross-linkers, this decay is much faster.  They also calculate the shear modulus by evaluating the auto-correlation of the Rouse modes \citep{Doi} and observe that the modulus decays with time. They also show that plateau modulus calculated from the above study is almost $67\%$ higher than the same, predicted using the affine model without trapping. 

\citet{Edward} have studied the effect of preservation of topology of the entanglement in systems of polymer loop. They calculate the contribution of the entanglements to the reduced stress in the network. From this analysis they find that cross-links in their approximation result in a term which is independent of the stretch, whereas entanglements have the contribution which depends on stretch. They also point out that the modulus rapidly increases at a very high extension. Slipping links due to entanglements cause softening because a stretched polymer gains more freedom before the links can slip-up to the permanent cross-links. Furthermore, from their theory, they conclude that contribution of the entanglement to the stress also increases with density.

An interesting study of the adhesive property of the highly cross-linked polymer network and effect on the cross-linker functionality on the same was done by \citet{Tsige2004a}. They consider a highly cross-linked polymer network bonded to a solid surface and study the fracture behavior of the system at various functionalities of cross-linkers --- $f=3,4 \mbox{ and } 6$. At full interfacial bond density the failure mode is cohesive and cohesive failure stress is similar for shear and tensile mode. Decreasing the number of interfacial bonds results in cohesive to adhesive transition. They also find that failure stress decreases with small values of $f$ while failure strain increases. Further study of the effect of cross-linkers on the mechanical properties of polymers was done by \cite{Tsige2004b} wherein they observed the effect of mixed functionality and degree of curing on the stress-strain behavior of highly cross-linked polymer networks. They vary the average value of the functionality $f_{\mathrm{av}}$ from $3$ to $6$ by mixing the cross-linkers of different functionalities. They have determined the range of strain of the plateau region in the stress strain curve and the failure strain $\epsilon_f$. The failure stress $\sigma_f$ for a fully cured network has a power law dependence on $f_{\mathrm{av}}$ as $f_{\mathrm{av}}^\alpha$. They also showed that the failure stress has two distinct regions. For $f_{\mathrm{av}}^\alpha < 4$, where $\alpha=1.22$, the failure stress  increases with increase in $f_{\mathrm{av}}$. In this regime, the work to failure is constant. For $f_{\mathrm{av}}^\alpha \ge 4$, the system fails interfacially, $\sigma_f$ becomes constant, and work to failure decreases with $f_{\mathrm{av}}$. With decrease in percentage of curing, failure stress decreases and failure strain increases. 

A study by \citet{Lyulin} also suggests that entanglements results in an increase in the stress. A comparison of molecular dynamics (MD) results with analytical theory for a permanent set of cross-linking networks was done by \citet{Rottach2006}. In this study they introduce cross-links in both unstrained and strained networks. A permanent set in the strain was observed after relaxation of the network to the state where the stress is zero. An uniaxial strain experiment for bead-spring polymer network suggests that slip-tube and double-tube models, which incorporate entanglement effects, are in agreement with the simulation. In their subsequent study, \cite{Rottach2007} measure the equilibrium stress before and after removing some or all of the cross-links introduced in the unstrained state and conclude that fractional stress reduction upon such removal of cross-linkers could be accurately calculated from the modified slip tube model of \cite{Rubinstein} in which a theoretical transfer function of \cite{Fricker} is used. A detailed account of the various models in polymer melt rheology can be found in \citet{Larson}. Another study on the stress-strain behavior for the deformation of polymeric networks was done by \cite{Bergstrom}, where they compared the result of the molecular simulation with the classical statistical models of the rubber elasticity. 

The stress generating mechanisms in cross-linked polymers is not very well understood. Especially in the case of time-varying or dynamic strain loading on the polymer. The evolution of micro-structural parameters such as mean-square bond length, radius of gyration, and similar variables with time could provide an insight into the mechanisms that create the stress. Further, the interplay of internal and external control variables such as chain length, functionality of the cross-linkers, temperature, density, and strain rate are known to have significant influence on the stress generated. However, what is little known is the influence of the control variable on the dynamics of the micro-structure that result in the stress response. This then is the subject of this research presented in this report.


\section{Mathematical Model}\label{sec:model}

Use of MD simulation in the canonical NVT ensemble requires a modification in the equations of motion. This is achieved by introducing an additional degree of freedom \citep{Nose, Hoover} representing kinetic mass in the equations of motion. For a system consisting of $N$ polymer chains, each having $n$ united atoms, the equations of motion can be written as

\begin{equation}
\label{eq:EOM}
m_i \frac{d^2\mathbf{r}_i}{dt^2} = -\frac{\partial U}{\partial \mathbf{r}_i} - \eta \mathbf{v}_i,\ i \in [1,2,\ldots,N],
\end{equation}

\nomenclature[A]{$m_i$}{mass of the united atom $i$}
\nomenclature[A]{$\mathbf{r}_i$}{position of united atom $i$}
\nomenclature[A]{$U$}{Potential function}
\nomenclature[A]{$\mathbf{v}_i$}{velocity of united atom $i$}
\nomenclature[A]{$\mathbf{a}_i$}{acceleration of united atom $i$}
\nomenclature[G]{$\eta$}{dynamical parameter to control the temperature}
\nomenclature[A]{$N$}{number of chains}
\nomenclature[A]{$n$}{number of united atoms in a chain}
\nomenclature[A]{$N_c$}{number of cross-linkers}
\nomenclature[A]{$f$}{functionality of cross-linkers}
\nomenclature[S]{$i$}{index of polymer chain, $i = 1, 2, \ldots N$}
\nomenclature[S]{$j$}{index of atom in a polymer chain, $j = 1, 2, \ldots n$}

where $m_i$ is the mass of the united atom $i$, and $\mathbf{r}_i$ is its position, $\mathbf{v}_i$ is its velocity, and $U$ is the total interaction potential influencing it. The parameter $\eta$ evolves as

\begin{equation}
\label{eq:etadot}
\dot\eta = \frac{1}{Q}\left[\sum_{i=1}^Nm_i\mathbf{v}_i^2 - f_dk_BT\right],
\end{equation}

\nomenclature[A]{$f_d$}{numbers of degrees of freedom}
\nomenclature[A]{$k_B$}{Boltzmann's constant}
\nomenclature[A]{$T$}{Temperature}
\nomenclature[A]{$Q$}{Inertia parameter used to control the temperature}

where $f_d$ is the number of degrees of freedom of the system and $Q$ is the kinetic mass.  From Equation \ref{eq:etadot}

\begin{equation}
\label{eq:etaddot}
\ddot{\eta} = \frac{2}{Q}\left[\sum_{i=1}^N m_i\mathbf{v}_i\cdot\mathbf{a}_i\right],
\end{equation}

where $\mathbf{a}_i$ is the acceleration of the $i^{\text{th}}$ united atom. The velocity Verlet algorithm is the most commonly used algorithm to integrate the equations of motion in MD and is given by

\begin{eqnarray}
\label{eq:velverpos}
\mathbf{r}_i(t+\delta t) & = & \mathbf{r}_i(t) + \delta t \mathbf{v}_i(t) + \frac{\delta t^2}{2}\mathbf{a}_i(t), \\
\label{eq:velvervel}
\mathbf{v}_i(t +\delta t) & = & \mathbf{v}_i(t) + \frac{\delta t}{2}[\mathbf{a}_i(t) + \mathbf{a}_i(t + \delta t)].
\end{eqnarray}

Fixing the temperature at a constant value in the simulation imposes a constraint on the direct use of the velocity Verlet algorithm. This is because the velocity at the current time step depends on the acceleration at the current time step which in turn depends on the current velocity, which is an unknown. Hence the velocity-Verlet algorithm is modified in the following way. Equation \ref{eq:EOM} is written as

\begin{equation}
\label{eq:discreteEOM}
m_i \mathbf{a}_i(t + \delta t) = -\frac{\partial U(t+\delta t)}{\partial \mathbf{r}_i} - \eta(t + \delta t) \mathbf{v}_i(t + \delta t).
\end{equation}

Substituting Equation \ref{eq:discreteEOM} in Equation \ref{eq:velvervel} we get

\begin{equation}
\mathbf{v}_i(t+\delta t) = \mathbf{v}_i(t) + \frac{\delta t}{2}\left[\mathbf{a}_i(t) -\frac{1}{m_i}\left\{\frac{\partial U(t+\delta t)}{\partial \mathbf{r}_i} + \eta(t + \delta t) \mathbf{v}_i(t+\delta t)\right\}\right].
\end{equation}

Simplification of the above equation results in

\begin{equation}
\label{eq:modvelvervel}
\mathbf{v}_i(t+\delta t) = \left(1 + \frac{\eta(t+\delta t)\delta t}{2m_i}\right)^{-1}\left\{\mathbf{v}_i(t) - \frac{\delta t}{2m_i}\frac{\partial U(t+\delta t)}{\partial \mathbf{r}_i} + \frac{\delta t}{2}\mathbf{a}_i(t)\right\}.
\end{equation}

Then the equation of motion can be integrated as: \citep{Bergstrom}

\begin{enumerate}[(1) ]
\item Start with the initial configuration, velocity and acceleration
\item Calculate the $\mathbf{r}_i(t + \delta t)$ by Equation \ref{eq:velverpos}
\item calculate $\eta(t + \delta t)$ by
$\eta(t+\delta t) = \eta (t) + \delta t \dot{\eta}(t) + \frac{\delta t^2}{2}\ddot{\eta}(t)$
\item Calculate $\mathbf{v}_i(t + \delta t)$ by Equation \ref{eq:modvelvervel}
\item Calculate $\dot{\eta}(t+\delta t)$ by Equation \ref{eq:etadot}
\item Calculate $\mathbf{a}_i(t + \delta t)$ by Equation \ref{eq:discreteEOM}
\item Calculate $\ddot{\eta}$ by Equation \ref{eq:etaddot}
\item Go to step (2)
\end{enumerate}  

While applying the strain on the system, all the atom positions are scaled in the same proportion as that of the cell dimensions. Here we assume the material to be incompressible. So, if the stretch ratio in the $x$ direction is $\lambda$ then the stretch ratio in the $y$ and $z$ direction would be $\frac{1}{\sqrt{\lambda}}$ because of the symmetry.

The equations of motion for the monomers which are involved in cross-linking is the same as that derived for the monomers belonging to the linear polymer chain \cite{Srivastava2010}. An extra term is included in the forces to account for the contribution from the cross-linkers.

\section{Simulation}\label{sec:sim}

Molecular dynamics simulation is performed on a polymeric network with cross-linkers present in it. The starting point of the simulation is to generate the initial equilibrated melt of the polymer chains. Uniformly spaced points in the cell are selected as center of mass of each chain. Chain growth takes place by generating series of random bond vectors.  A random bond vector in three dimensions is generated by using a random dihedral angle between $0^\circ$ to $360^\circ$ while keeping the bond angle and bond length of the chain at their equilibrium values. This helps in generating an initial structure of the polymeric system very close to the one corresponding to the minimum energy configuration. The position of the new monomer is obtained by adding these bond vectors to the previously generated sites. While chain growth takes place, a check is performed for the occupancy of the newly generated site. In other words, if the new generated position for the monomer is already occupied we go back one step and generate a new site position. A partial overlap of the monomers is allowed during the chain growth which is later removed by subjecting the polymeric system to a soft repulsive potential to ensure that all the monomers are separated by a distance confirming the absence of overlap. After generating the series of linear polymer chains, we add stoichiometric number of cross-linkers with functionality $f$ in the system by randomly placing them in the simulation cell. Checks are performed for the occupancy of the site. The monomers at the ends of the chains are the active sites which are linked with the cross-linker when it comes within the reaction radius of these active sites. Before the cross-linking takes place, the linkers interact with other monomers through a Lennard-Jones (LJ) potential. Once the cross-linking takes place, we add a bonded potential, commonly known as finitely extended non-linear elastic (FENE) potential, between the cross-linked members. 

\begin{figure}
\begin{center}
\includegraphics[scale=0.5]{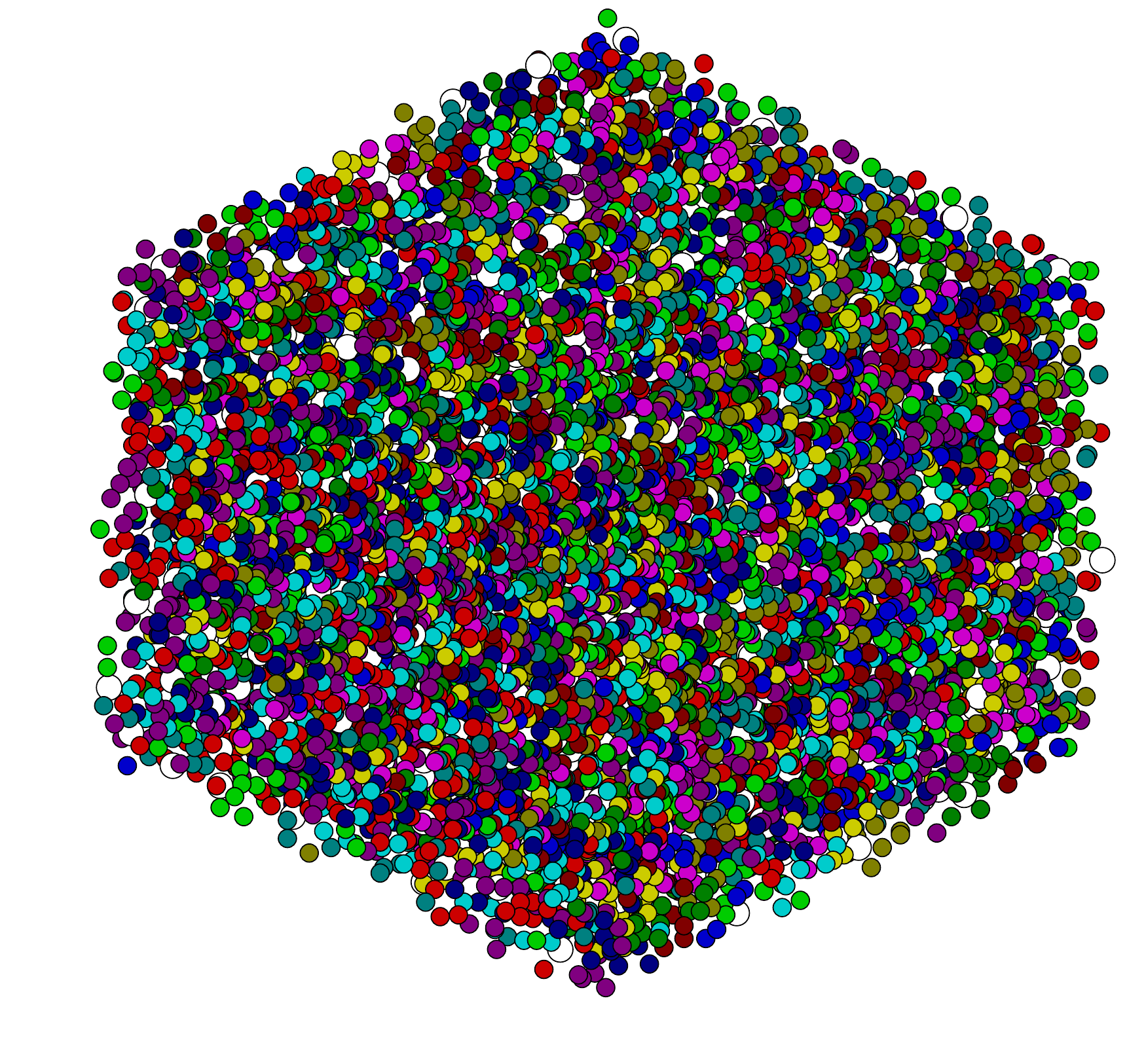}
\caption{Initial configuration}
\label{fig:init}
\end{center}
\end{figure}

Figure~\ref{fig:init} shows the initial structure of the polymer system generated by the above described method. Cross-linkers are shown in white color. 

Once the initial structure is ready, we perform an MD simulation by subjecting the system to the bonded and non-bonded potential which describe the inter-atomic interaction. To model the van dar Walls forces, we use the shifted and truncated Lennard-Jones potential which acts between all pairs of the united atoms.

\begin{equation}
U_{LJ}(r) = \begin{cases}
4\epsilon_{LJ}\left[\left(\frac{\sigma}{r}\right)^{12}-\left(\frac{\sigma}{r}\right)^6-\left(\frac{\sigma}{r_c}\right)^{12}+\left(\frac{\sigma}{r_c}\right)^6\right], & \text{if} \quad r < r_c = 2.5\sigma. \\
0 & \text{otherwise}.
\end{cases}
\end{equation} 

\nomenclature[G]{$\sigma$}{Lennard-Jones length parameter}
\nomenclature[G]{$\epsilon$}{Lennard-Jones energy parameter}
\nomenclature[A]{$r_c$}{Cut-off distance for short range potentials}

In the above, $\sigma = 0.401$ nm is the LJ length parameter, and $\epsilon = 0.468$kJ/mol is the LJ energy parameter which is related to the well depth of the LJ potential. To describe the other interactions namely, different kind of bonded interaction, we use bond stretching, bond bending and bond torsional potentials. 

\begin{figure}
\centering
\includegraphics[scale = 0.8]{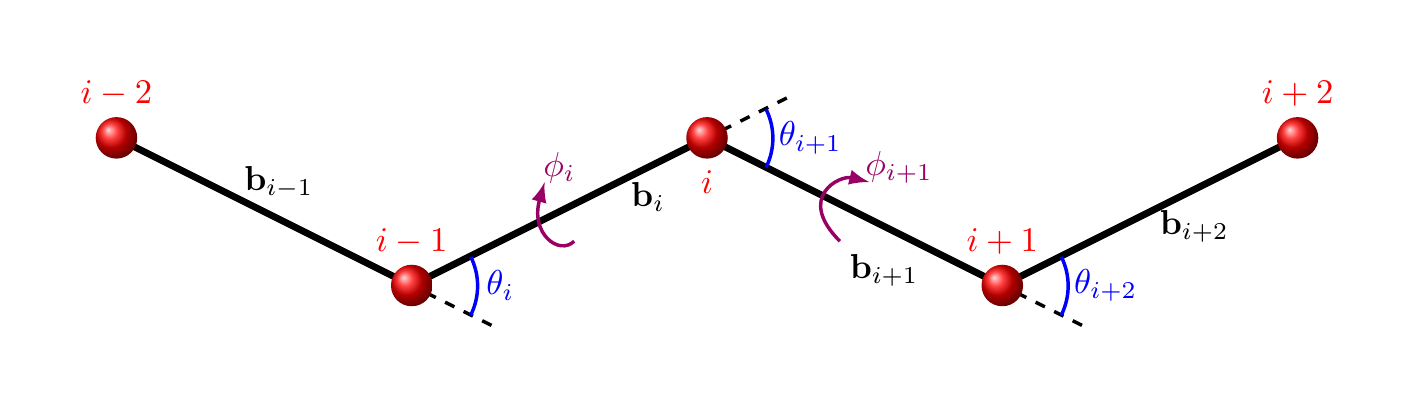}
\caption{Chain structure}
\label{fig:chain}
\end{figure}

Stretching of the bond is modeled through the FENE potential. 

\begin{equation}
U_{FENE}(r) = -\frac{1}{2}KR_0^2\ln\left[1-\left(\frac{r}{R_0}\right)^2\right].
\end{equation}

\nomenclature[A]{$K$}{Bond stretch stiffness}
\nomenclature[A]{$R_0$}{Maximum bond length}

Here $K = 70\epsilon/\sigma^2$ is the stiffness of the FENE potential.  This potential acts between the monomers which are directly bonded to each other in a chain and also between the monomers which are cross-linked with each other. 

The bond bending potential is a function of the bond angle $\theta$ as

\begin{equation}
U_{bending}(\theta) = \frac{1}{2}K_{\theta}(\cos\theta-\cos\theta_0)^2,
\end{equation}

\nomenclature[A]{$K_{\theta}$}{Bond bending stiffness}
\nomenclature[G]{$\theta_0$}{Mean bond angle}
\nomenclature[G]{$\theta$}{Bond angle}

where $K_\theta = 520$ kJ/mol is the bond bending stiffness and $\theta_0 = 110^\circ$. The bond angle $\theta$ is defined as the angle between the two contiguous covalent bonds

\[
\cos\theta_i = \frac{\bond_{i-1}\cdot\bond_i}{|\bond_{i-1}||\bond_i|},
\]

where $\bond_i = \vecr_i - \vecr_{i-1}$ is the bond vector between the atoms $i-1$ and $i$.

\nomenclature[A]{$\mathbf{b}_i$}{Bond vector between the united atoms $i-1$ and $i$}

The torsional potential is a four body potential and is a function of the dihedral angle $\phi$ and is of the form

\begin{equation}
U_{torsion}(\phi) = \displaystyle\sum_{l=0}^{3} K_{\phi l} \cos^{l}\phi,
\end{equation}

\nomenclature[A]{$K_{\phi l}$}{Bond torsion stiffness parameters, $l = 0,\ 1,\ 2,\ 3$}
\nomenclature[G]{$\phi$}{Dihedral angle}

where $K_{\phi 0} = 14.477\ \textrm{kJ/mol},\ K_{\phi 1} = -37.594\ \textrm{kJ/mol},\ K_{\phi 2} = 6.493\ \textrm{kJ/mol}\ \textrm{and}\ K_{\phi 3} = 58.499\ \textrm{kJ/mol}$. The dihedral angle is defined as the angle between the plane formed by the first two bond vectors and the plane formed by the last two bond vectors from sets of three consecutive bonds such that

\[
\cos\phi_i = \frac{(\bond_{i-1}\times\bond_i)\cdot(\bond_i\times\bond_{i+1})}{|\bond_{i-1}\times\bond_i||\bond_i\times\bond_{i+1}|}.
\]

In the present simulation we use the time step of $2.2 \times 10^{-15}$ seconds for the numerical integration of the equations of motion which are sufficient for the stability of the integration scheme and maintaining the accuracy of the same. In order to simulate the constant strain rate uniaxial deformation, the dimensions of the simulation cell in $\mathbf{e}_1$ direction, $L_1$, was increased at a fixed rate. Since rubber is almost incompressible, we apply the appropriate contraction in the transverse direction of the simulation cell. All the atom positions are scaled in the same proportion as the dimension of the simulation cell changes. The average of the properties are taken over $200$ steps of the simulation run. During the simulation, the non-dimensional parameters used are listed in Table~\ref{tb:ndp}.

\begin{table}
\setlength{\itemsep}{0cm}%
\setlength{\parskip}{0.5cm}%
\begin{center}
\begin{tabular}{ll}
\hline
time & $\bar{t}$ = $\displaystyle\frac{t}{\sigma}\left(\frac{\epsilon}{m}\right)^\frac{1}{2}$ \\[0.5cm]
number density & $\bar\rho$ = $\displaystyle\sigma^3\rho$ \\[0.5cm]
pressure & $\bar{p}$ = $\displaystyle\frac{\sigma^3p}{\epsilon}$ \\[0.5cm]
temperature & $\bar{T}$ = $\displaystyle\frac{k_BT}{\epsilon}$ \\
\hline
\end{tabular}
\end{center}
\caption{Scaled parameters}
\label{tb:ndp}
\end{table}

In Table~\ref{tb:ndp}, $m$ is the mass of the united atom --- in this case $14\ \mathrm{amu}$. $\sigma$ and $\epsilon$ denote length and energy scale parameters, respectively, and their values are the same as for the Lennard-Jones potential. From here onwards, we drop the bar used for scaled symbols, and symbols without bar should be considered as denoting the scaled parameters. 

An elastomer is a collection of polymers that are cross-linked. Cross-linking could take place either at the end atoms of each chain or anywhere in the chain. Let there be $N$ number of chains, each with $n$ united atoms, in a polymeric system. The chains are cross-linked by adding a cross-linker. A cross-linker is an united atom by itself and its role is to connect united atoms in different chains together through chemical bonds. A cross-linker with functionality $f$ implies that this cross-linker would join $f$ united atoms, in different or the same chains, together. A stoichiometric number $N_c$ of cross-linkers links all the chains together, including chain ends --- $N_c = 2N/f$. 

In order to study the effect of parameters on the stress and structure, we choose a reference system with $N=1458$ polymer chains, each chain having $n = 10$ united atoms. Stoichiometric number of cross-linkers with functionality $f=4$ was added to the system. This choice would lead to a highly cross-linked network of polymer chains. The temperature of this reference system is maintained at $T = 4.0$ in a thermal bath. The simulation cell has a volume consistent with a number density $\rho = 1.0$. This reference model was strained at a strain rate of $\dot\epsilon = 1.0\times 10^9/\mathrm{sec}$.

\nomenclature[A]{$\mathbf{f}_{ij}$}{force on united atom}
\nomenclature[A]{$T_{ext}$}{external temperature imposed on the system}
\nomenclature[X]{$\dot{(\,)}$}{derivative with respect to time $t$}
\nomenclature[G]{$\lambda$}{stretch ratio}
\nomenclature[G]{$\boldsymbol{\tau}$}{virial stress tensor}
\nomenclature[A]{$\mathbf{I}$}{identity tensor}
\nomenclature[A]{$V$}{volume of the simulation cell}
\nomenclature[A]{$R$}{end to end length of the polymer chain}
\nomenclature[A]{$R_G$}{radius of gyration of the polymer chain}
\nomenclature[A]{$G_{xy}$}{element of mass distribution tensor}
\nomenclature[A]{$g_k$}{eigen values of the mass distribution tensor, $k=1,\,2,\,3$}
\nomenclature[G]{$\psi$}{chain angle, angle between end to end vector of chain and loading direction}

\section{Stress and its dependence on polymer structure}

\begin{figure}
\centering
\subfloat[Stress-strain curve]{\label{fig:fit_stress_strain}\includegraphics[width=0.5\textwidth]{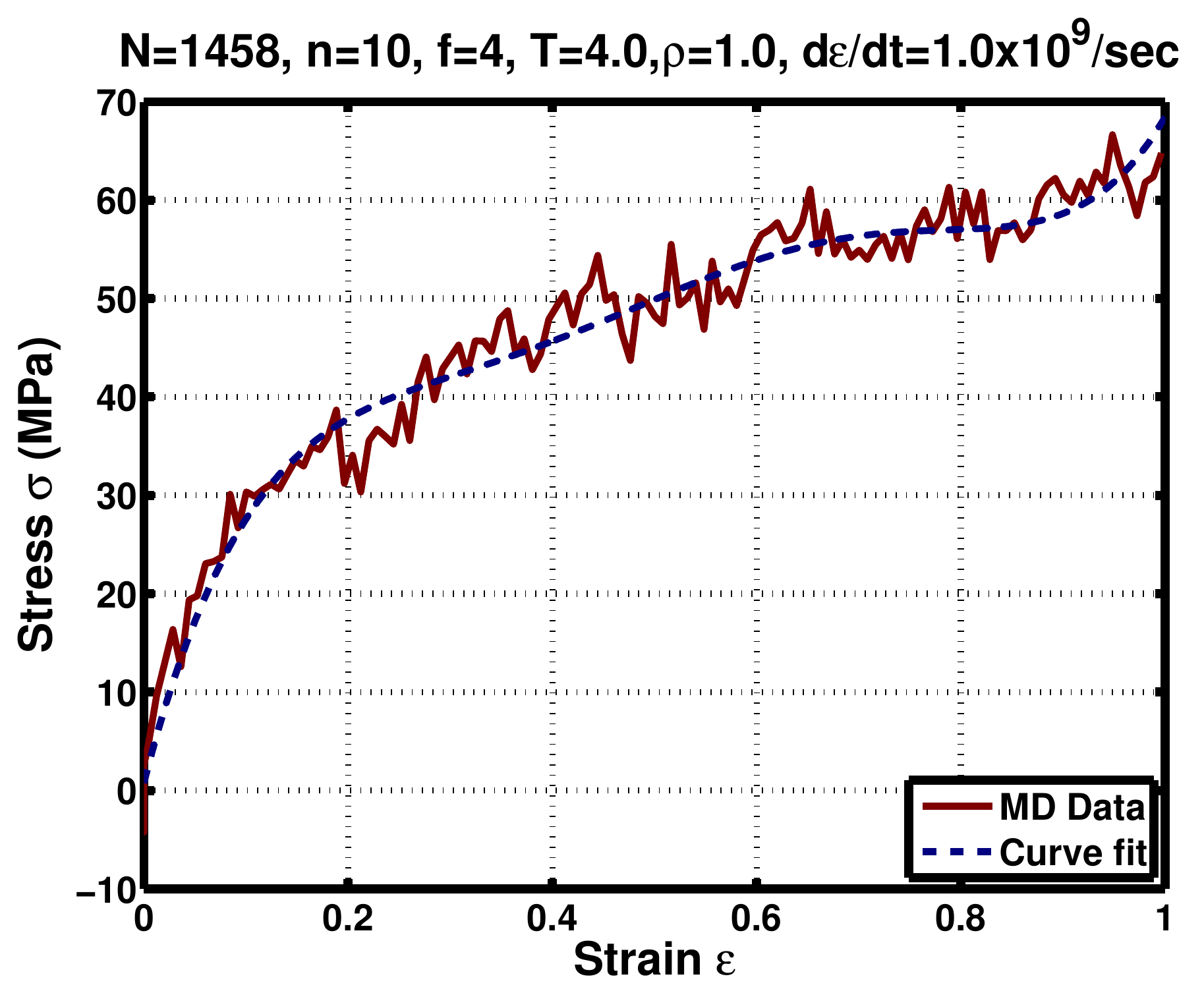}}
 \subfloat[Modulus vs. strain]{\label{fig:modulus}\includegraphics[width=0.5\textwidth]{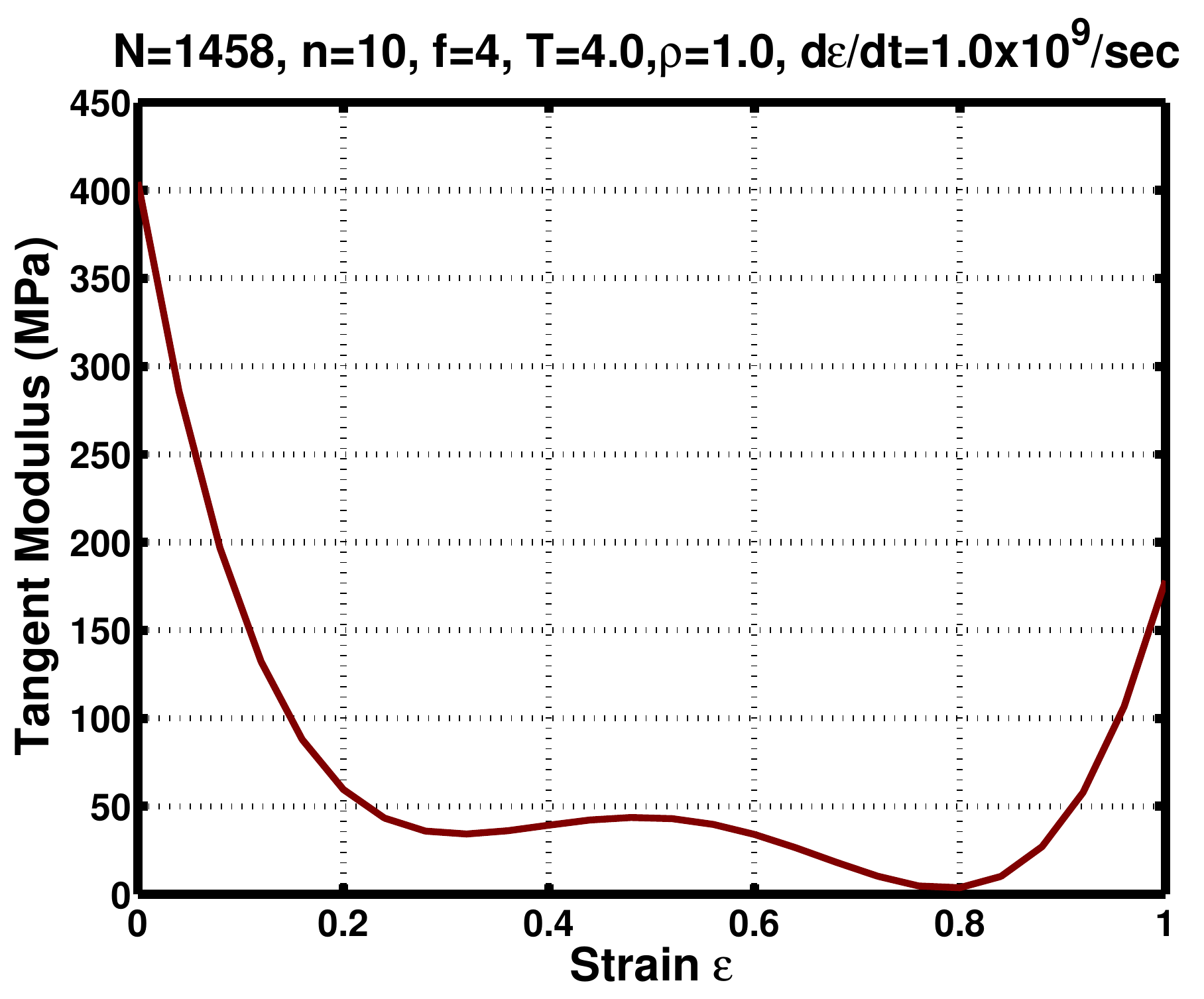}}
\caption{Stress response under uniaxial constant strain rate loading}
\label{fig:StressStrainMod}
\end{figure}

The system is subjected to a uniaxial strain at a constant strain rate. Figure~\ref{fig:fit_stress_strain} shows the response of the cross-linked polymer chains for the same loading. The stress response is highly nonlinear and this nonlinearity increases at higher strains. A power law curve fit on the MD data is also shown in the same figure. The tangent modulus is obtained by differentiating this expression with respect to strain. The variation of the tangent modulus with strain, thus obtained, is shown in Figure~\ref{fig:modulus}. The stress initially increases very rapidly with strain up to $\epsilon=0.2$, and subsequently the rate of increase reduces with increase in strain. This is also reflected in the modulus curve wherein the modulus is very high in the beginning of the loading and decreases rapidly with strain. After $\epsilon=0.2$, there is no significant change in the modulus, except for a little fluctuation caused by the fluctuations present in the MD data. 

Further insight can also be obtained by observing the effect of the deformation on the internal structure of the elastomer. In this regard, we note the variation of different parameters representing the internal structure and size and shape of the elastomers, namely, mean-square bond length, mean bond angle, mean-square end-to-end length, mean-square radius of gyration, mass ratios, and mean chain angle. 
 
\begin{figure}
 \centering
 \subfloat[Mean-square bond length]{\label{fig:MeanBondLenStrain}\includegraphics[width=0.5\textwidth]{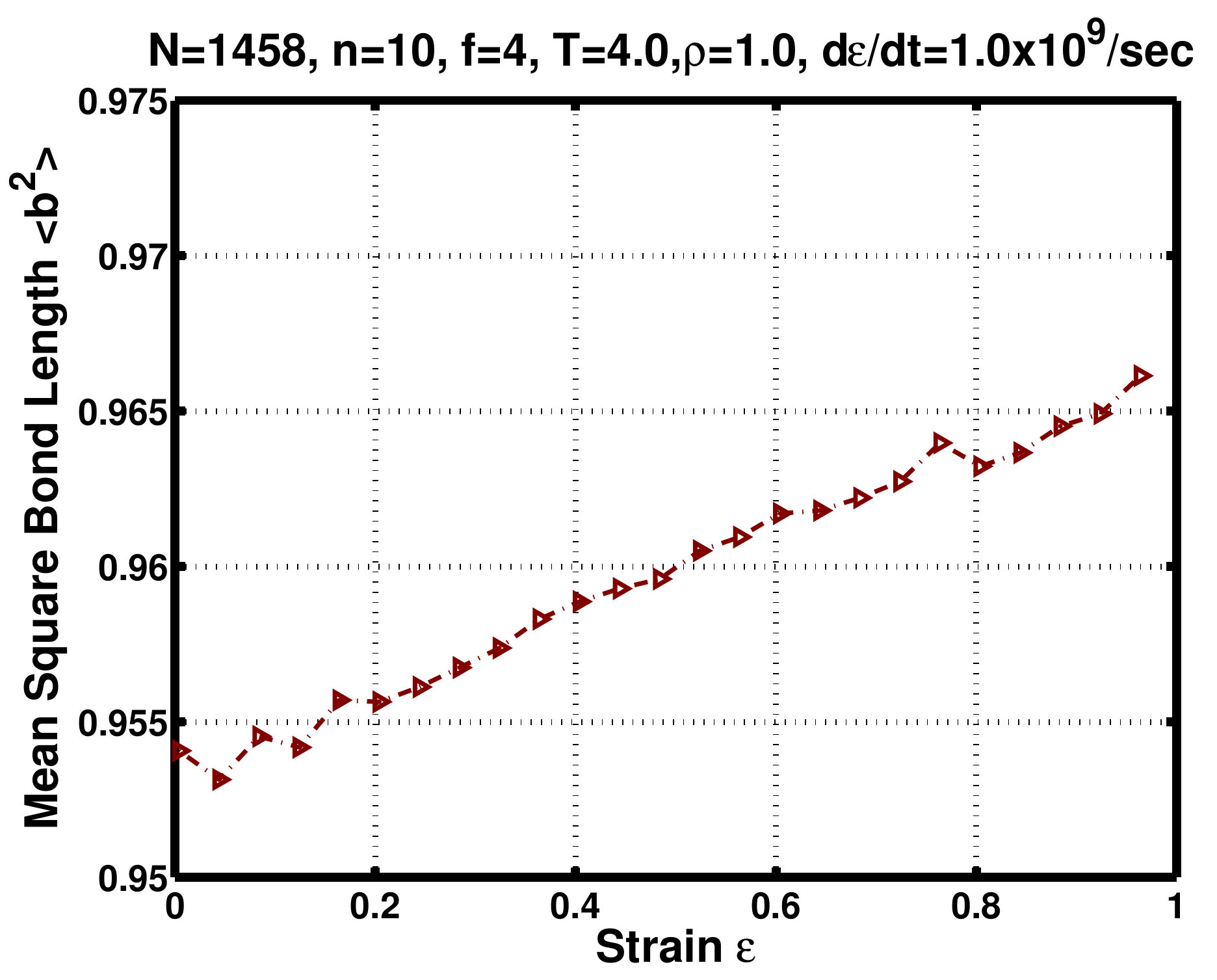}}
 \subfloat[Mean bond angle]{\label{fig:BondAngle}\includegraphics[width=0.5\textwidth]{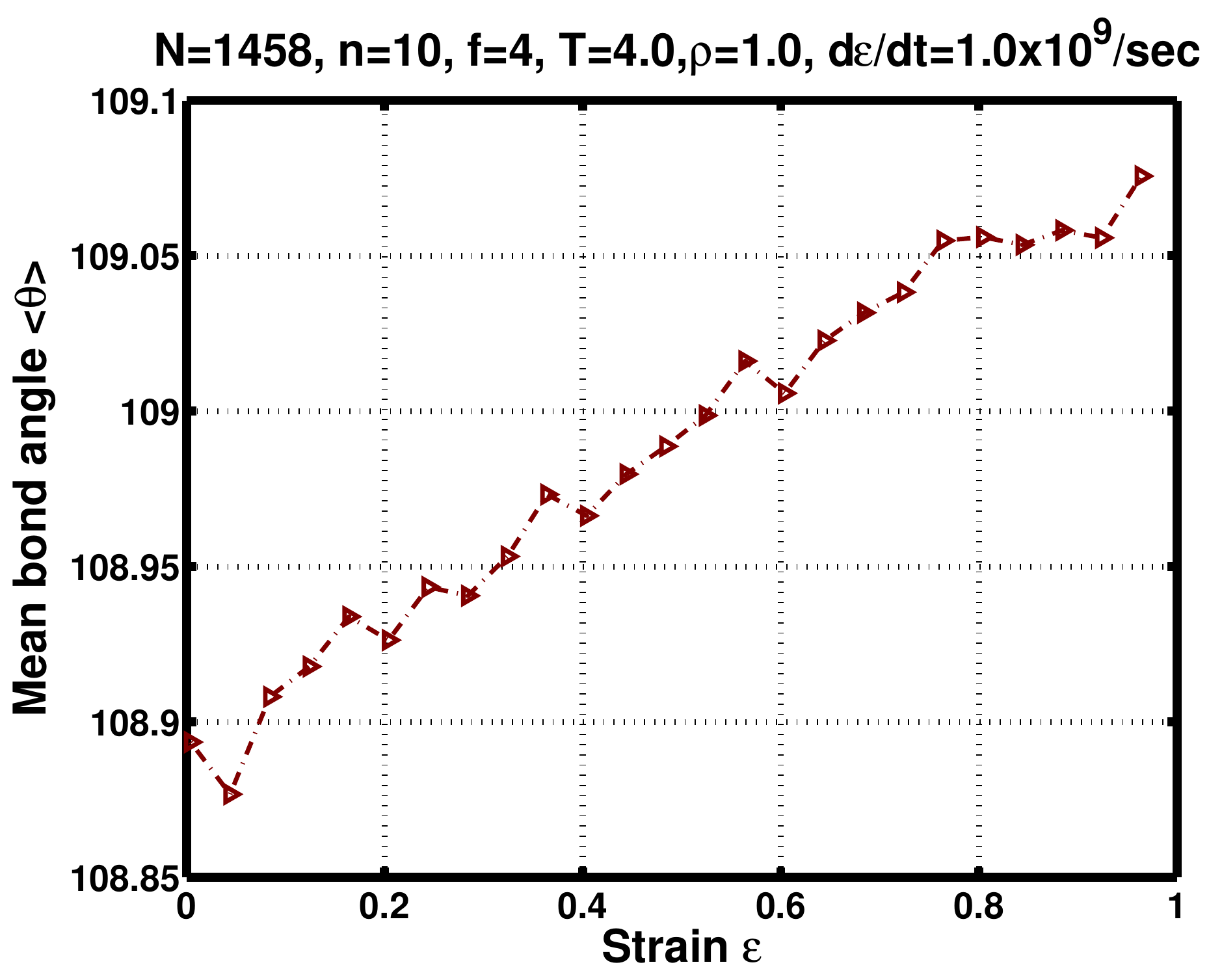}}\\
 \subfloat[End-to-end length]{\label{fig:EndToEnd}\includegraphics[width=0.5\textwidth]{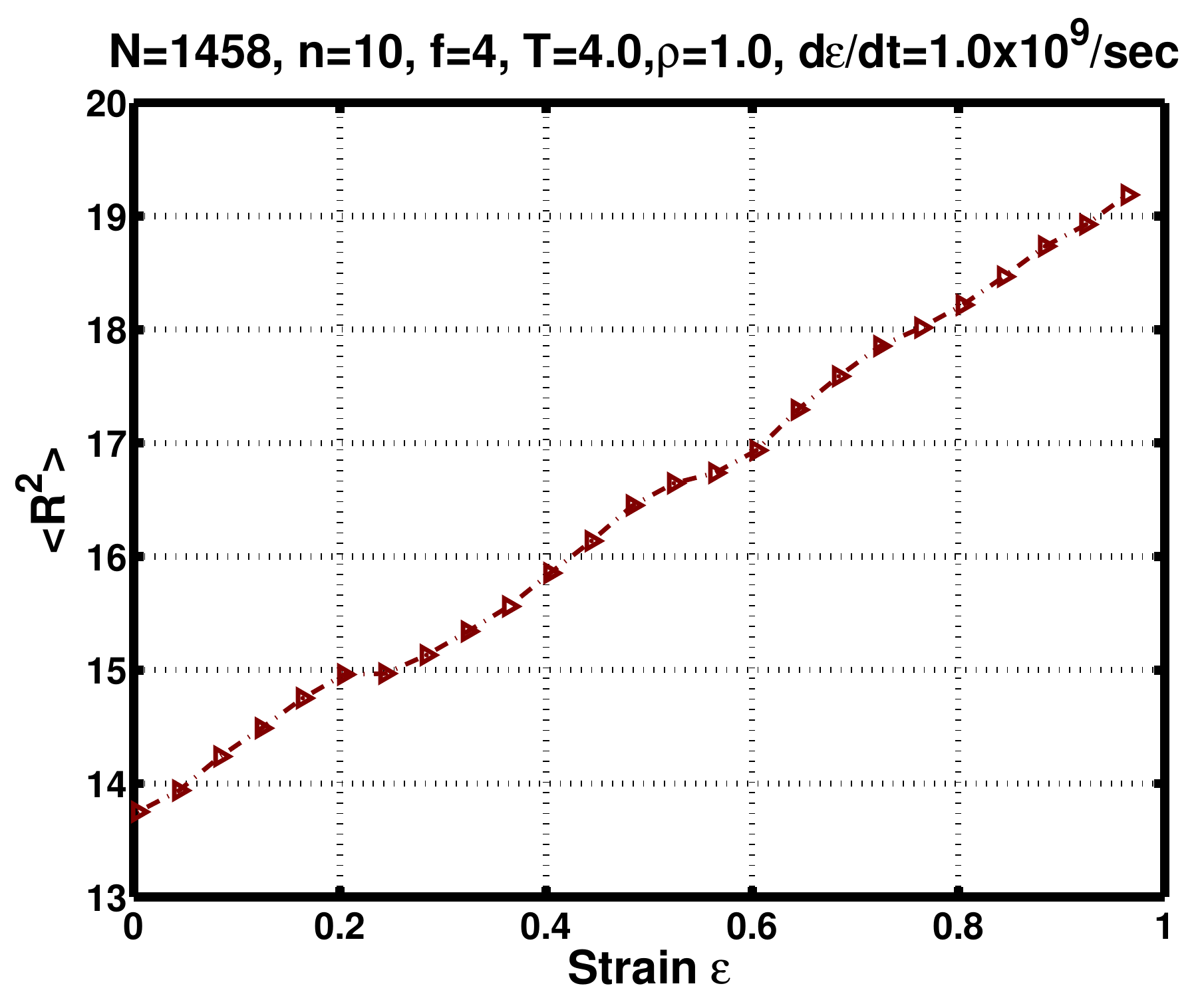}}
 \subfloat[Radius of gyration]{\label{fig:RadGyr}\includegraphics[width=0.5\textwidth]{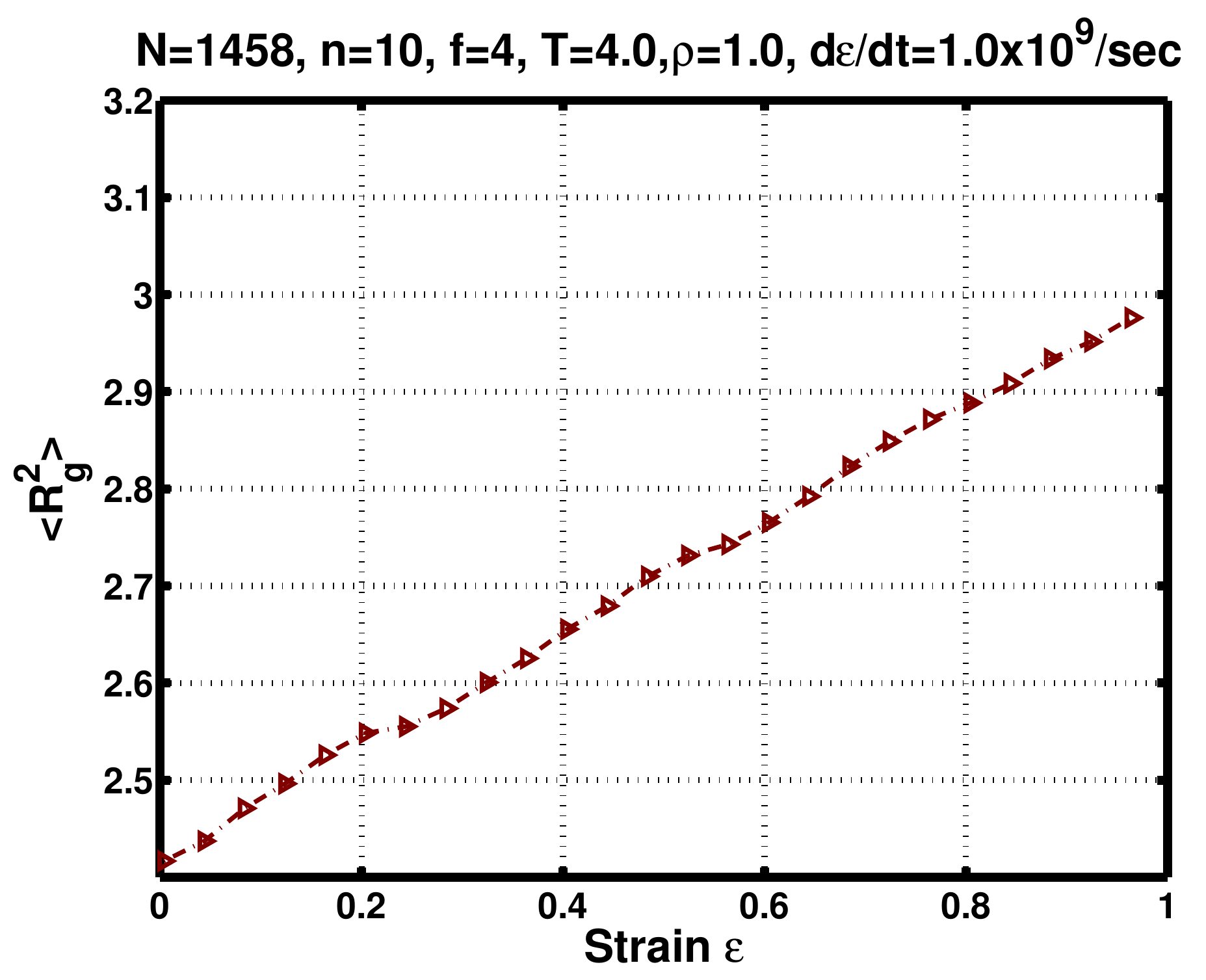}}\\
 \subfloat[Mass ratio]{\label{fig:MassRatio}\includegraphics[width=0.5\textwidth]{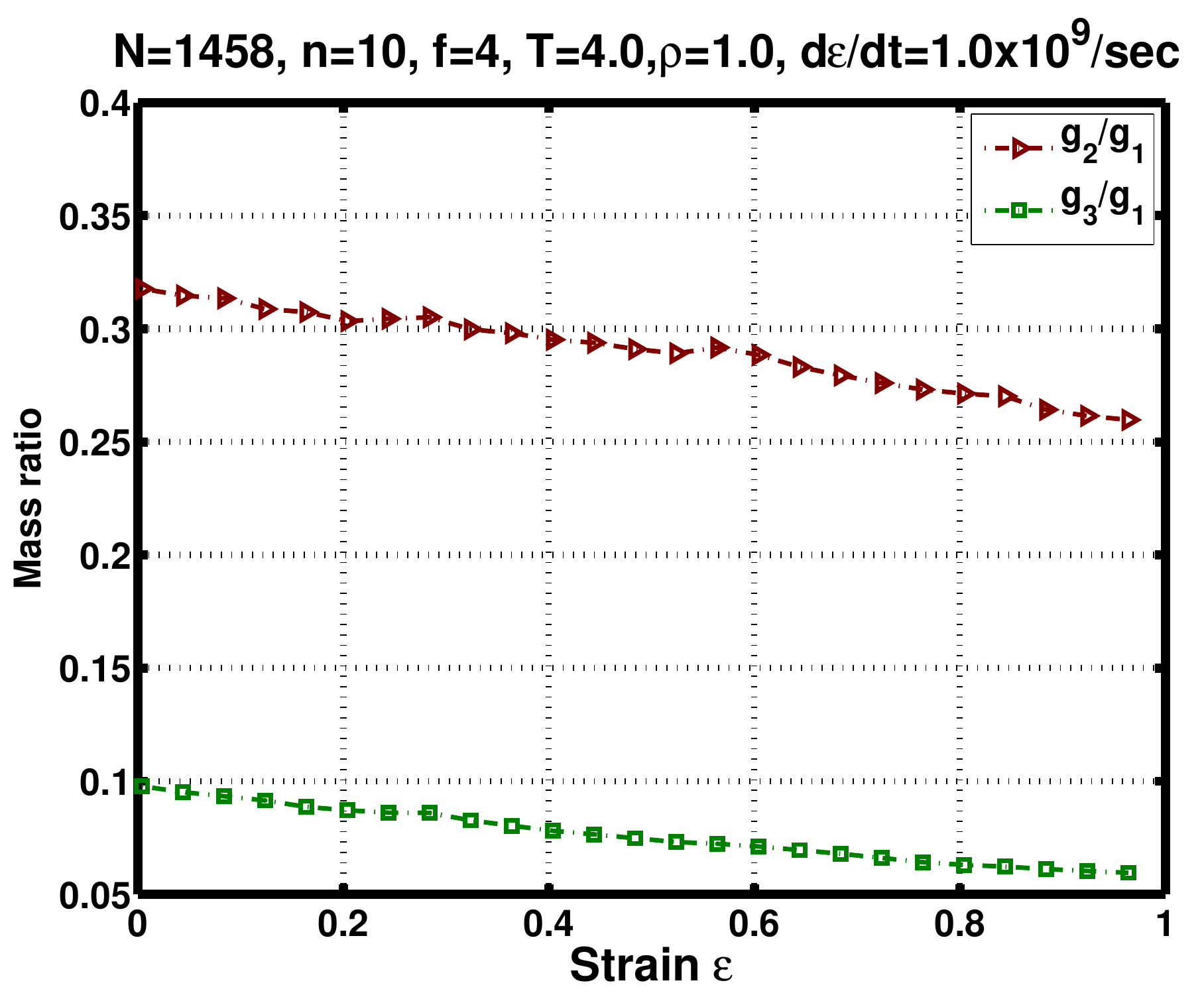}} 
  \subfloat[Chain angle]{\label{fig:ChainAngle}\includegraphics[width=0.5\textwidth]{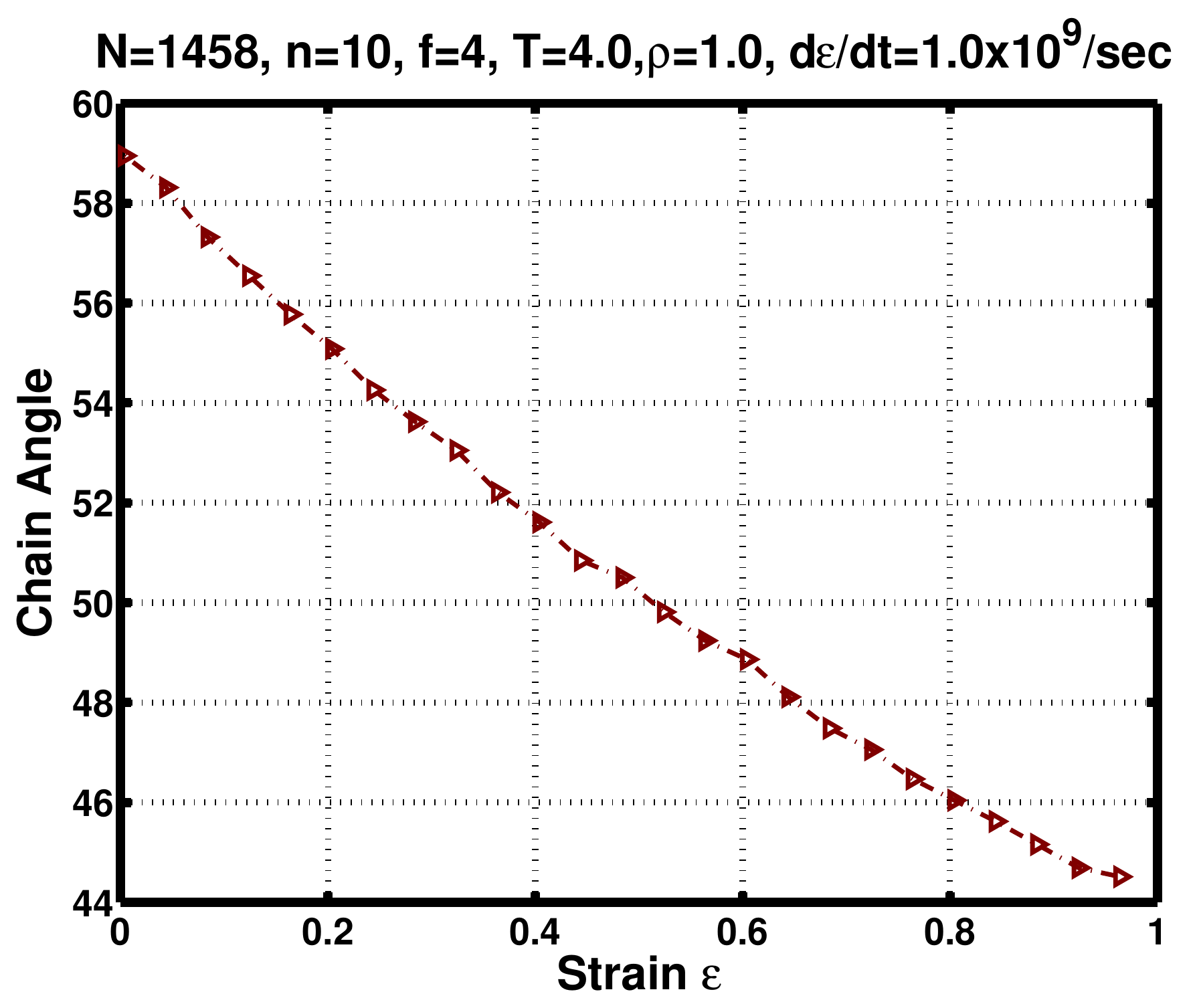}}
 \caption{Variation of structural properties for uniaxial constant strain rate loading}
 \label{fig:PropertiesA}
\end{figure}

The variation of the mean bond length with strain is shown in Figure~\ref{fig:MeanBondLenStrain}. The mean bond length increases almost linearly with strain. This affirms that bond stretching is the mechanism which contributes significantly to the stress. 

The mean bond angle varies almost linearly with strain as shown in Figure~\ref{fig:BondAngle}. Some fluctuations are observed from this linear variation. Also, it is to be noted that the overall change in the bond angle during the strain loading is only $0.2^\circ$ which is very small. Moreover, bond bending is not as stiff as bond stretch. Hence, even though there is contribution to anisotropic stress from it, it is not as significant as bond stretch. 

The variation of the mean-square end-to-end length of the chain and mean-square radius of gyration are shown in Figure~\ref{fig:EndToEnd} and Figure~\ref{fig:RadGyr}, respectively. These are also found to vary linearly with strain. A linear variation represents uniform uncoiling of the chains with strain. The reason for this linear variation is the short length of the polymer chains considered in the study and the presence of the cross-linkers. Note that in a very long chain polymer, the end-to-end length and radius of gyration variation will show a nonlinear increase with strain as shown in Section~\ref{subsec:control_para_chain_length} ---  Figure~\ref{fig:EndToEnd_ChainLen} and Figure~\ref{fig:RadGyr_ChainLen}. 

Chain alignment can be studied by observing two parameters, namely, mass ratios and the chain angle. The variation of the mass ratios with strain is shown in Figure~\ref{fig:MassRatio}. From this figure it can be observed that from the beginning, chains are not spherical as the mass ratios are very small compared to unity. Stretching of the system further reduces the mass ratios linearly which indicates that chains tend to take a one dimensional configuration with increase in strain. Very low values of $g_3/g_1$ indicates that chains take mostly a planar configuration. The chain angle decreases almost linearly with strain as shown in Figure~\ref{fig:ChainAngle}. This indicates that more and more number of end-to-end vectors of the chains align in the loading direction. 
 
\begin{figure}
 \centering
 \subfloat[Bond length]{\label{fig:LenBondLen}\includegraphics[width=0.5\textwidth]{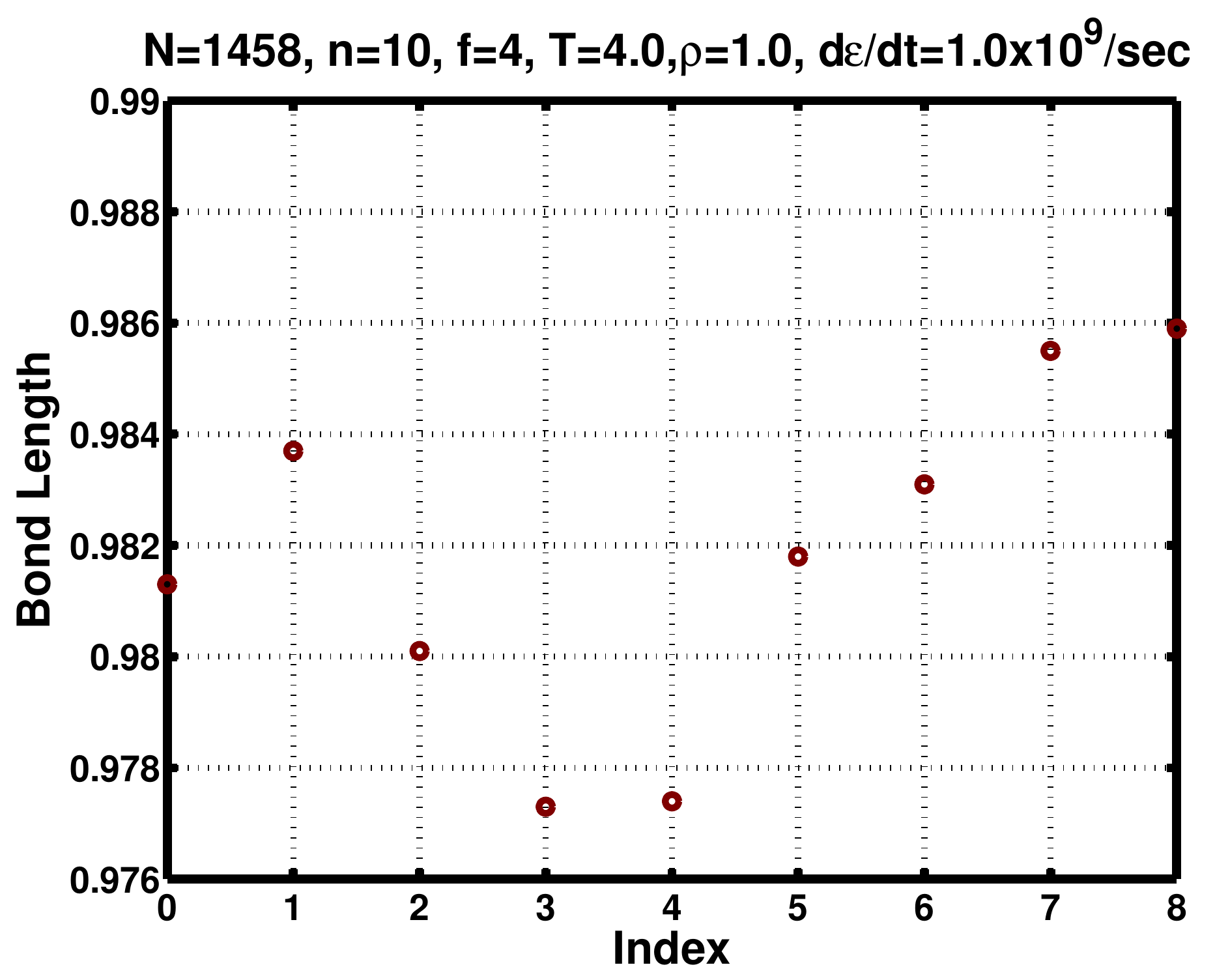}}
 \subfloat[Bond angle]{\label{fig:LenBondAng}\includegraphics[width=0.5\textwidth]{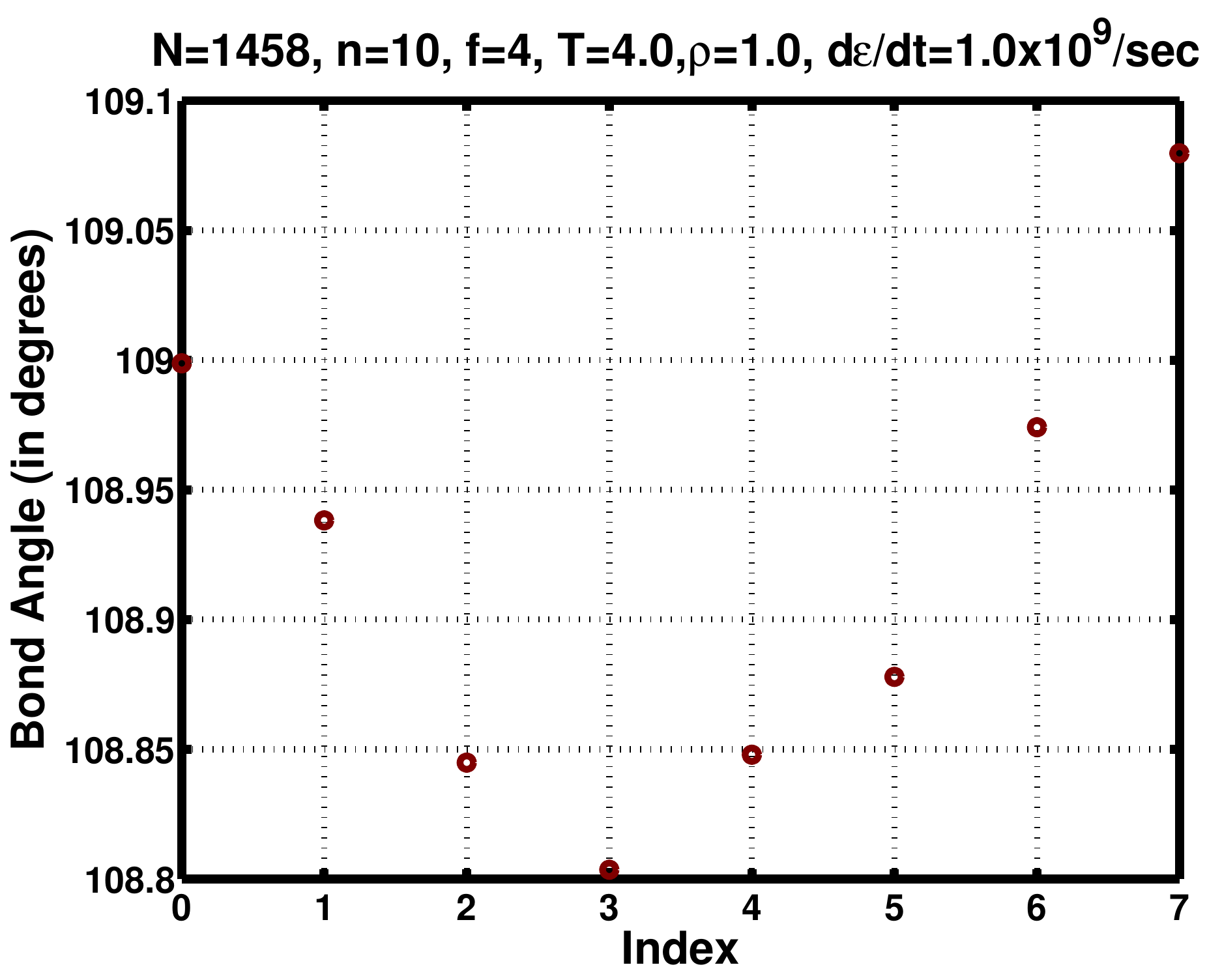}}\\
 \subfloat[Dihedral angle]{\label{fig:LenDihedAng}\includegraphics[width=0.5\textwidth]{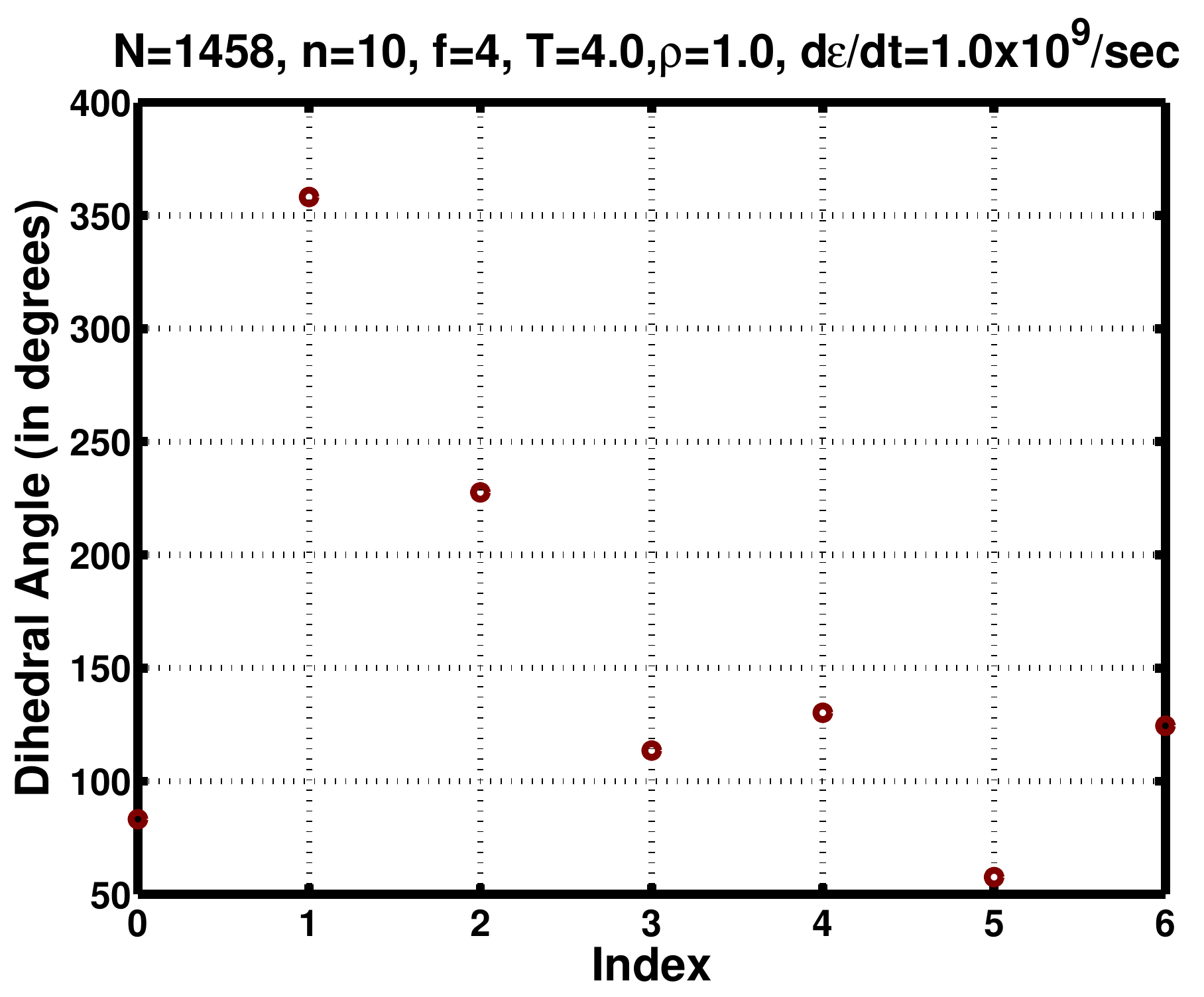}}
 \caption{Distribution of structural properties along the length of the chain for constant strain rate loading}
 \label{fig:LenDist}
\end{figure}

We also studied the variation of mean bond length, mean bond angle and dihedral angle along the length of the chain. The variation of the mean bond length along the length of the chain is shown in Figure~\ref{fig:LenBondLen}. Bond lengths at the ends of the chains are greater than the bond lengths at the middle of the chains. From one end, the bond length uniformly decreases towards the middle of the chain, and subsequently, in contrast, from the middle of the chain to the other end, it increases uniformly. This is because of the presence of the cross-linkers which try to pull the chains towards each other through the end atoms. A similar trend is also observed in the bond angle variation along the length of the chain as shown in Figure~\ref{fig:LenBondAng}. But in this case the variation is not uniform and large because the bending potential is not as stiff as bond stretching. In this case too, we attribute this type of behavior to the presence of cross-linkers The variation of the dihedral angle along the length of the chain is shown in Figure~\ref{fig:LenDihedAng}. We find that many of the dihedral angles are near $113^\circ$ and some of them are close to  $360^\circ$ and $247^\circ$ which are the equilibrium dihedral angles for the torsional potential used in this study.

\section{The dependence of stress and structure on control variables}
\label{sec:control_para_stress_strain_resp}

We now study the effect of control variables that are imposed on the system and their effect on the stress and structure of the polymeric system when it is subjected to dynamic strain. Some of the control parameters studied are the stoichiometry of the cross-linkers, temperature, density, strain rate, and chain length. We also study the effect of these parameters on the evolution of the structural properties and their correlation with the stress response since this should give insight into the observed stress-strain behavior of the cross-linked polymer.

\subsection{Effect of cross-linkers}
\label{subsec:control_para_cross-linkers}

The effect of cross-linkers is investigated by preparing the system with cross-linkers of different functionalities. Stoichiometric number of cross linkers --- $N_c=2N/f$ --- were added to the system. For this study, three functionalities of the cross-linkers, namely, $f=0,2,4$ were considered. Here $f=0$ represents no functionality of the cross-linker, hence, no cross-linker was added to the system and as such represents an exceptional case to the stoichiometric formula. A system prepared with $f=0$ is simply a system of linear polymer chains. In the other two cases, stoichiometric number of cross-linkers were added to the system.

\begin{figure}
\centering
\includegraphics[width=0.45\textwidth]{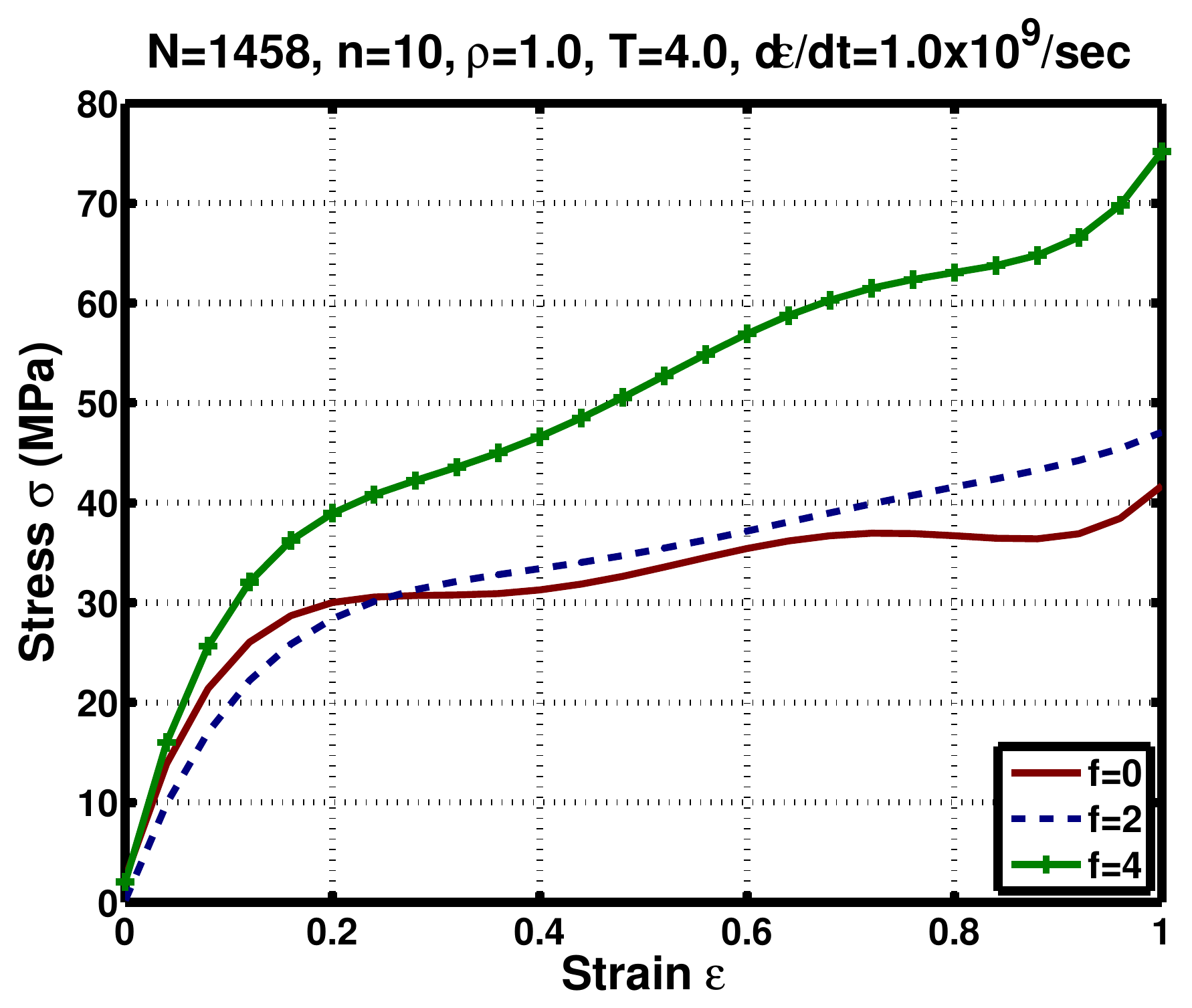}
\caption{Effect of functionality of cross-linker on stress response}
\label{fig:fit_stress_strain_cross}
\end{figure}

The stress response of the elastomeric system with different functionalities of the cross-linkers is shown in Figure~\ref{fig:fit_stress_strain_cross}. We observe that there is no significant difference between the response of the polymer at $f=0$ and $f=2$. Functionality $f=2$ implies that a cross-linker with this functionality will add two chains together. It also implies the cross-linking will result in a increase in the length of the polymer chain. There is no significant difference in stress for the cases $f=0$ and $f=2$, although at high strains the stress levels for  $f=2$ is marginally higher. Cross-linkers with functionality $f=4$ result in a highly cross-linked system. This puts an additional constraint on the motion of monomers. As a result, we note that the stress level increases significantly from the other two cases. In a study by \cite{Tsige2004b}, it was reported that failure stress increases with increasing functionality of the cross-linkers below $f_{\mathrm{av}}^\alpha < 4$. $f_{\mathrm{av}}$ is the average functionality of the cross-linkers as cross-linkers of different functionalities between $3$ to $6$ are added into the system. They also report that above $f_\mathrm{av}^\alpha = 4$, the failure mechanism itself changes, and hence the failure stress decreases with increasing functionality. Since we have varied the functionality of cross-linkers from $0$ to $4$, we observe an increase in the stress with increasing functionality of the cross-linkers.


\begin{figure}
 \centering
 \subfloat[Mean-square bond length]{\label{fig:MeanBondLenStrain_cross}\includegraphics[width=0.45\textwidth]{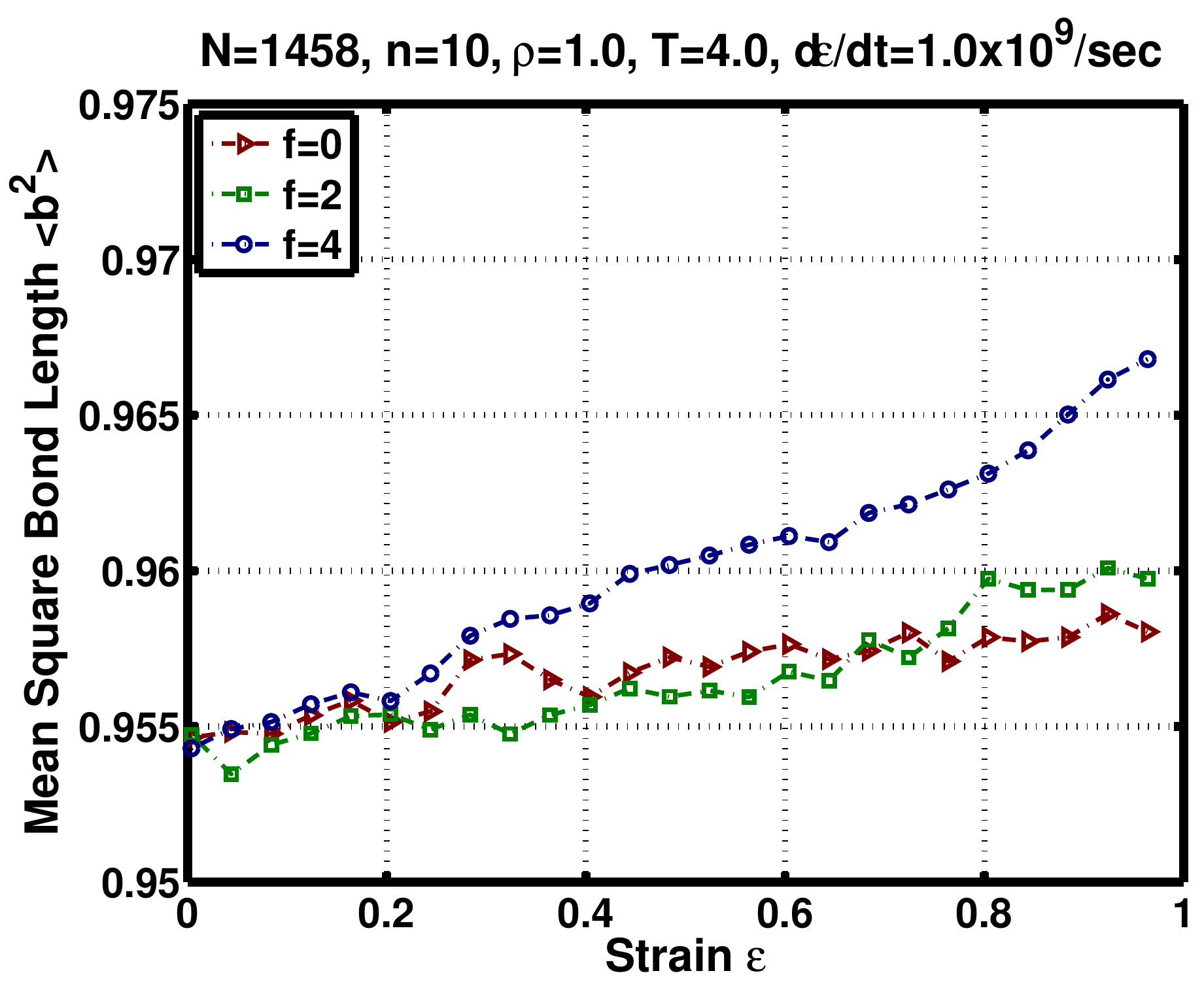}}
 \subfloat[Mean bond angle]{\label{fig:BondAngle_cross}\includegraphics[width=0.45\textwidth]{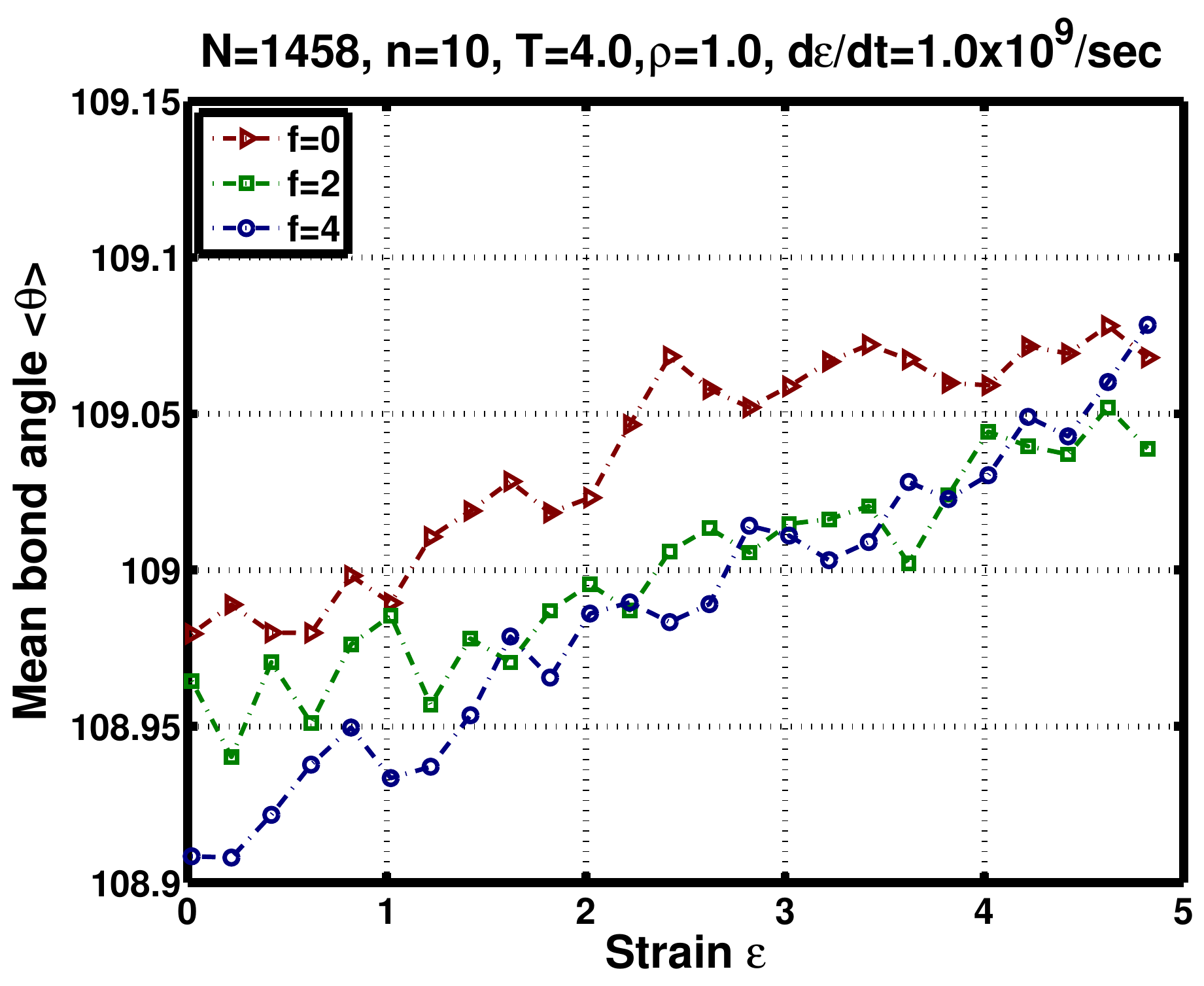}}\\
 \subfloat[End-to-end length]{\label{fig:EndToEnd_cross}\includegraphics[width=0.45\textwidth]{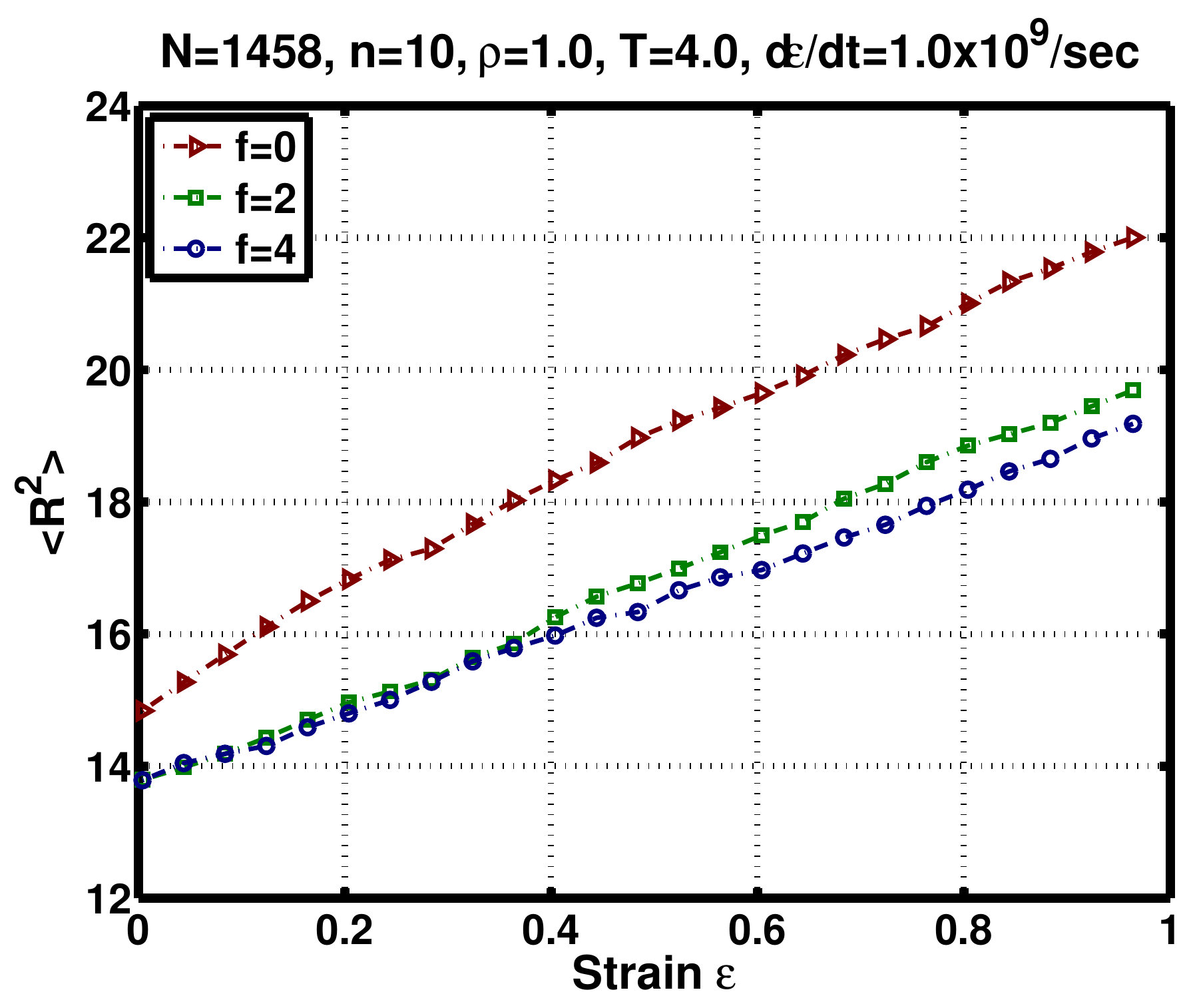}}
 \subfloat[Radius of gyration]{\label{fig:RadGyr_cross}\includegraphics[width=0.45\textwidth]{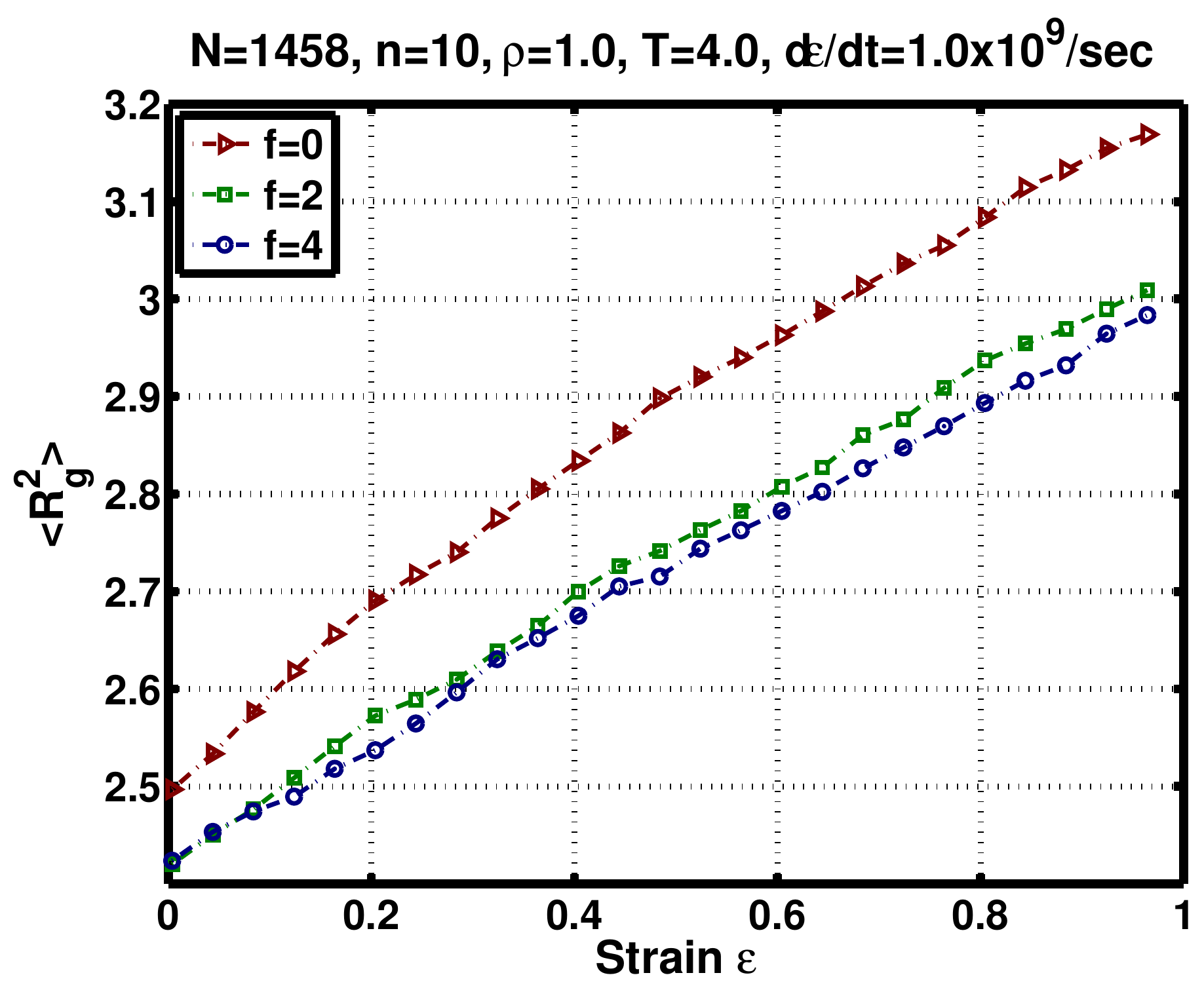}}
 \caption{Variation of structural properties for uniaxial constant strain rate loading: Effect of cross-linkers}
 \label{fig:Properties_cross}
\end{figure}

The variation of the micro-structural parameters of the elastomer with the functionality of the elastomer is correlated with the stress-strain behavior observed in Figure~\ref{fig:fit_stress_strain_cross} in order to gain an understanding of the stress response. The variation of the mean-square bond length with strain is shown in Figure~\ref{fig:MeanBondLenStrain_cross} for different functionalities of the cross-linkers. We observe that the mean-square bond length is significantly different at $f=4$ as compared to when $f=0$ and $f=2$. Mean-square bond length at $f=0$ and $f=2$ remains almost similar with very small variation. The variation of mean bond angle at different functionalities of cross-linkers is shown in Figure~\ref{fig:BondAngle_cross}. We note that with increasing functionality of the cross-linkers, the mean bond angle is lower at zero strain. During the axial stretch, the mean bond angle in all the cases tend to arrive at the same value, and therefore the change in mean bond angle is greater in the case of high functionalities of the cross-linkers. 
 
The effect of functionalities of the cross-linkers on the mean-square end-to-end length and mean-square radius of gyration is shown in Figure~\ref{fig:EndToEnd_cross} and \ref{fig:RadGyr_cross}, respectively. The mean-square end-to-end length and mean-square radius of gyration are distinctly higher when no cross-linker is present from the cases when cross-linkers are present. This indicates that the chains open more freely when the cross-linkers are absent --- an expected result. There is a slight difference in the values of these parameters when the functionality is increased from $f=2$ to $f=4$ though we observe slightly lower values at $f=4$ that is indicative of somewhat higher constraints on the motion of the monomers when functionality is increased. 

\begin{figure}
 \centering
 \subfloat[$g_2/g_1$]{\label{fig:MassRatiog2_cross}\includegraphics[width=0.45\textwidth]{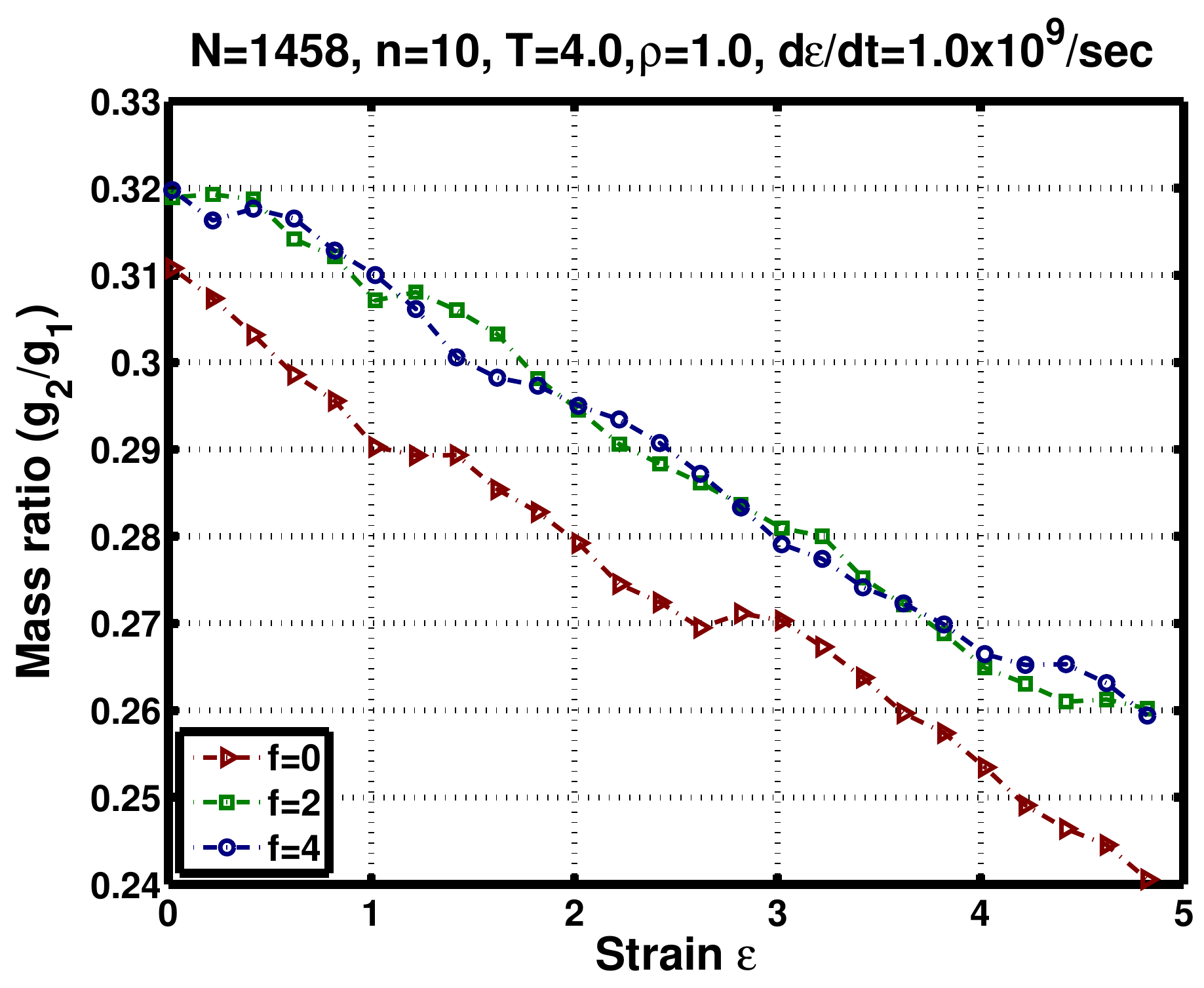}}
 \subfloat[$g_3/g_1$]{\label{fig:MassRatiog3_cross}\includegraphics[width=0.45\textwidth]{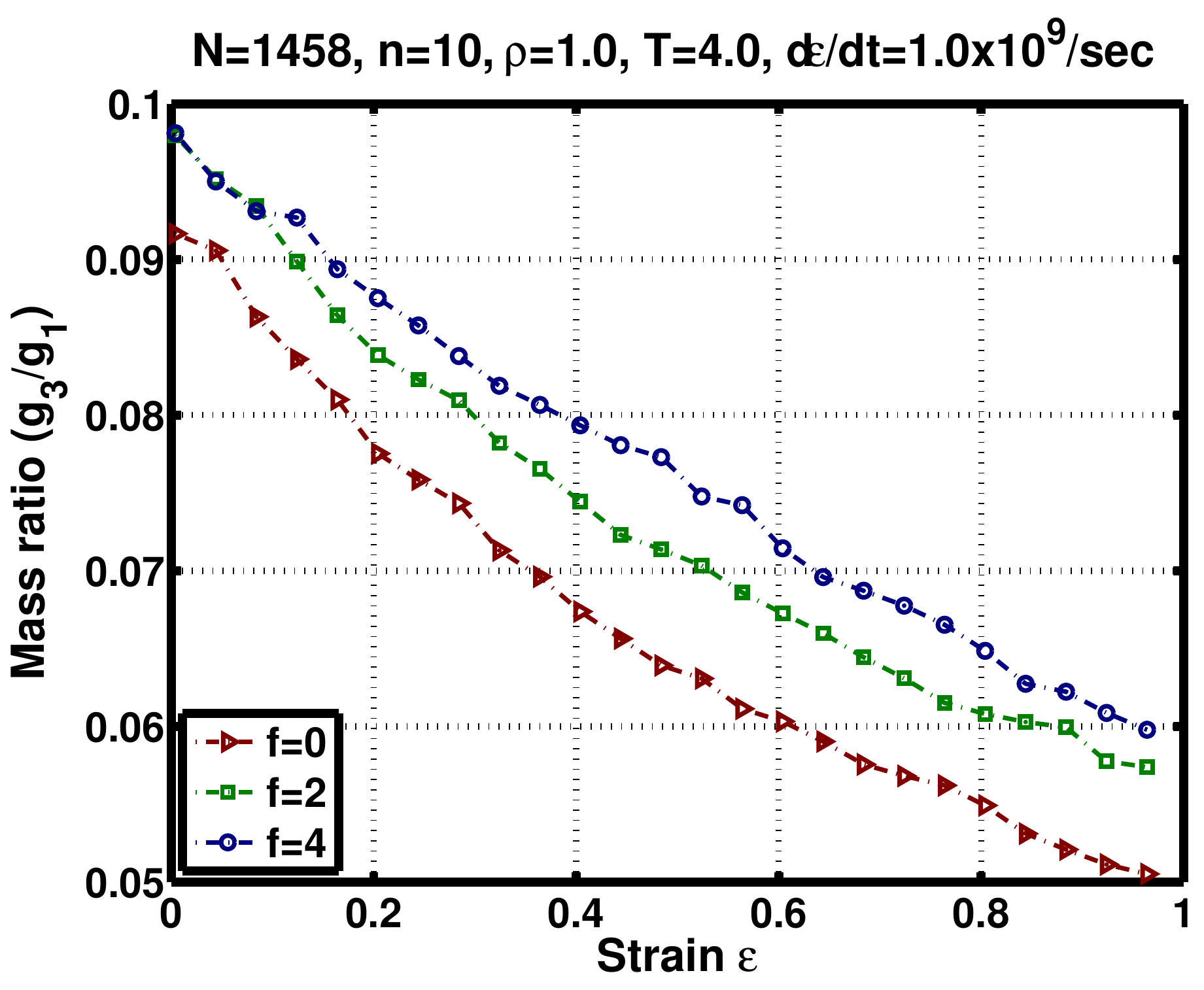}}
 \caption{Effect of cross-linkers on mass ratio}
 \label{fig:MassRatio_cross}
\end{figure}

The variation of mass ratios at various functionalities of cross-linkers is shown in Figure~\ref{fig:MassRatio_cross}. Both $g_2/g_1$ and $g_3/g_1$ are lower when there is no cross-linker indicating that the chains without cross-linkers are less spherical. As the functionality of the cross-linkers is increased we observe a more spherical structure of the chains. Under axial strain, the chains loose their sphericity in all the cases. However, the rate at which they lose their sphericity is lower with increase in the functionality of the cross-linkers. 

\begin{figure}
 \centering
 \includegraphics[width=0.5\textwidth]{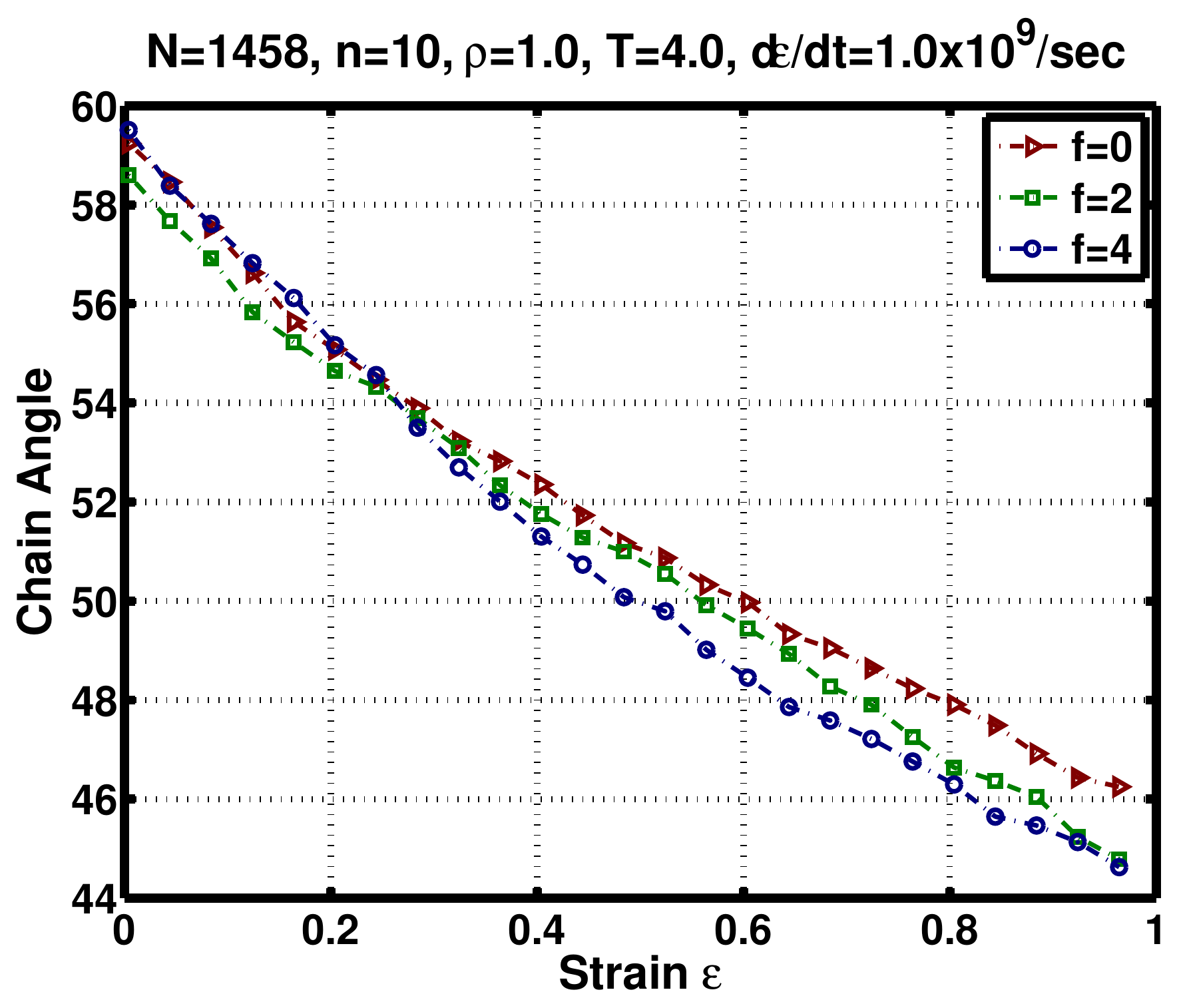}
 \caption{Effect of cross-linkers on chain angle}
 \label{fig:ChainAng_cross}
\end{figure}

The functionality of the cross-linkers have a very small effect on the chain angle as shown in Figure~\ref{fig:ChainAng_cross}. Note that chains with cross-linkers align in the loading direction slightly more as compared to those without cross-linkers. 

\begin{figure}
 \centering
 \subfloat[Bond length]{\label{fig:LenBondLen_cross}\includegraphics[width=0.45\textwidth]{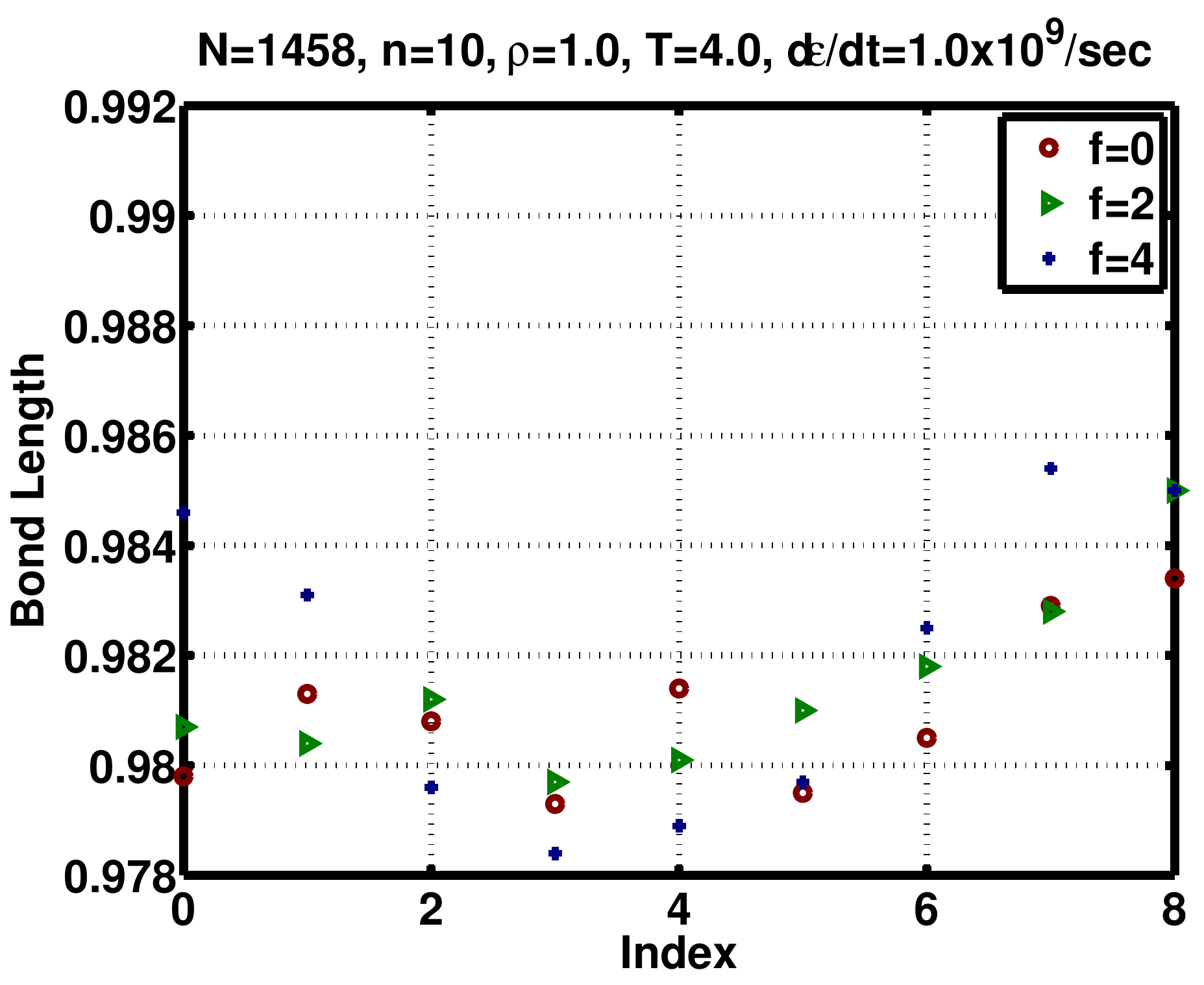}}
 \subfloat[Bond angle]{\label{fig:LenBondAng_cross}\includegraphics[width=0.45\textwidth]{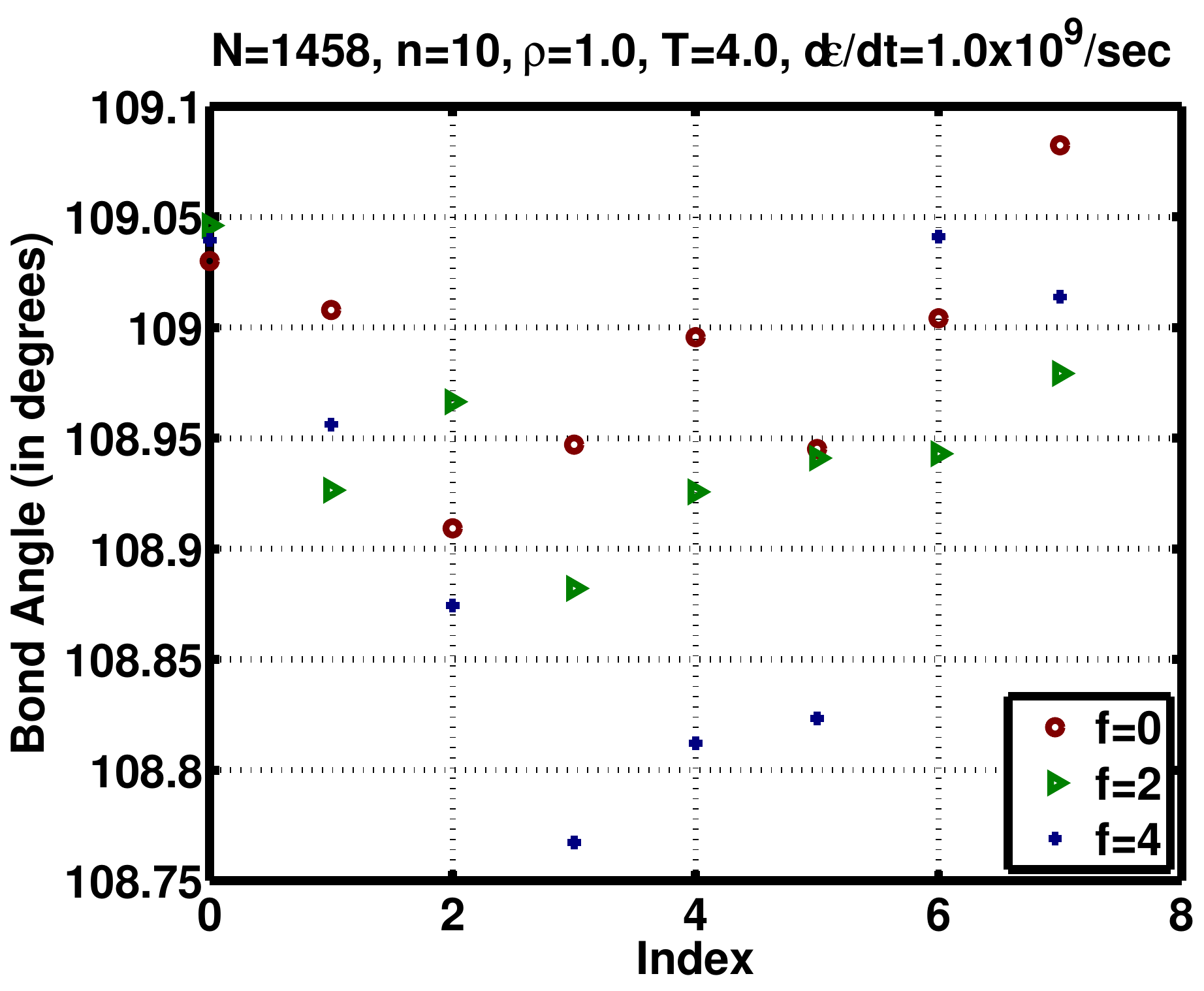}}\\
 \caption{Effect of functionality of cross-linkers: Distribution of structural properties along the length of the chain for constant strain rate loading}
 \label{fig:LenDist_cross}
\end{figure}

The variation of the mean bond length along the length of the chain is shown in Figure~\ref{fig:LenBondLen_cross}. We observe that bonds are stretched more, in the presence of cross-linkers, towards the end of the chains. When the cross-linkers are not present, $f=0$, and at $f=2$, one observes that bond length variation along the chains is almost uniform with a very small increase towards the end of the chains. At $f=4$, though, when the chains are highly cross-linked, the increase in bond length towards the end of the chain is quite significant. At $f=4$, as we go along the chain, the mean bond length decreases first at the middle and subsequently starts increasing as we move towards the other end. 

The effect of cross-linkers on mean bond angle along the length of the chain is shown in Figure~\ref{fig:LenBondAng_cross}. Note that chains are more inclined towards the end. At $f=0$ and $f=2$, though, this effect is very mild, whereas at $f=4$, we observe a relatively large variation in the mean bond angle along the length of the chain with chain angles dipping down in the middle.

 Mean bond length and mean bond angle at equilibrium before the application of strain are $b = 0.89$ and $\theta_0 = 108.86^\circ$. From the above two plots we see that throughout, along the length of the chain, bonds are stretched. But the bond stretching in the middle of the chain is less as compared to that at the end of the chain. This additional stretching at the end is because of the constraints due to cross-linkers. Different chain ends which were earlier free to move are now connected with each other through a cross-linker. As a result they start affecting the motion of each other and give additional forces to the bonds near the chain ends. This is the reason for higher bond lengths as we move towards the end of the chain. Since the chains are short --- $n=10$ --- the effect of the end motion gets diffused along the length of the chain and hence instead of an abrupt change in the bond length at the end, there is a uniform increase in the bond length from the middle of the chain towards either end. In the case of bond angles, we have similar trends, but without cross-linkers and with cross-linker of functionality $f=2$, we have a more uniform increase in the bond angle along the length of the chain. At $f=4$ there is large variation in the bond angles. Also, bond angles at the middle of the chain are even smaller than those at $f = 0$ and $f=2$. These angles are very close to the equilibrium bond angle indicating a very small change in the middle of the chain. This indicates that the effect of the strain in low functionality cross-linked network is reflected uniformly in all the bonds along the length of the chain, where-as at high functionality, the deformation is not uniform and is rather dominated by few bond angles participating in the deformation. At $f=4$, we also notice abrupt change in the bond angles. Hence, we can also conclude that the effect of cross-linker along the length of the chain is less diffusive in bond angle as compared to the bond length.

\subsection{Effect of temperature}
\label{subsec:control_para_temperature}

\begin{figure}
  \centering
  \subfloat[Stress-strain curve]{\label{fig:fit_stress_strain_temp}\includegraphics[width=0.5\textwidth]{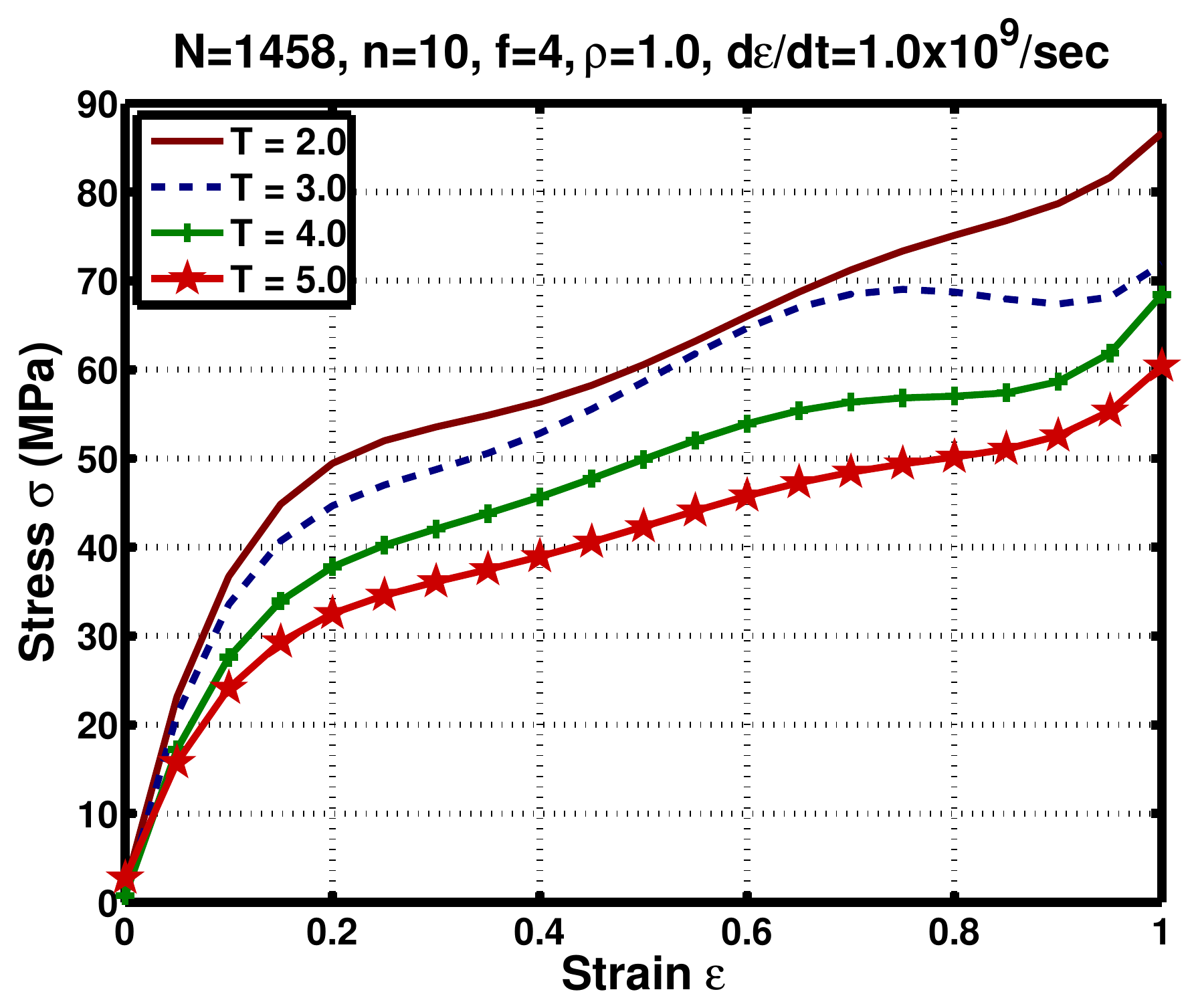}}
  \subfloat[Modulus vs. strain]{\label{fig:modulus_temp}\includegraphics[width=0.5\textwidth]{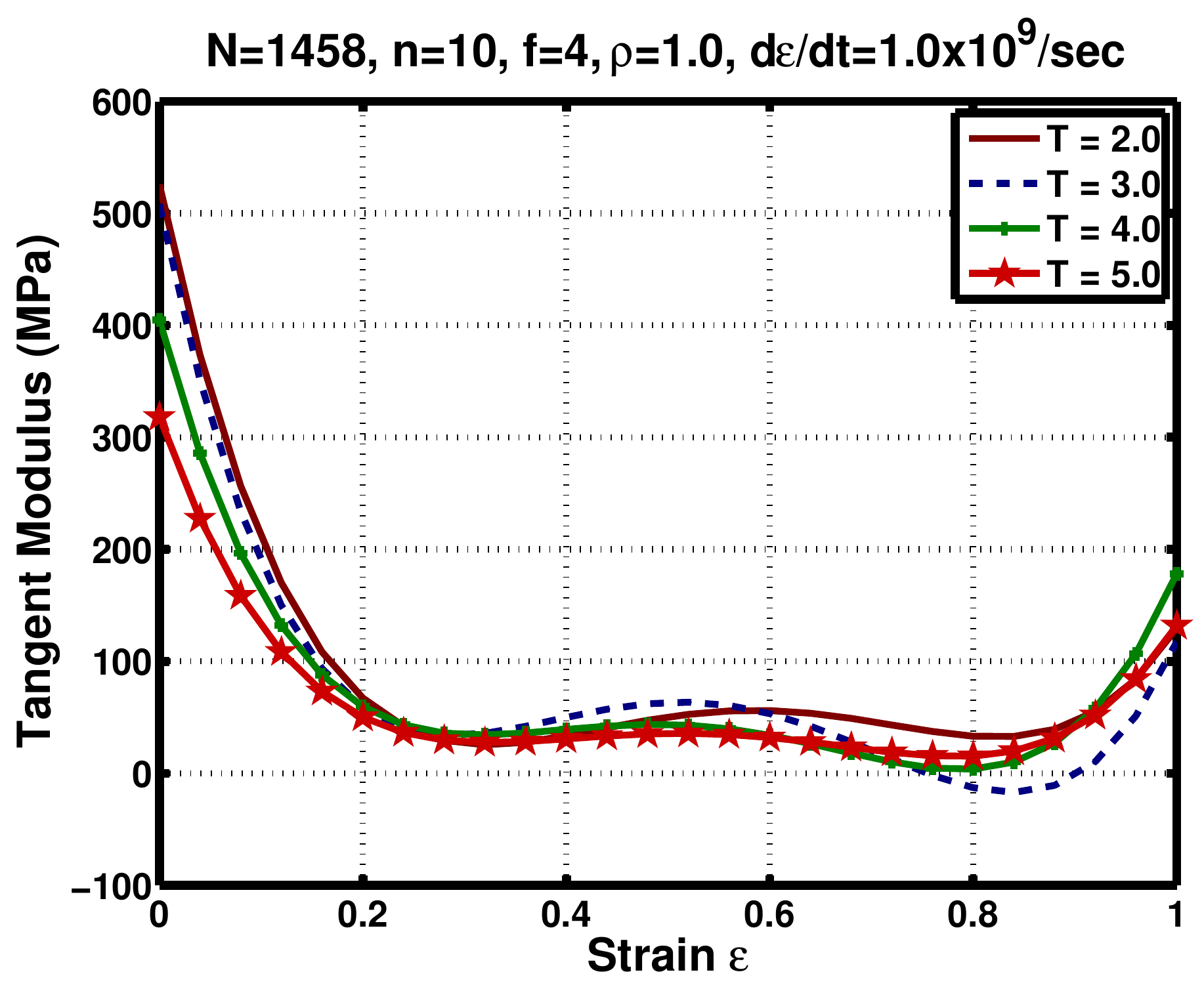}}
  \caption{Effect of temperature on stress response}
  \label{fig:StressStrainMod_temp}
\end{figure}

\begin{figure}
 \centering
 \subfloat[Mean-square bond length]{\label{fig:MeanBondLenStrain_temp}\includegraphics[width=0.45\textwidth]{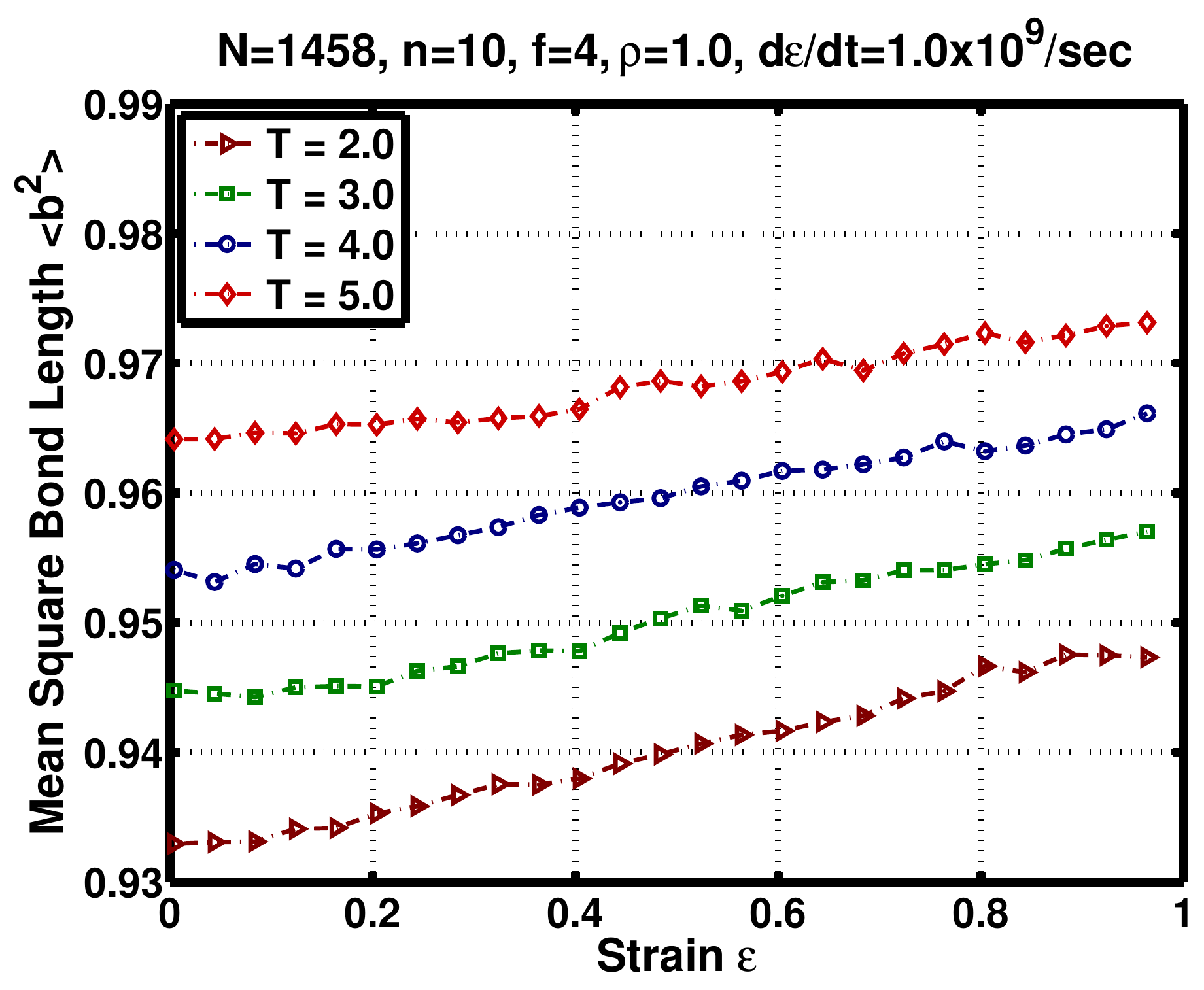}}
 \subfloat[Mean bond angle]{\label{fig:BondAngle_temp}\includegraphics[width=0.45\textwidth]{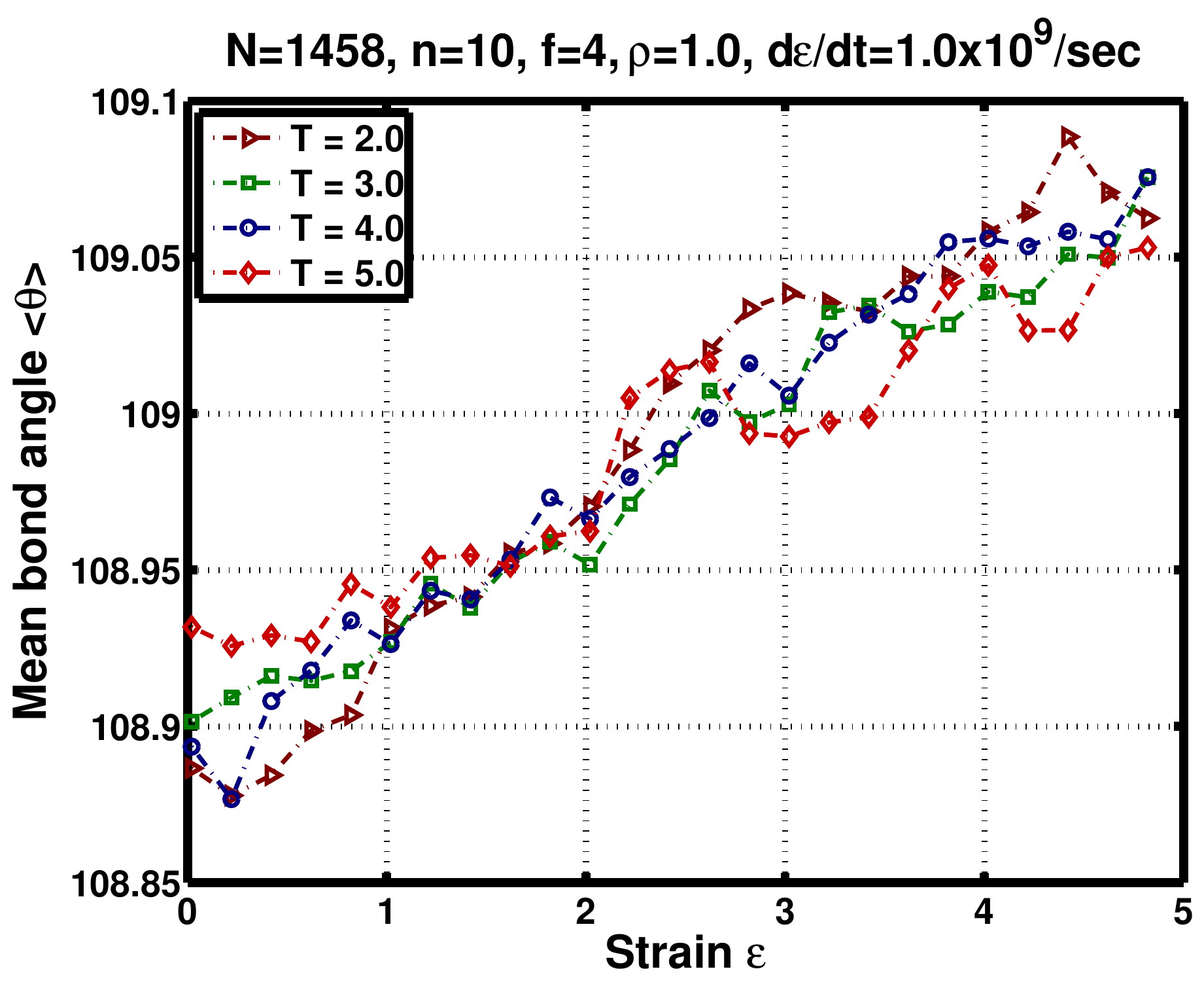}}\\
 \subfloat[End-to-end length]{\label{fig:EndToEnd_temp}\includegraphics[width=0.45\textwidth]{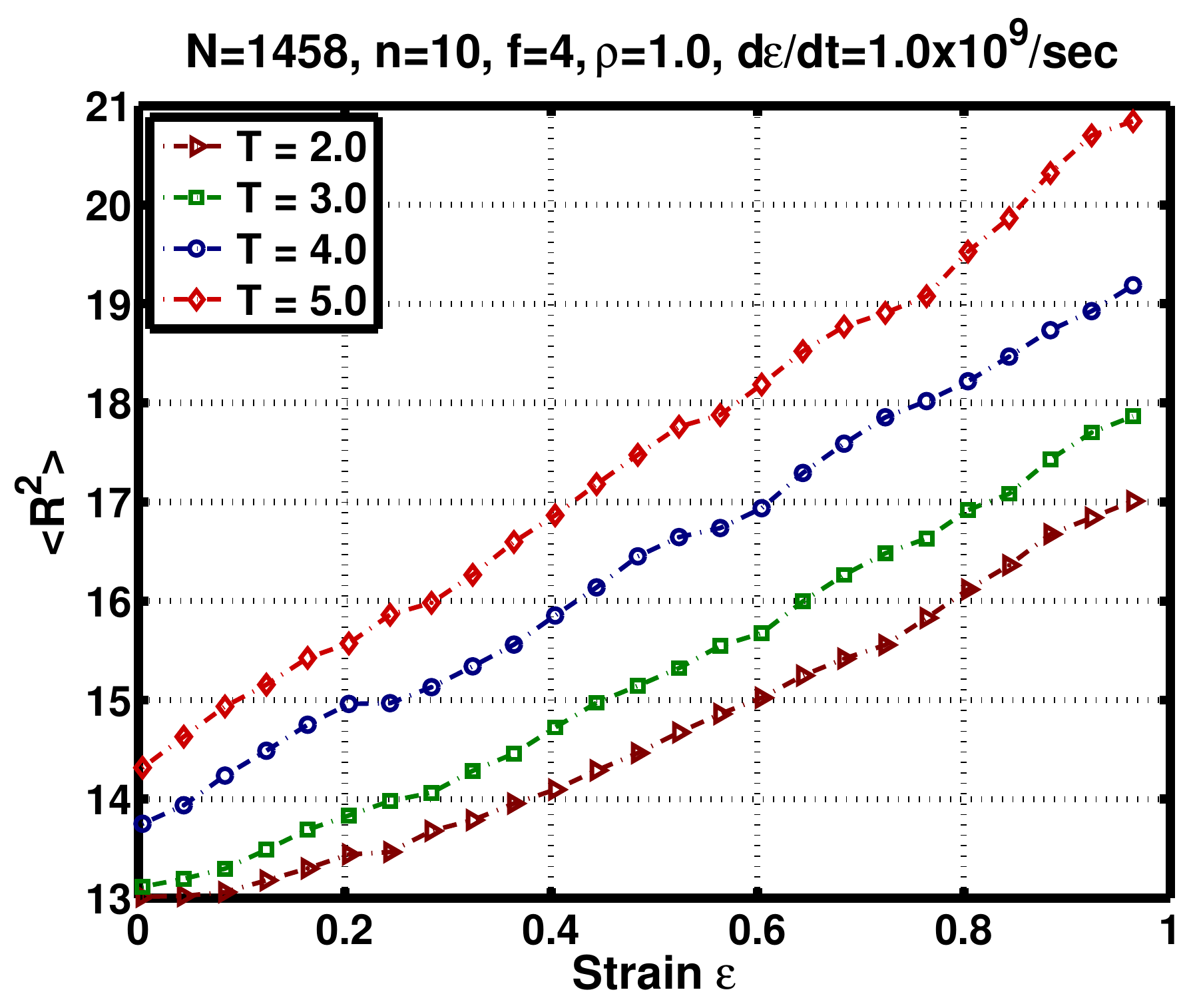}}
 \subfloat[Radius of gyration]{\label{fig:RadGyr_temp}\includegraphics[width=0.45\textwidth]{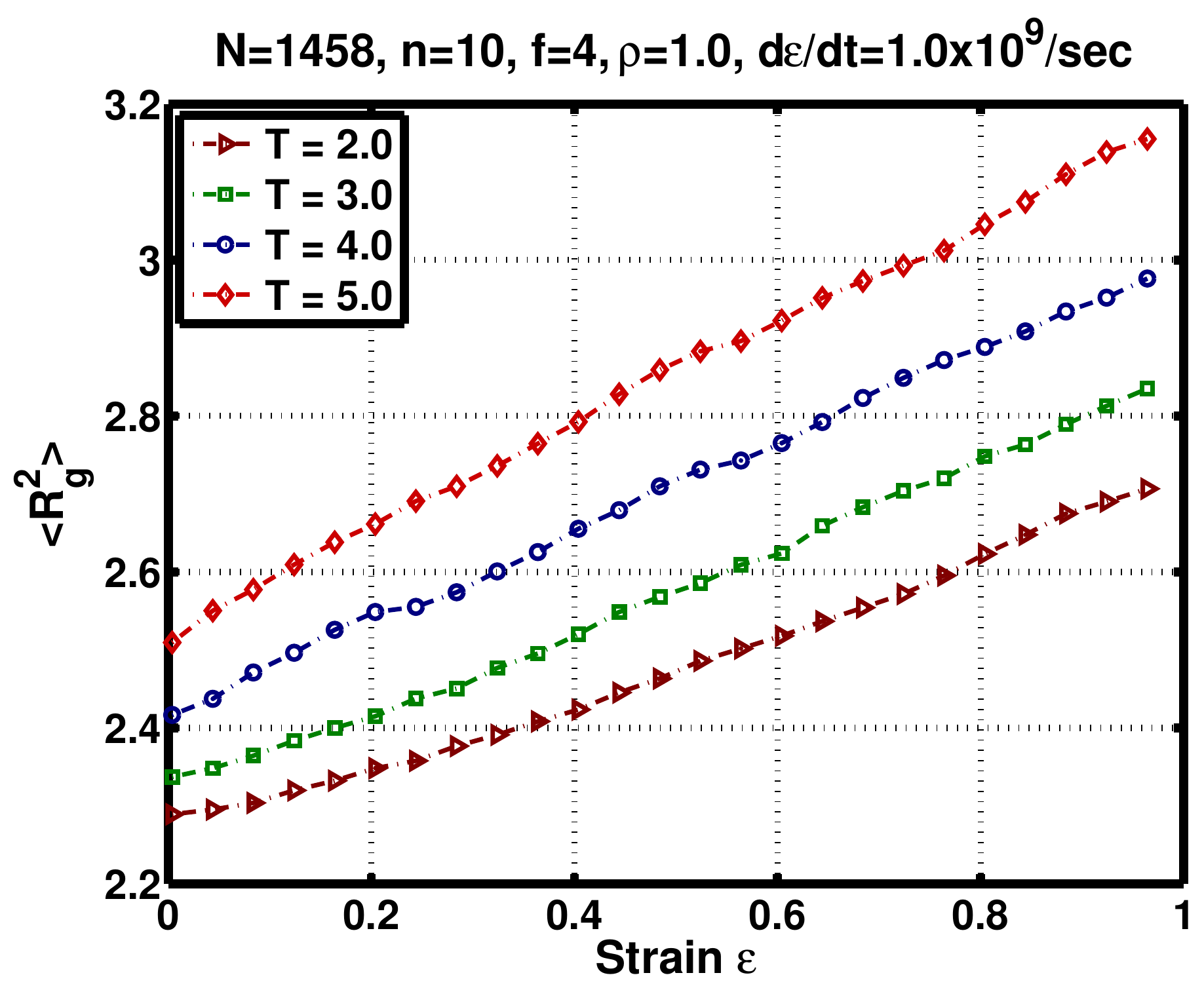}}
 \caption{Variation of structural properties for uniaxial constant strain rate loading: Effect of temperature}
 \label{fig:Properties_temp}
\end{figure}

Th stress response of the system at various temperatures is shown in Figure~\ref{fig:fit_stress_strain_temp}. As we increase the temperature, the stress induced in the system reduces significantly. At high temperatures the system relaxes faster, but we also observe longer bond lengths at higher temperatures, as shown in Figure \ref{fig:MeanBondLenStrain_temp}. Longer bond lengths should typically increase the stress level in the polymer. But what is observed here is a decrease in the stress that is contrary to our expectation. The resultant stress is a contribution of bond stretching, bond bending, bond twisting as well as excluded volume interaction. At higher temperatures, excluded volume interaction is less because of the better relaxed structure. Hence, as a net effect we find reduction in the stresses at higher temperatures. In a Monte-Carlo study, \citet{Chui} report a reduction in the induced stress at higher temperatures. They also note that strain softening is due to intermolecular contribution whereas strain hardening is due to intramolecular interactions. A plot of the tangent modulus, obtained by computing the derivative of the stress response with respect to strain, is shown in Figure~\ref{fig:modulus_temp}. From this figure, we observe that at values of strain below $\epsilon = 0.2$, the value of the modulus is low at higher temperatures. Thereafter, the modulus is very close to zero for almost all the temperatures that we have simulated. It increases by a very small amount at strains very close to $\epsilon=1$. The increase in the modulus towards the end of the loading could be because of entanglements as well as alignment of the chains in the loading direction.  

We now try and correlate the stress response with the evolution in time of the various structural parameters characterizing the deformation of the chains. Figure~\ref{fig:MeanBondLenStrain_temp} shows the variation of the mean-square bond length with strain at different temperatures. As expected, the mean-square bond length is greater at higher temperatures. At high temperatures, the kinetic energy of the atoms increases in comparison to the potential energy associated with the bond between them and hence the atoms vibrate with a larger amplitude at higher temperatures. Also, note that the mean-square bond length increases almost linearly with strain at all temperatures. 

The bond angle variation at different temperatures is shown in Figure~\ref{fig:BondAngle_temp}. We observe that there is no significant difference in the bond angle at different temperatures. Further, the bond angle also varies linearly with strain at all temperatures. Figure~\ref{fig:EndToEnd_temp} describes the variation of the mean-square end-to-end length with strain. Higher the temperature greater the mean-square end-to-end length. At higher temperatures, the atoms have high kinetic energy which helps them to cross some of the potential energy barriers which otherwise might not be possible. As a result, the mean-square end-to-end length is larger at higher temperatures. This also implies that higher temperatures help in uncoiling of the chains. A similar behavior is also observed in the variation of the mean-square radius of gyration, as shown in Figure~\ref{fig:RadGyr_temp}. The cause of this behavior is the same argument given above for the behavior of the mean-square end-to-end length. Note that, once again, the mean-square end-to-end length and the radius of gyration vary almost linearly with strain at all temperatures. 
 
\begin{figure}
 \centering
 \subfloat[$g_2/g_1$]{\label{fig:MassRatiog2_temp}\includegraphics[width=0.45\textwidth]{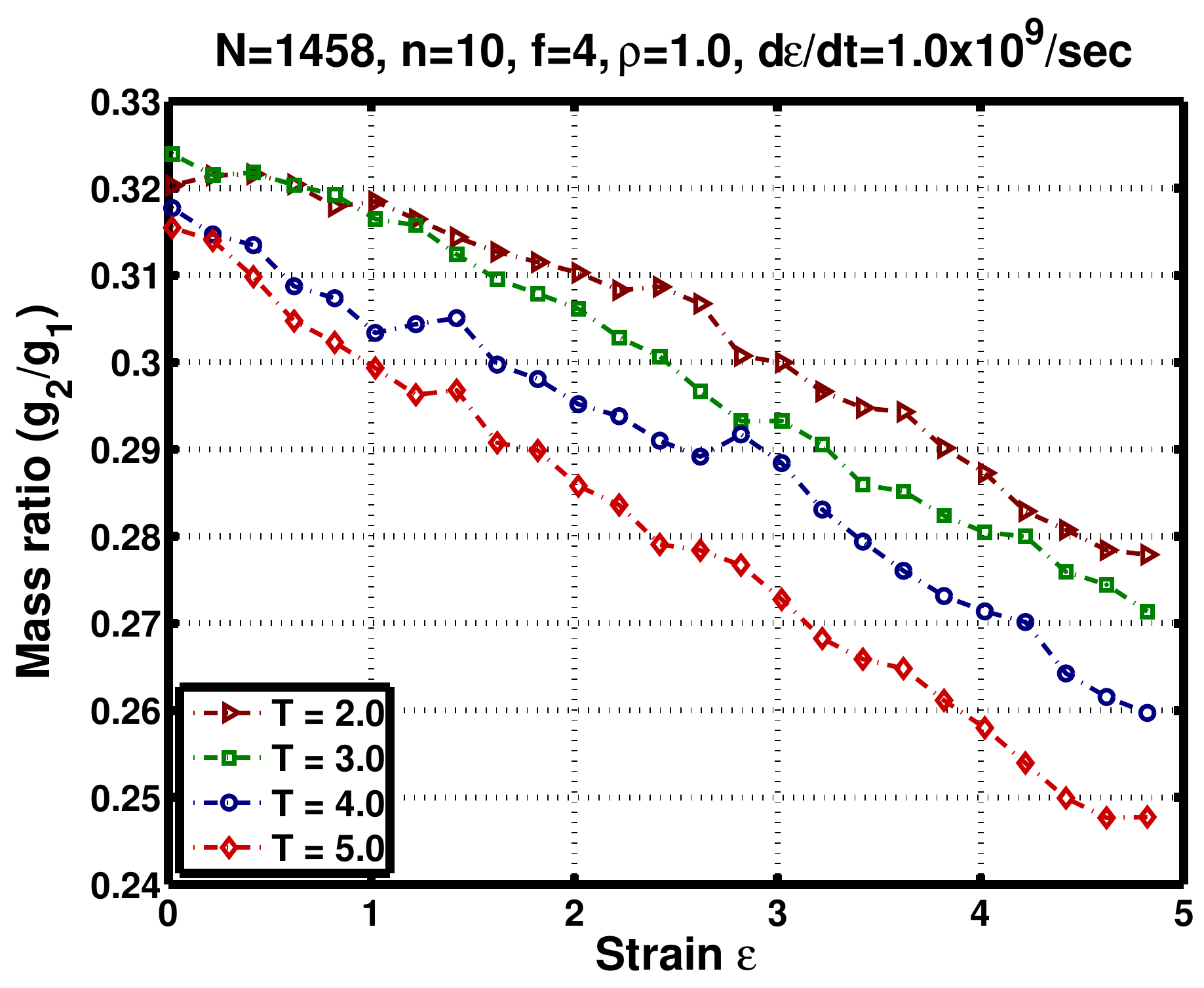}}
 \subfloat[$g_3/g_1$]{\label{fig:MassRatiog3_temp}\includegraphics[width=0.45\textwidth]{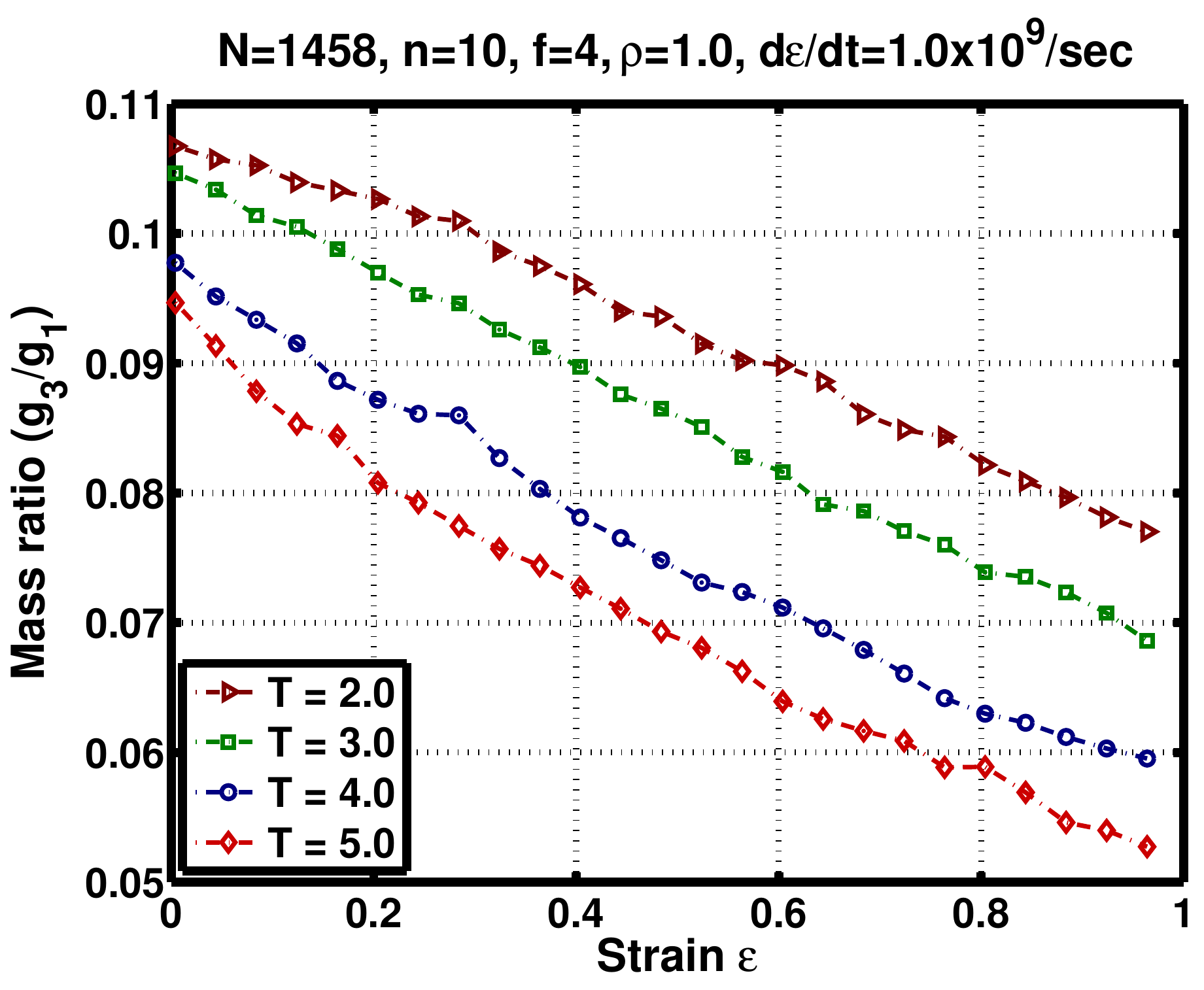}} 
 \caption{Effect of temperature on mass ratios}
 \label{fig:MassRatios_temp}
\end{figure}

The variation in the mass ratios at different temperatures with strain is shown in Figure~\ref{fig:MassRatios_temp}. At higher temperatures, the mass ratios decrease as the system is strained. This is  indicative of the fact that chains tend to take a one-dimensional configuration under uniaxial tension, and this effect is enhanced at higher temperatures. The values of the mass ratios at $\epsilon=1.0$ indicate that chains have almost taken a planar configuration as the contribution of $g_3$ is very low. 

\begin{figure}
 \centering
 \includegraphics[width=0.45\textwidth]{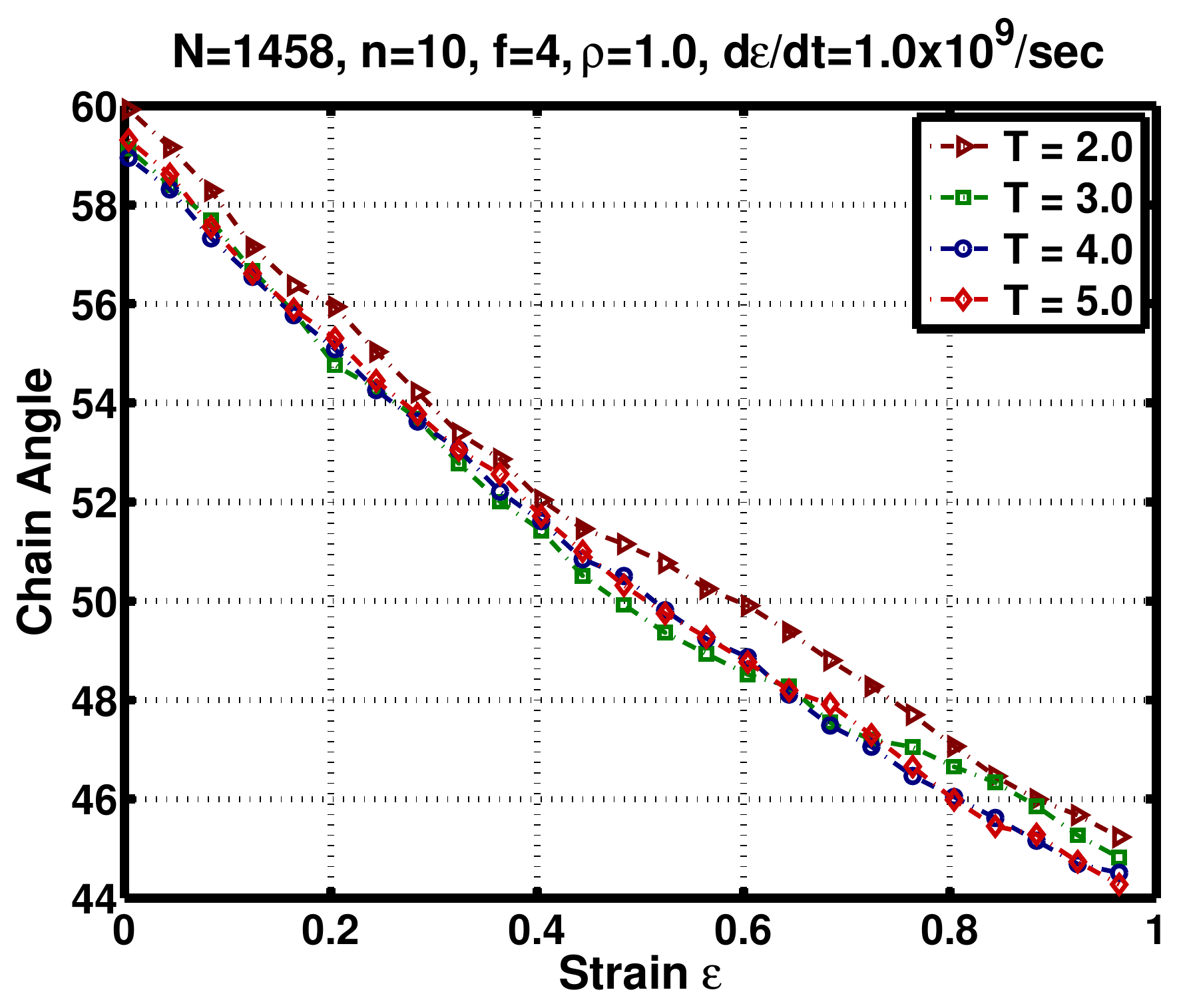}
 \caption{Effect of temperature on chain angle}
 \label{fig:ChainAng_temp}
\end{figure}

The effect of temperature on the chain angle is shown in Figure~\ref{fig:ChainAng_temp}. From this figure we observe that temperature has almost no effect on the alignment of the chains in the loading direction, as the variation in the chain angle is very small with temperature. 

In summary, we observe a large change in bond length with temperature, while at the same time, better relaxation of the structure is also observed. Figures \ref{fig:EndToEnd_temp} and \ref{fig:RadGyr_temp} lead us to make these observations. But this increase in the mean-square bond length with temperature is not reflected in the response in terms of increment in stress. This is because of the dominance of excluded volume effects which result in lower stresses at higher temperatures. Therefore, even though there is internal deformation of the polymer chains, the response is dominated by the excluded volume effects.
 
\begin{figure}
 \centering
 \subfloat[Bond length]{\label{fig:LenBondLen_temp}\includegraphics[width=0.45\textwidth]{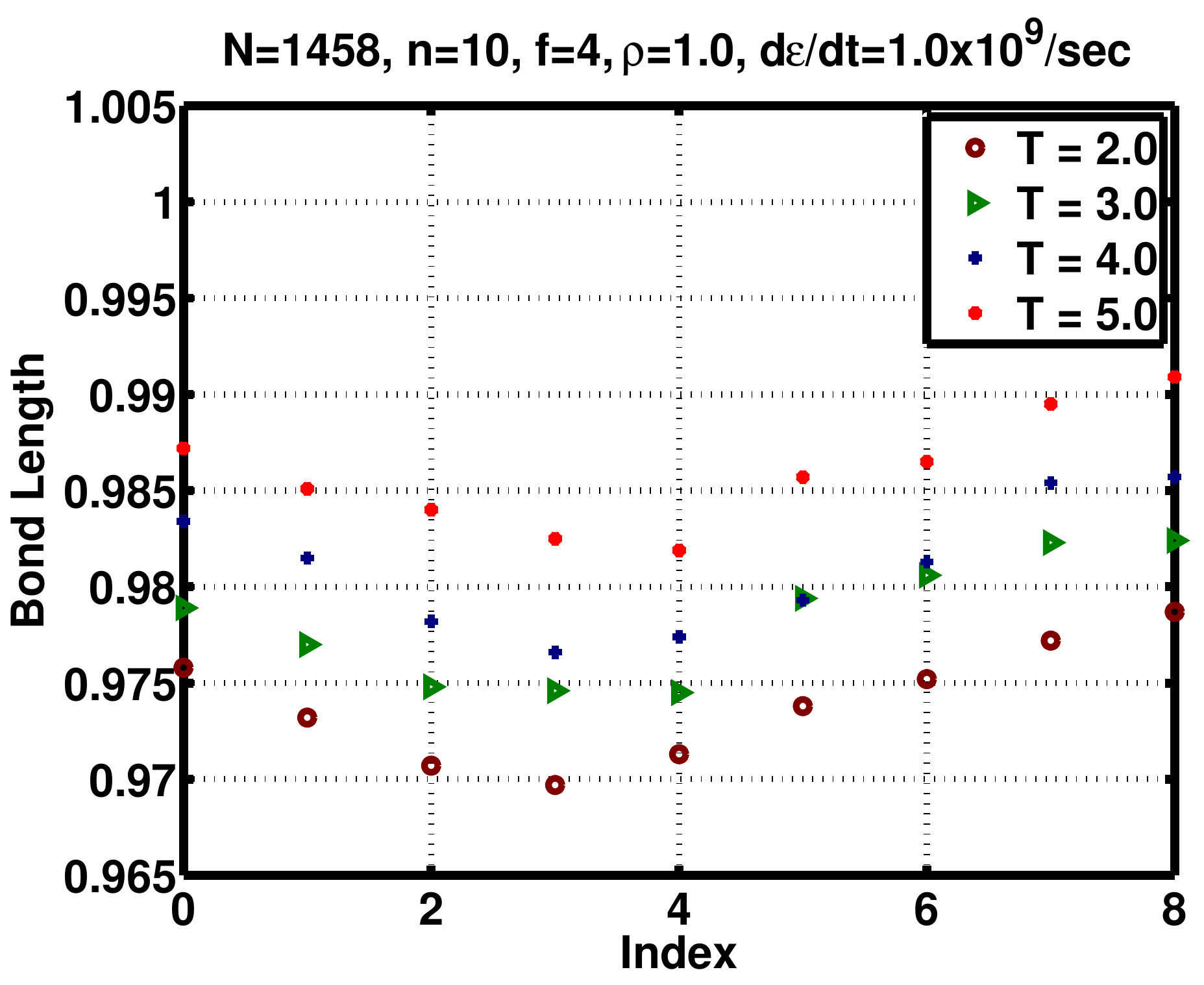}}
 \subfloat[Bond angle]{\label{fig:LenBondAng_temp}\includegraphics[width=0.45\textwidth]{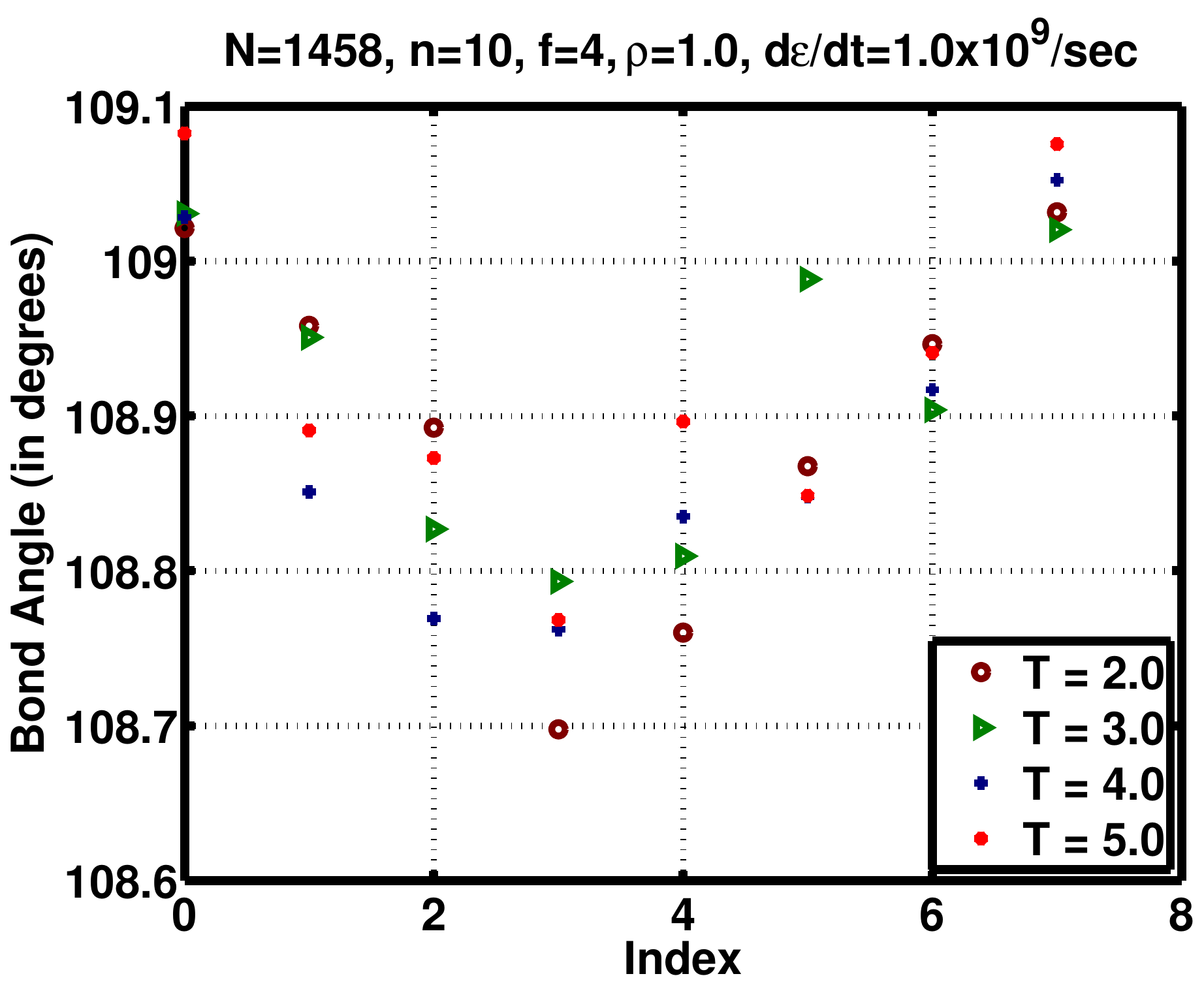}}\\
 \caption{Effect of temperature: Distribution of structural properties along the length of the chain for constant strain rate loading}
 \label{fig:LenDist_temp}
\end{figure}

Now we take a look at the variation of some internal degrees of freedom, namely bond length and bond angle, along the length of the chain. Bond length variation along the length of the chain is shown in Figure~\ref{fig:LenBondLen_temp}. Bonds at the end of the chains are stretched more as compared to the those falling in the middle of the chain at all temperatures.  Bond length along the length of the chain uniformly decreases from one end of the chain to the middle of the chain and subsequently starts increasing again towards the other end. This is because of the presence of cross-linkers that offer constraints to the motion of monomers at the chain ends. With increase in temperature, we find that the bond lengths of the chain are greater, with very little difference between bonds at the edges and the middle of the chain. This is due to the higher kinetic energy of the monomers that helps in resisting the bond stretching potential. A similar variation is also found in the bond angles along the length of the chain as shown in Figure~\ref{fig:LenBondAng_temp}. Again we note that bond angles are larger at higher temperatures which is because of the higher kinetic energy of the monomers. But then the effect of temperatures is not so significant on the bond angles as it is on bond lengths of the chain. 

\subsection{Effect of density}
\label{subsec:control_para_density}

\begin{figure}
  \centering
  \subfloat[Stress-strain curve]{\label{fig:fit_stress_strain_dens}\includegraphics[width=0.5\textwidth]{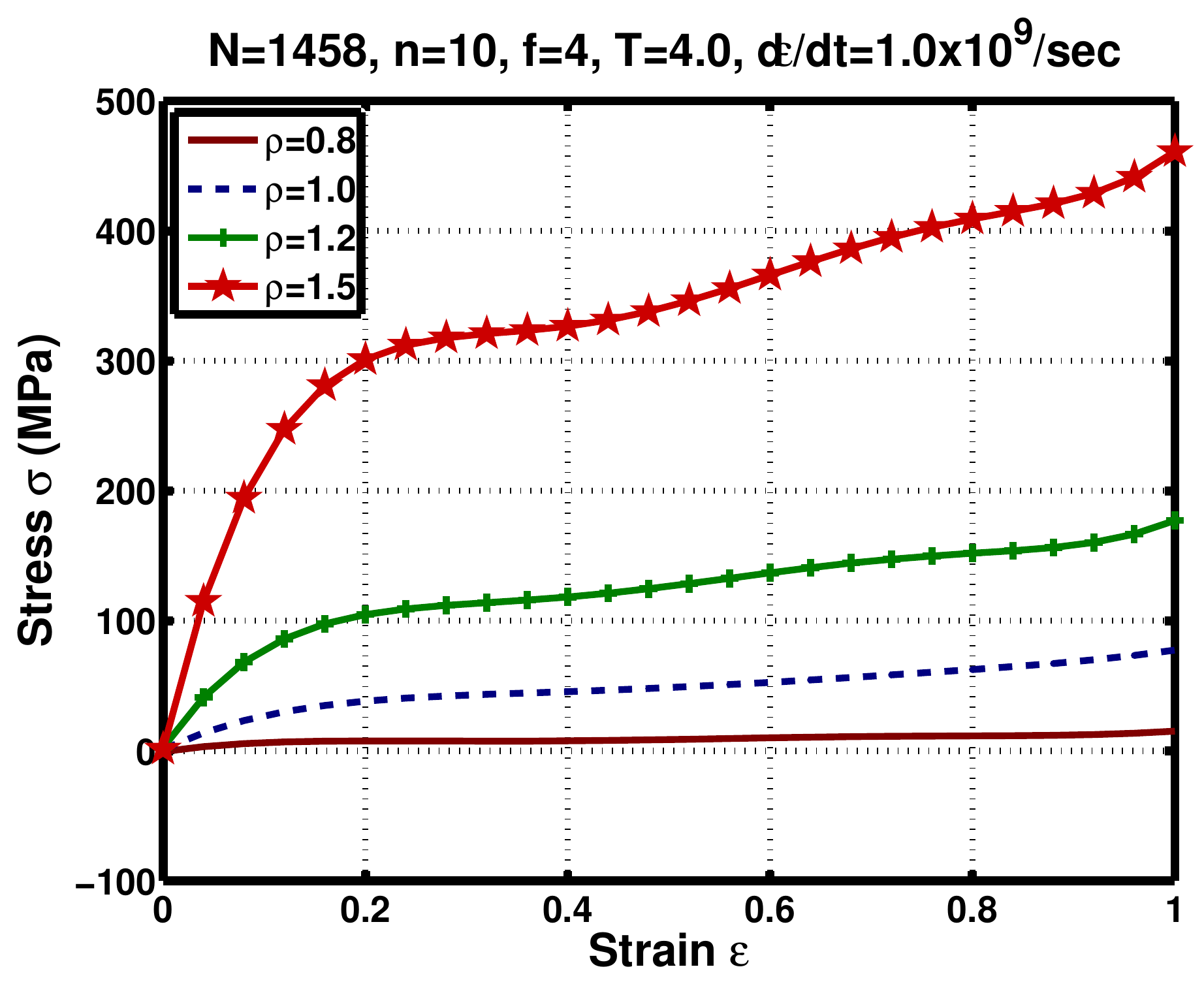}}
  \subfloat[Modulus vs. strain]{\label{fig:modulus_dens}\includegraphics[width=0.5\textwidth]{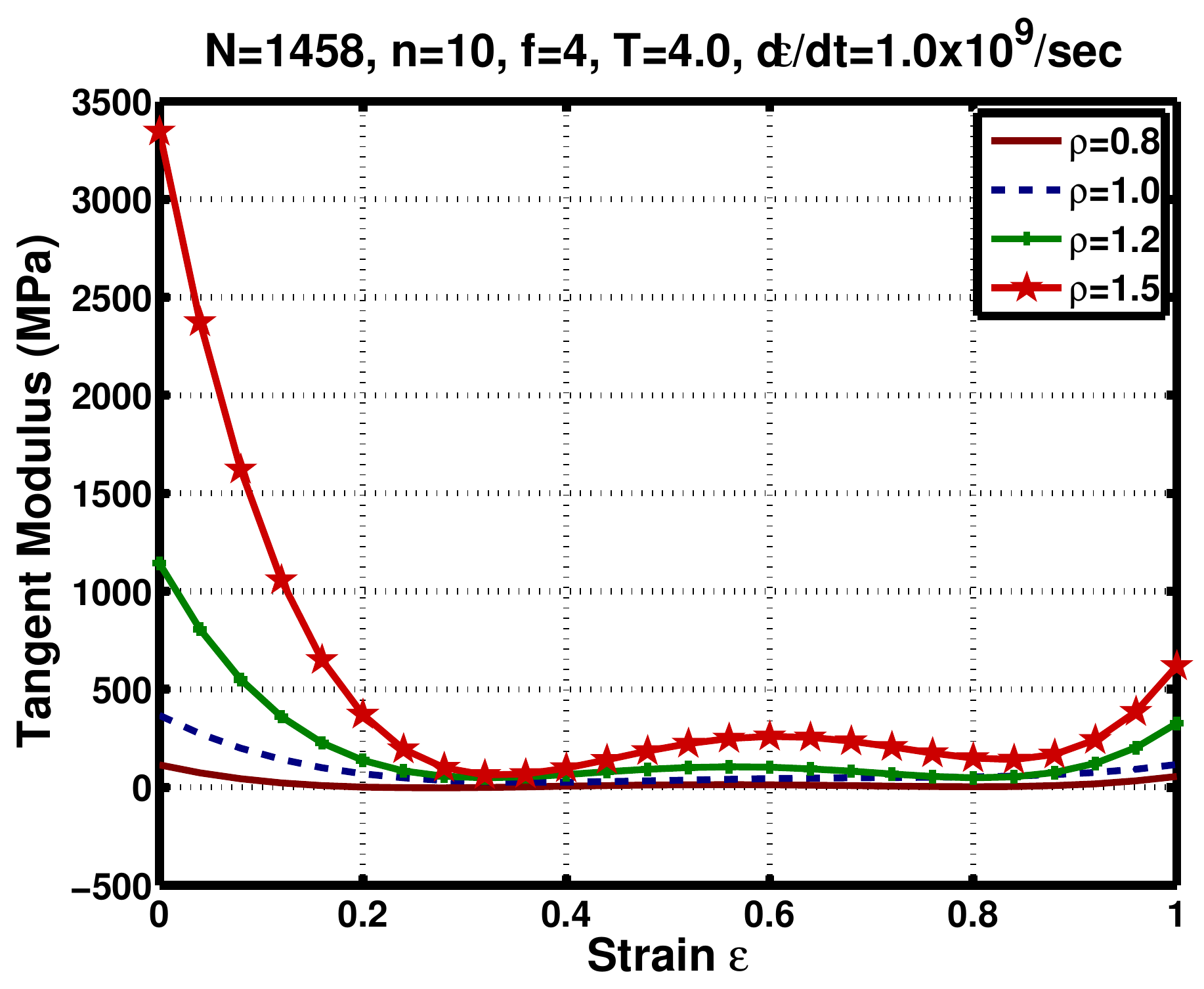}}
  \caption{Effect of density on stress response}
  \label{fig:StressStrainMod_dens}
\end{figure}

The density of the polymer is now varied in order to study its effect on stress and structure. The stress response of the system at various values of the density is shown in Figure~\ref{fig:fit_stress_strain_dens}. There is a significant change in the stress level as the density of the system is increased from $\rho=1.2$ to $\rho=1.5$. \cite{Bower2006} also report a large jump in stress from $\rho=1.0$ to $\rho=1.5$ and they attribute this jump to glass transition. The tangent modulus is shown in Figure~\ref{fig:modulus_dens}. Note that here too there is an increase in the modulus with increase in density in the initial phase of the loading, that is $\epsilon < 0.3$. Beyond this value of the strain, the modulus is close to zero for all values of density, though as $\epsilon \rightarrow 1.0$, it again increases by a small amount. This increase is due to the alignment of the chains in the loading direction. At high density, due to close packing of the monomers, chains cannot move easily. Any external deformation, hence, is caused due to the internal deformations in the chain such as change in bond length and bond angle. Also, at very high densities, excluded volume effects are very significant since monomers are really very close to each other. Hence, the observed increase in stress is the result of chain deformation as well as excluded volume effect \citep{Bower2006}. Furthermore, at higher densities, contribution to the stress due to entanglement of the chains also increases \citep{Edward}. Entropic contribution to the stress also increases due to the increased packing entropy. 

\begin{figure}
 \centering
 \subfloat[Mean-square bond length]{\label{fig:MeanBondLenStrain_dens}\includegraphics[width=0.45\textwidth]{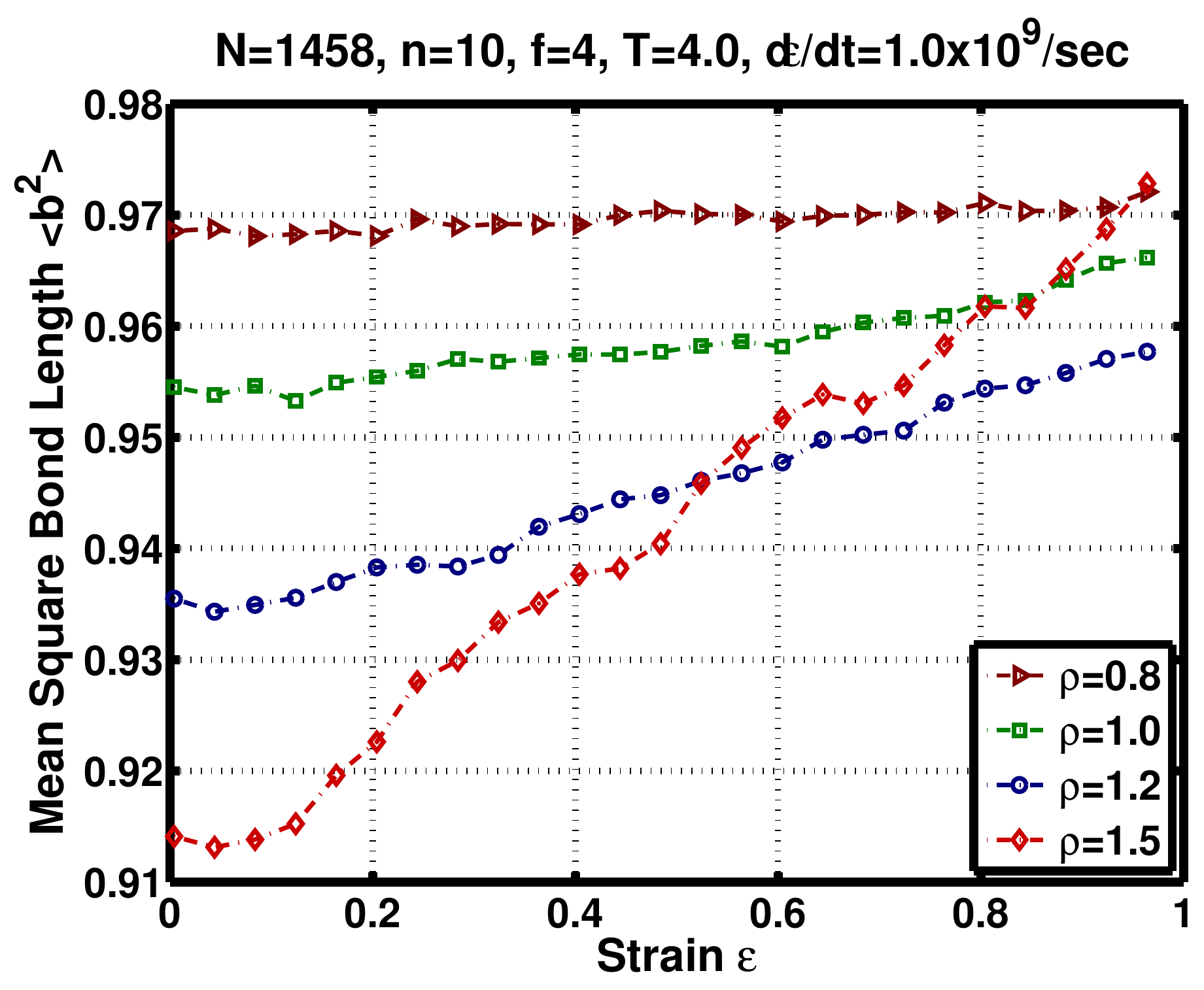}}
 \subfloat[Mean bond angle]{\label{fig:BondAngle_dens}\includegraphics[width=0.45\textwidth]{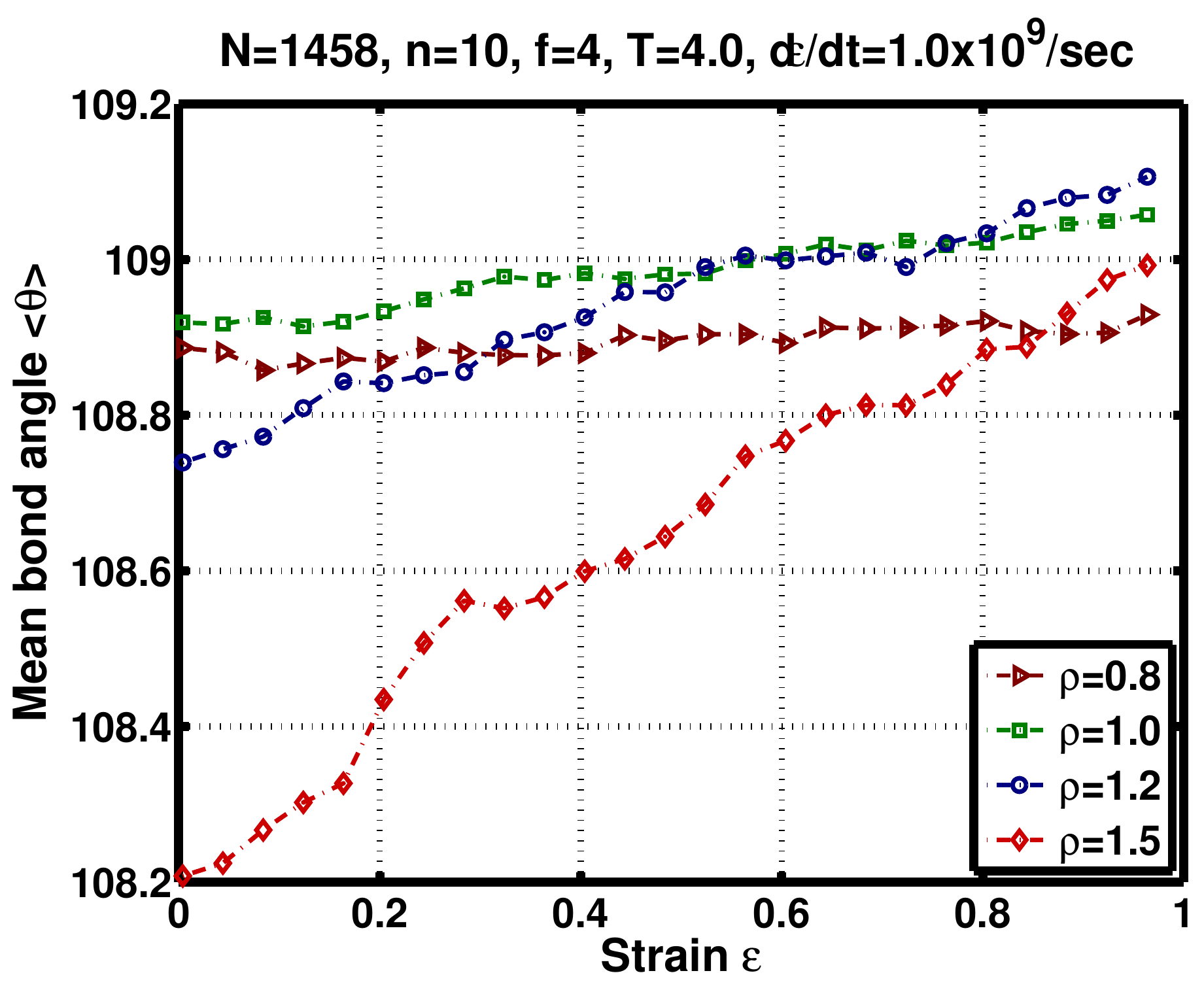}}\\
 \subfloat[End-to-end length]{\label{fig:EndToEnd_dens}\includegraphics[width=0.45\textwidth]{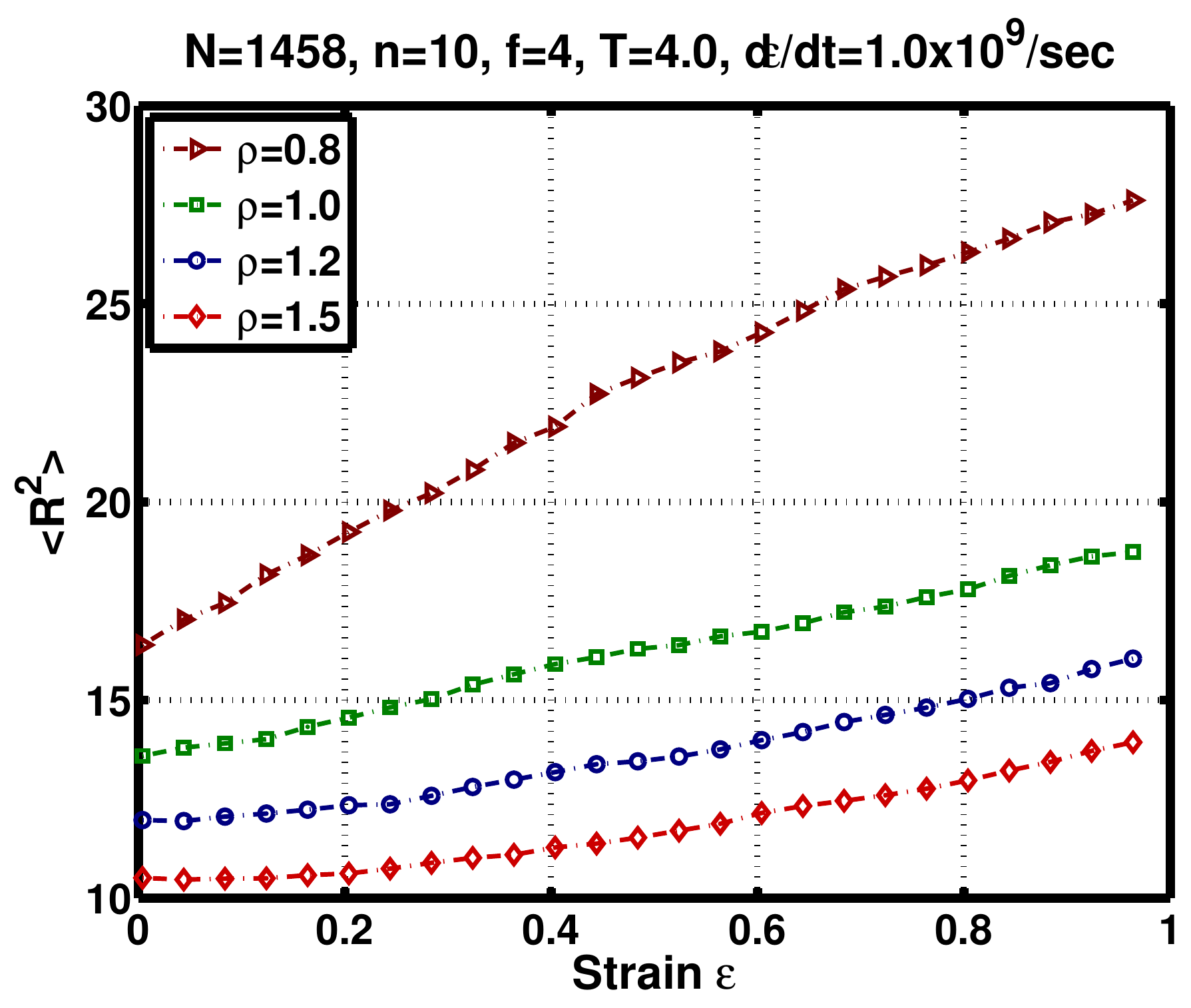}}
 \subfloat[Radius of gyration]{\label{fig:RadGyr_dens}\includegraphics[width=0.45\textwidth]{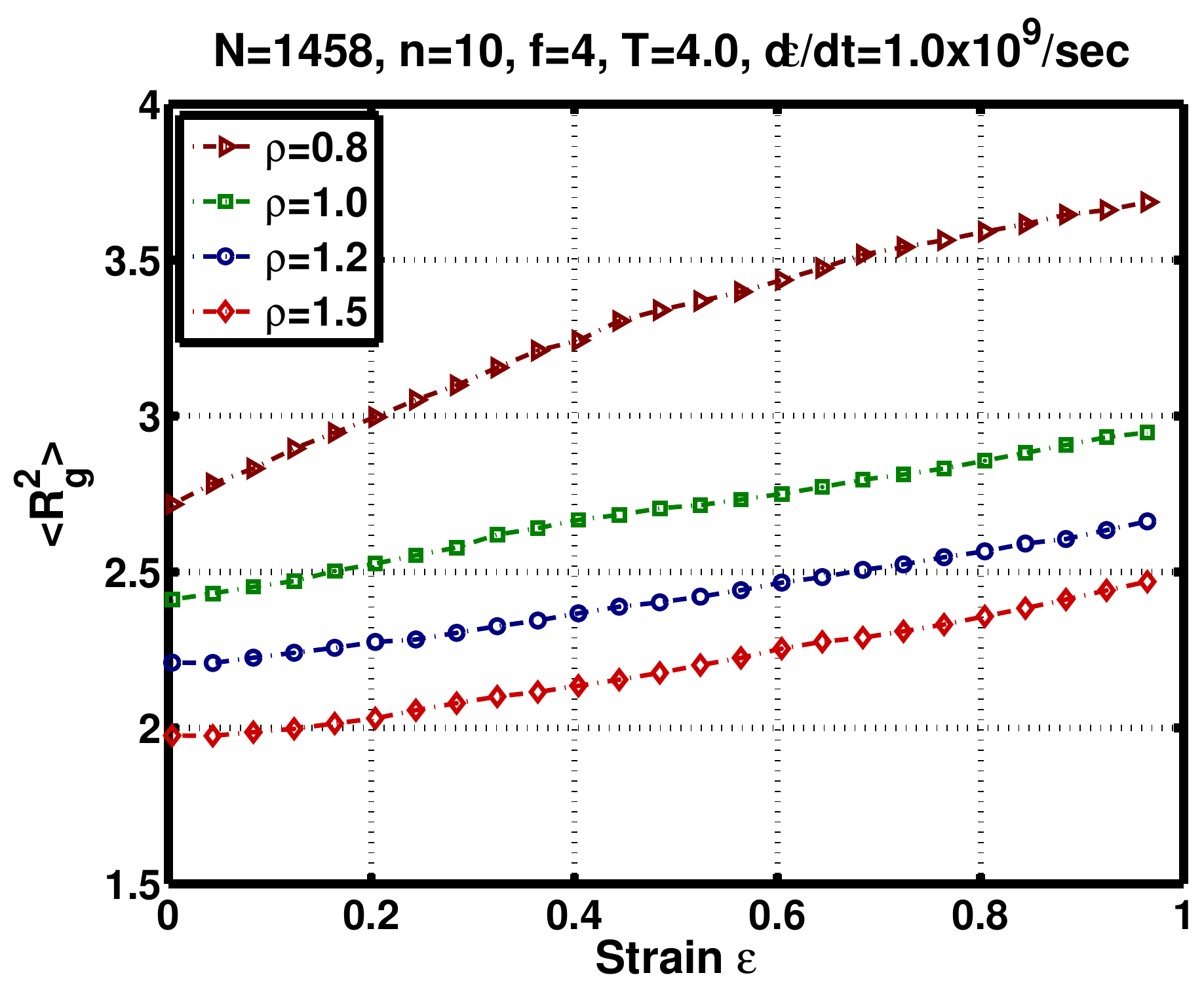}}
 \caption{Variation of structural properties for uniaxial constant strain rate loading: Effect of density}
 \label{fig:Properties_dens}
\end{figure}

A more detailed perspective into this behavior is obtained by observing some of the parameters representing the size, shape, and internal degrees of freedom of the elastomeric network. Figure~\ref{fig:MeanBondLenStrain_dens} shows the variation of the mean-square bond length at different densities as the system is stretched. The first thing one notices is that in the equilibrium configuration, that is, without any imposition of strain on the elastomer, the mean-square bond length decreases as the density is increased. This is due to the small space available at a high density values. Then the bond lengths are smaller resulting in a compact configuration. Now as the system is stretched, in general the mean-square bond length increases for all the values of density chosen for the simulation. However, note the behavior at the density value $\rho = 1.5$. The mean-square bond length increases at a higher rate and in fact attains a value greater than that for all other densities at the end of the simulation, that is $\epsilon = 1.0$. Contrast this with the variation for $\rho=0.8$ where it is observed that there is almost no change in the mean-square bond length. This is also reflected in the stress response in Figure~\ref{fig:fit_stress_strain_dens} for this value of the density. At high density, due to the compact configuration of the chain, there are constraints on the motion of the monomers that hinders the uncoiling of the chain. The deformation at such high densities is therefore mainly due to the deformation of the internal degrees of freedom of the chains such as bond length and bond angle rather than the change in shape and size of the chains. At low densities, though, there is enough space available that chains can uncoil freely resulting in mild internal deformation. 

The effect of density on mean bond angle is shown in Figure~\ref{fig:BondAngle_dens}. Similar to the mean-square bond length, the mean bond angle also decreases with density in the equilibrium configuration. At higher values of density, due to the small space available, the bonds are bent to give a compact configuration to the system. As the system is stretched, we observe that the rate at which the bond angle changes with density is high for values of density greater than or equal to $\rho=1.5$. This rate of change decreases as the density is made smaller and smaller. In fact, at $\rho=0.8$, there is almost no change in the mean bond angle with strain. 

The mean-square end-to-end length and radius gyration of the cross-linked chains at various densities is studied in Figure~\ref{fig:EndToEnd_dens} and Figure~\ref{fig:RadGyr_dens}, respectively. Naturally, the mean-square end-to-end length and mean-square radius of gyration are higher at lower densities since  enough space is available in the representative volume. However, both of these parameters show a decrease as the density is increased even at the equilibrium configuration. At $\rho=0.8$, we find a large change in both of the above parameters and at all other densities these changes are almost same as the system is stretched. All these observations indicate that at higher densities internal degrees of freedom of the chains dominate in contributing to the overall deformation of the system whereas at lower densities uncoiling of the chains is the dominant mechanism contributing to the deformation. 

\begin{figure}
 \centering
 \subfloat[$g_2/g_1$]{\label{fig:MassRatiog2_dens}\includegraphics[width=0.45\textwidth]{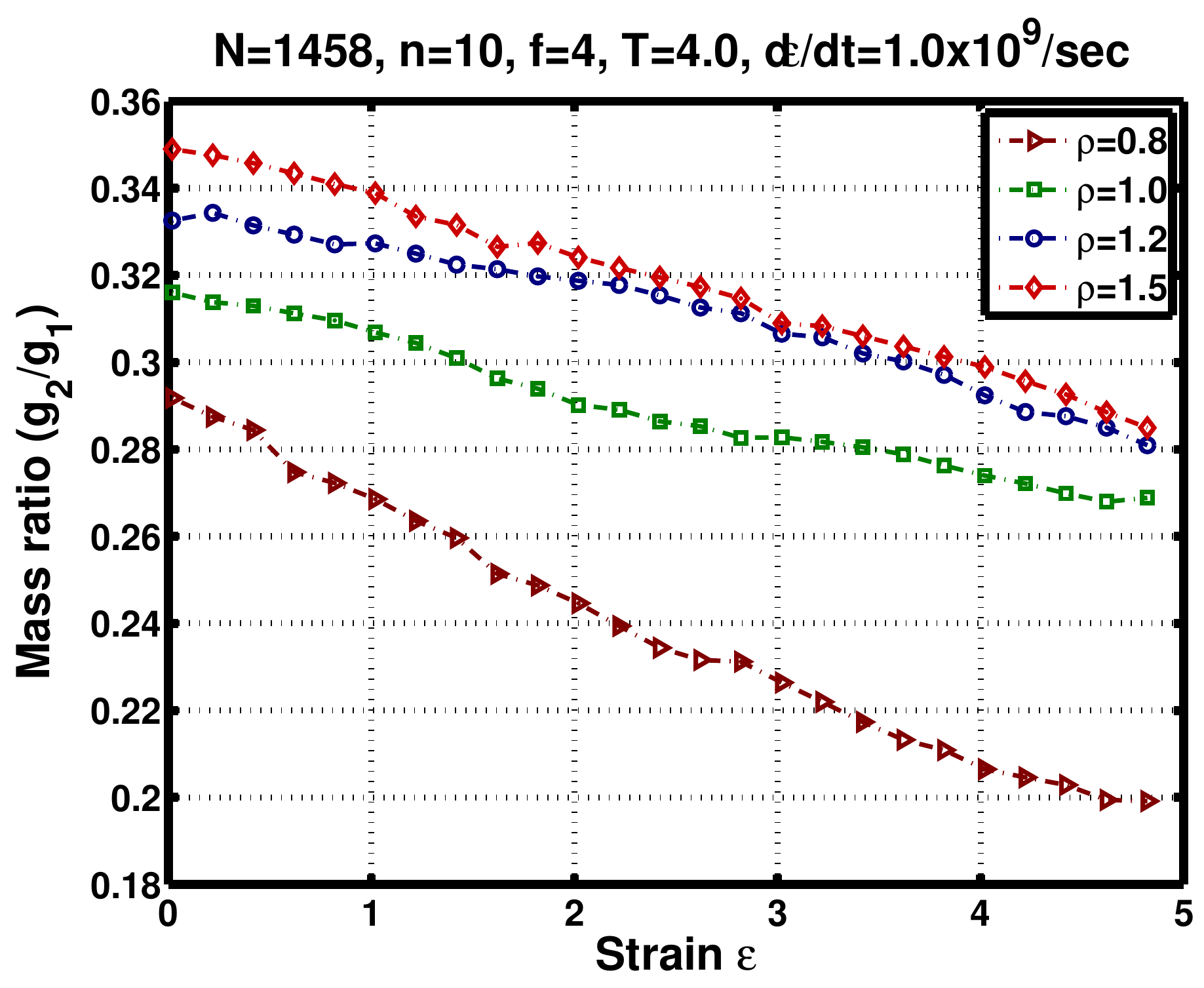}} 
 \subfloat[$g_3/g_1$]{\label{fig:MassRatiog3_dens}\includegraphics[width=0.45\textwidth]{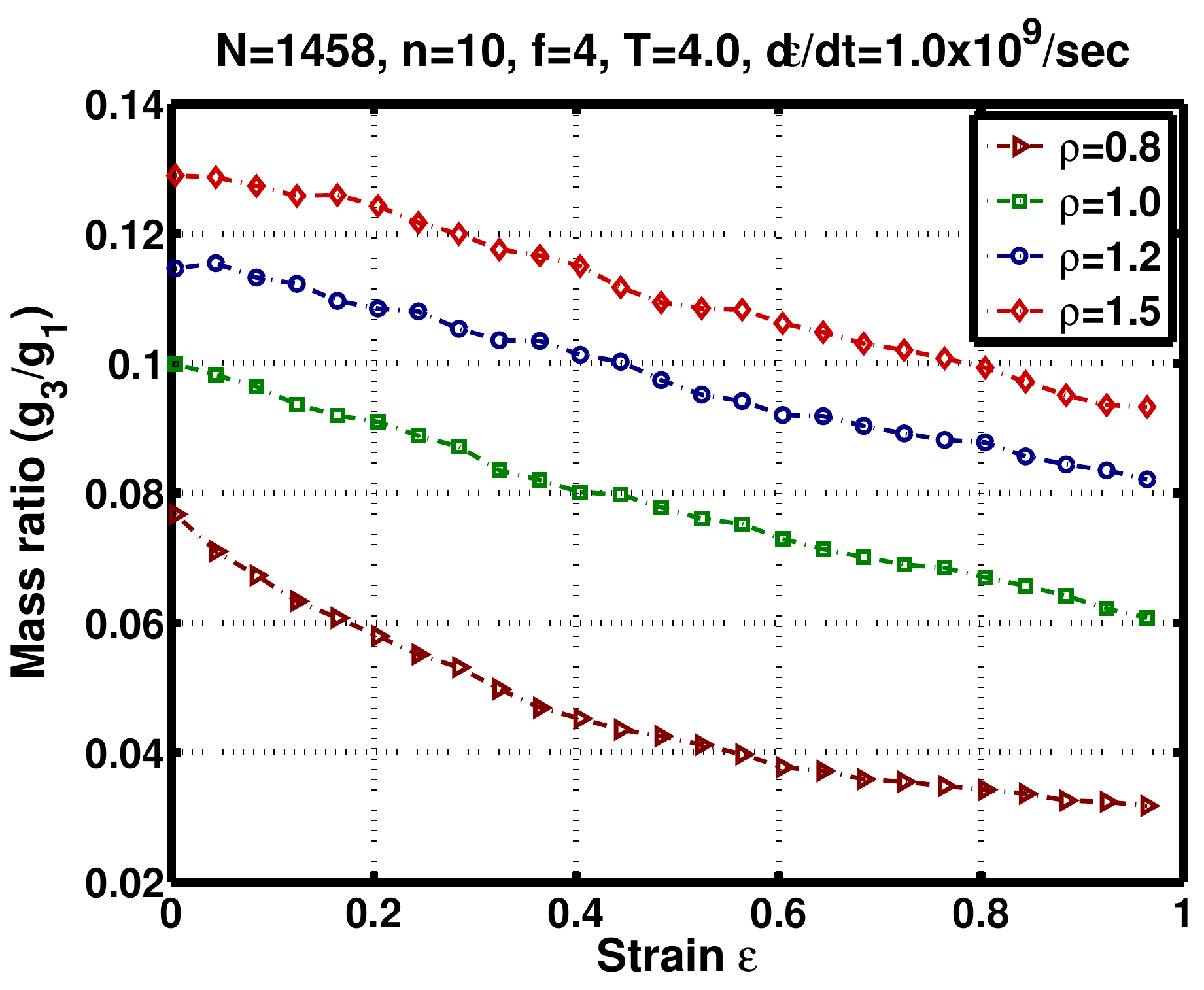}} 
 \caption{Effect of density on mass ratios}
 \label{fig:MassRatios_dens}
\end{figure}

The effect of density on mass ratios is shown in Figure~\ref{fig:MassRatios_dens}. Mass ratios increase with density indicative of a more spherical shape of the elastomeric chains. At a very low value of density --- $\rho = 0.8$ --- we observe that chains are almost two dimensional because the contribution of $g_3$ is very small. Also, reduction in the mass ratio with increasing strain is greater at $\rho = 0.8$ relative to that at higher values of density. From the above observations, we conclude that the mass distribution is less spherical at low values of density, and as the density is increased, the distribution becomes more and more spherical. 

\begin{figure}
 \centering
 \includegraphics[width=0.5\textwidth]{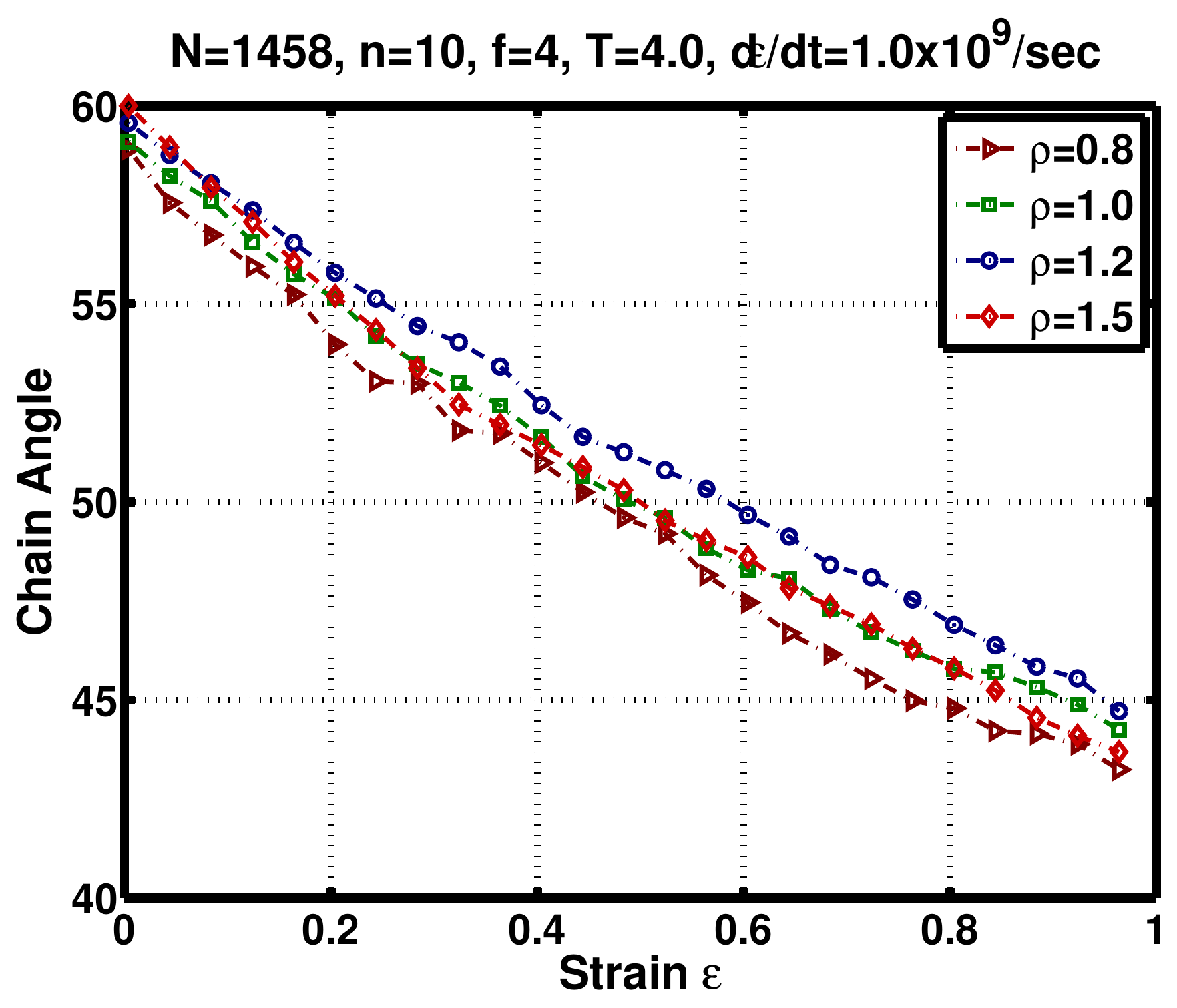}
 \caption{Effect of density on chain angle}
 \label{fig:ChainAng_dens}
\end{figure}

The effect of density on the alignment of the chain in the loading direction is also studied. Figure~\ref{fig:ChainAng_dens} shows the chain angle variation with strain loading at various densities. There is no significant effect of the density on the alignment of the chain. 

\begin{figure}
 \centering
 \subfloat[Bond length]{\label{fig:LenBondLen_dens}\includegraphics[width=0.45\textwidth]{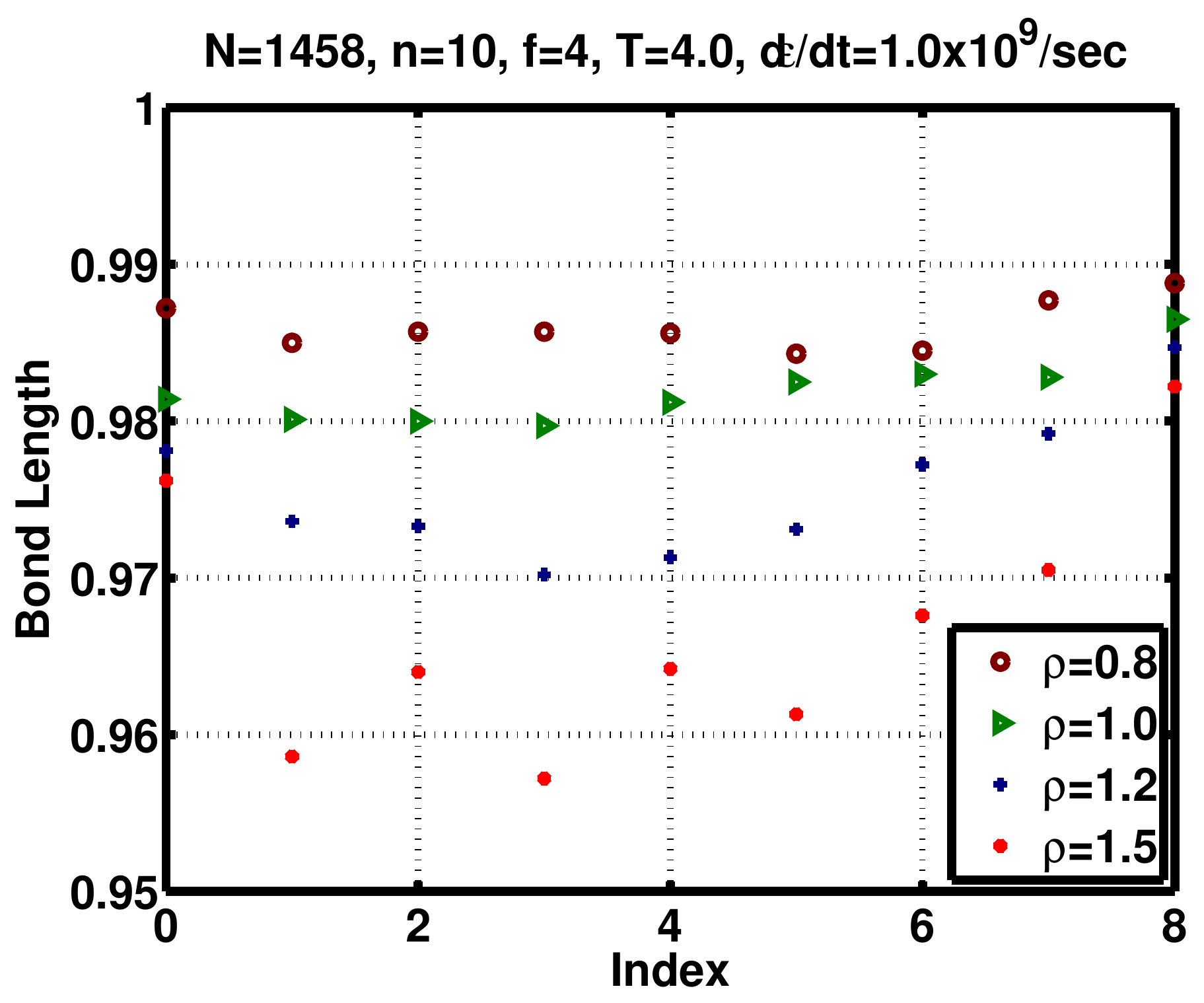}}
 \subfloat[Bond angle]{\label{fig:LenBondAng_dens}\includegraphics[width=0.45\textwidth]{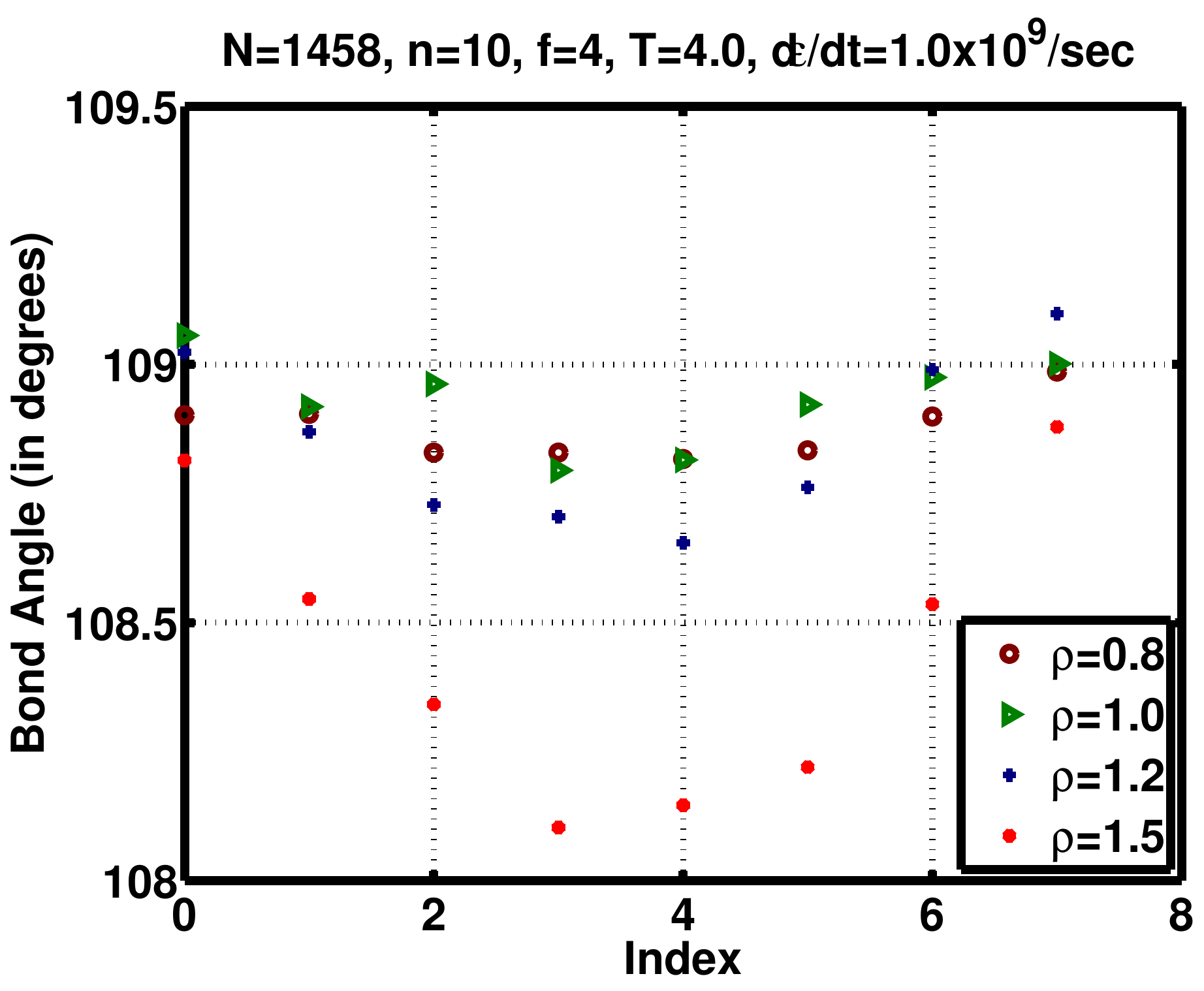}}\\
 \caption{Effect of density: Distribution of structural properties along the length of the chain for constant strain rate loading}
 \label{fig:LenDist_dens}
\end{figure}

The variation of the structure parameters along the length of the chain was determined at various values of  densities. As noted earlier, the bond length is longer towards the ends of the chain and uniformly decreases from both sides towards the middle of the chain, as shown in Figure~\ref{fig:LenBondLen_dens}. Now, this effect becomes more prominent at higher densities. At high densities, due to the limited availability of space, the  bonds are subject to compression, but, presence of cross-linkers at the end of the chains put additional constraints on these monomers and they cannot move as freely as the monomers in the middle of the chains. In order to keep all chains connected through a cross-linker together, bonds at the ends of the chains are stretched more. Mean bond angle also shows a similar variation as shown in Figure~\ref{fig:LenBondAng_dens}. Here also, we find that bond angles are higher at the end of the chains and decrease towards the middle from both ends. But unlike the case of mean-square bond length we find that this effect is more predominant at a very high value of density, namely $\rho=1.5$. At other densities too, the trend is observed, but there is no significant difference in the variation with the change in values of density.

\subsection{Effect of strain rate}
\label{subsec:control_para_rate}

\begin{figure}
  \centering
  \subfloat[Stress-strain curve]{\label{fig:fit_stress_strain_rate}\includegraphics[width=0.5\textwidth]{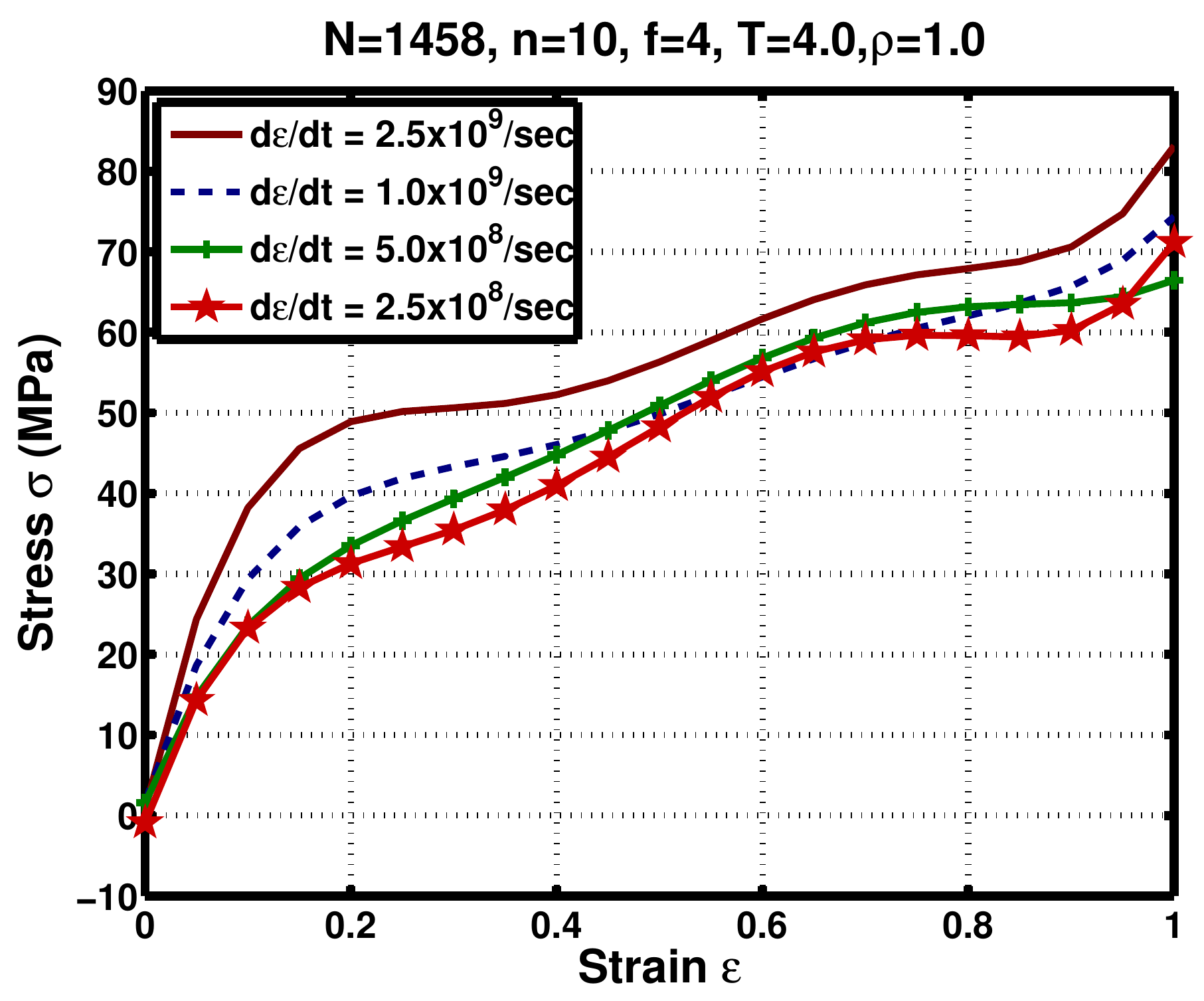}}
  \subfloat[Modulus vs. strain]{\label{fig:modulus_rate}\includegraphics[width=0.5\textwidth]{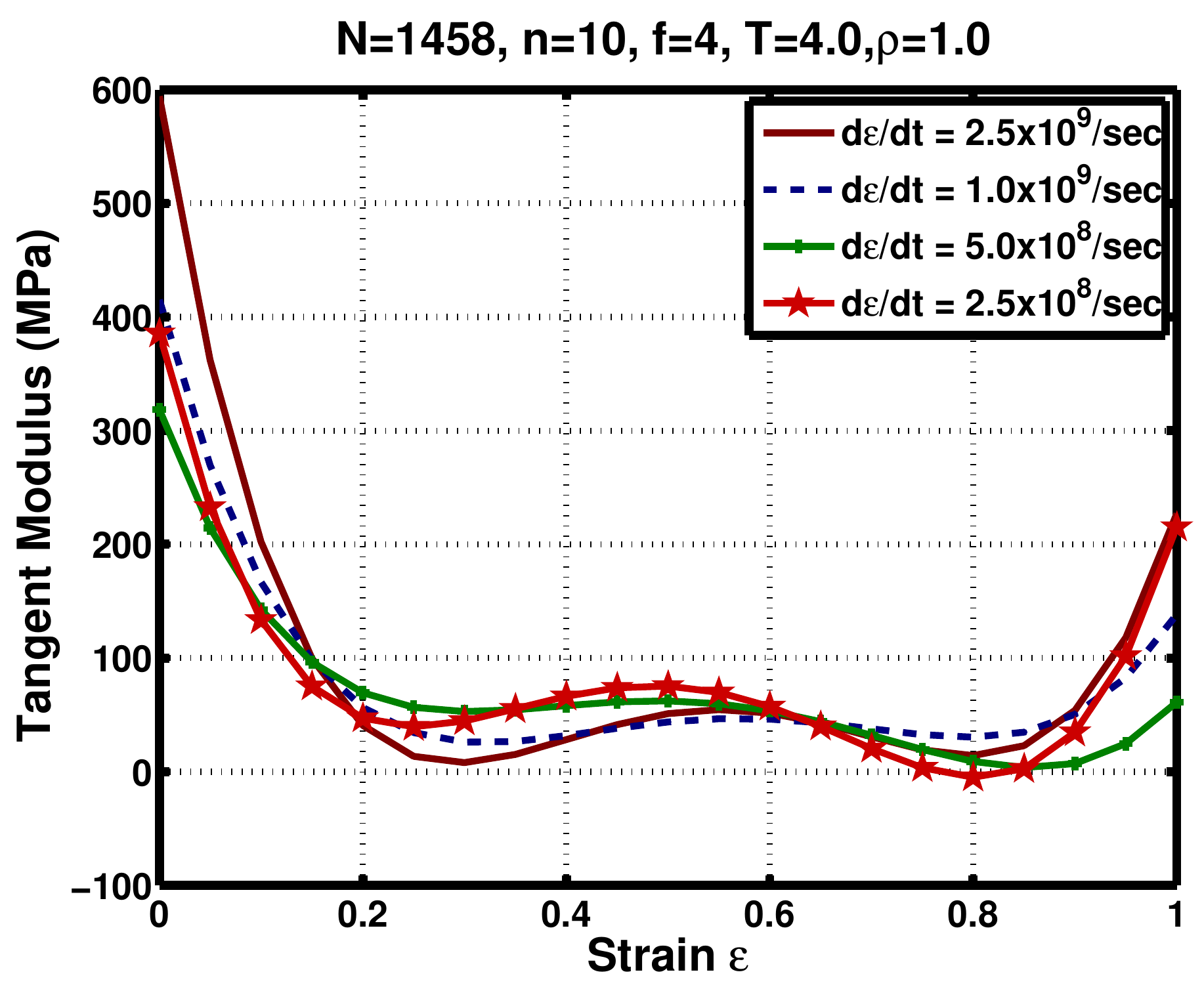}}
  \caption{Effect of strain rate on stress response}
  \label{fig:StressStrainMod_rate}
\end{figure}

\begin{figure}
 \centering
 \subfloat[Mean-square bond length]{\label{fig:MeanBondLenStrain_rate}\includegraphics[width=0.45\textwidth]{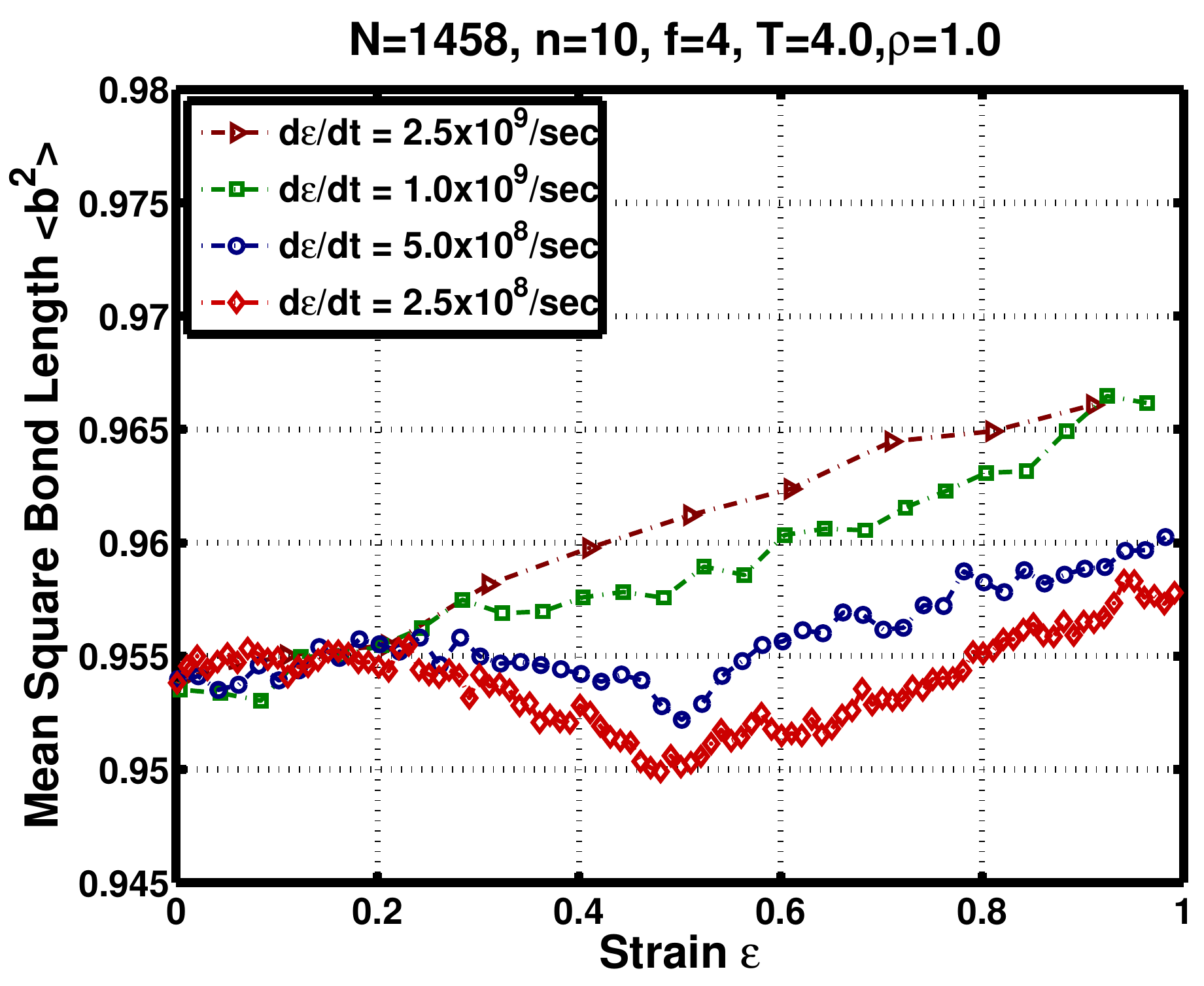}}
 \subfloat[Mean bond angle]{\label{fig:BondAngle_rate}\includegraphics[width=0.45\textwidth]{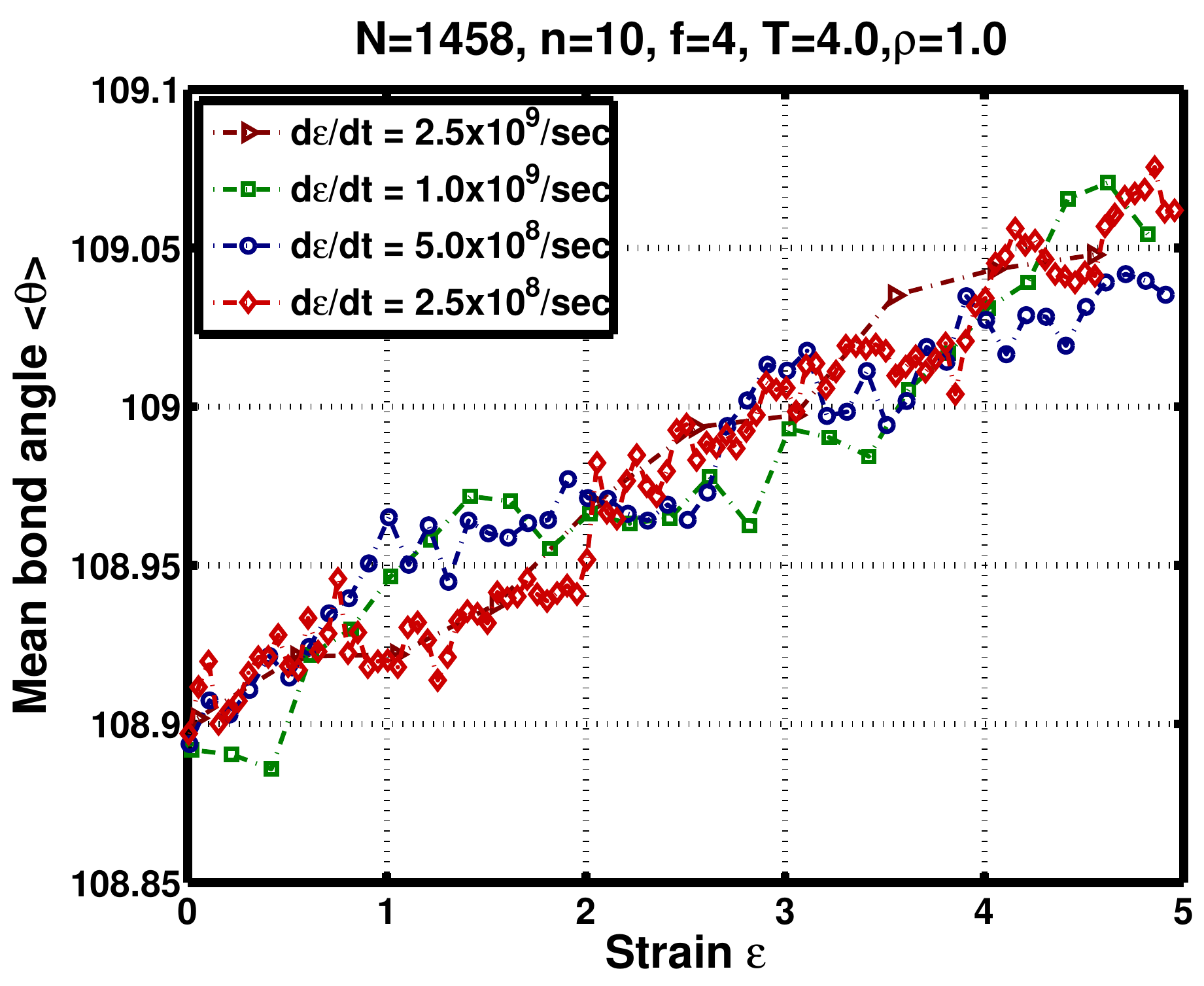}}\\
 \subfloat[End-to-end length]{\label{fig:EndToEnd_rate}\includegraphics[width=0.45\textwidth]{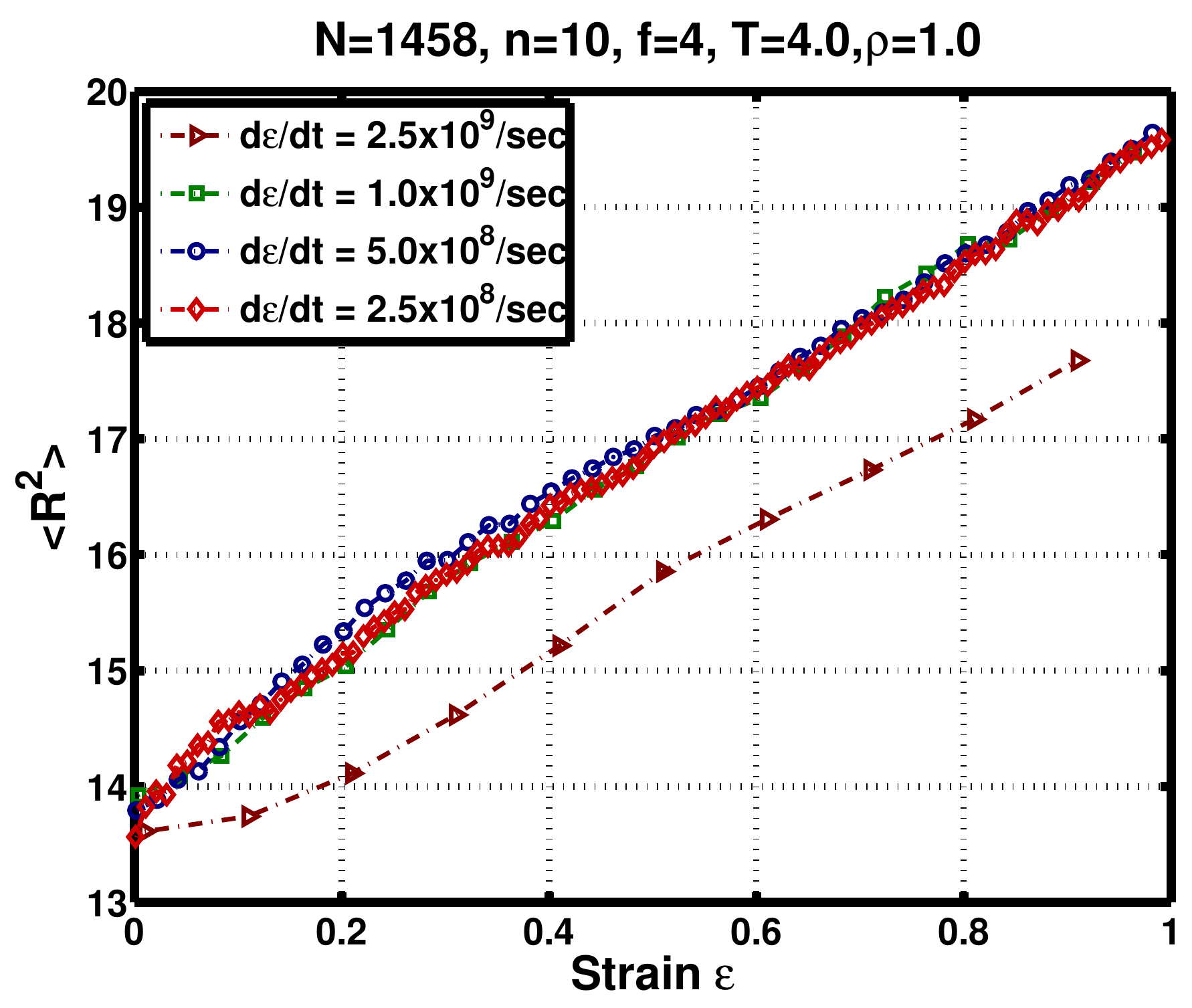}}
 \subfloat[Radius of gyration]{\label{fig:RadGyr_rate}\includegraphics[width=0.45\textwidth]{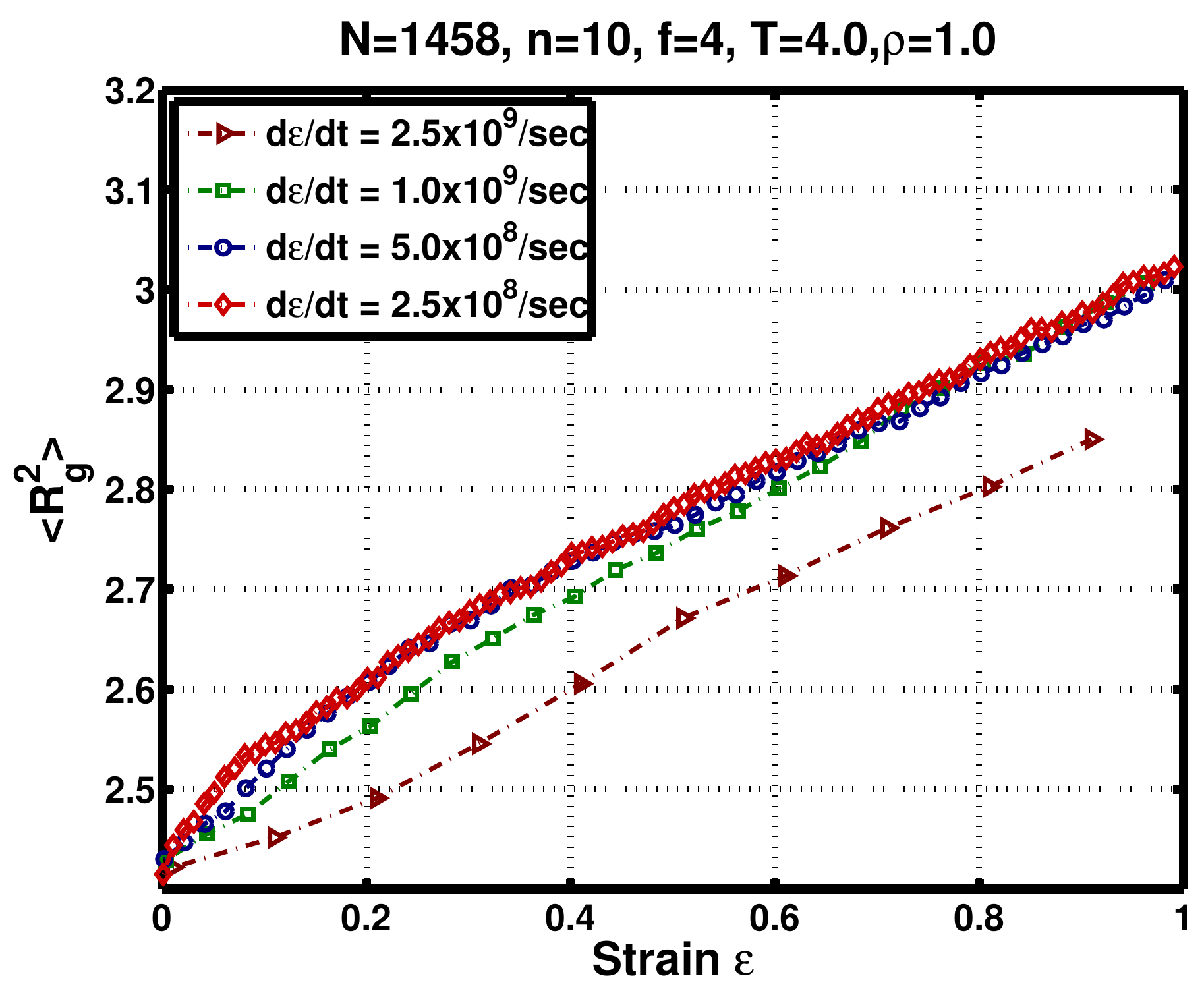}}
 \caption{Variation of structural properties for uniaxial constant strain rate loading: Effect of strain rate}
 \label{fig:Properties_rate}
\end{figure}

The polymeric system was loaded at different strain rates to study its effect on its stress response. Figure~\ref{fig:fit_stress_strain_rate} shows the stress response at different strain rates. We observe that faster the rate of loading, higher is the stress induced in the system, with a significant jump as we increase the strain rate from $1.0\times10^9$ to $2.5\times10^9$/sec. A similar observation is also reported by \cite{Chui} in their Monte-Carlo study of a cross-linked polymer. The corresponding tangent modulus shown in Figure~\ref{fig:modulus_rate} also indicates that at low values of strain, the system is more stiff at higher rates of loading. During an intermediate range of strain, $0.2\le\epsilon\le 0.8$, the tangent modulus at the strain rate of $2.5\times10^9$ is lower than that of the slower strain rate cases. This primarily happens because of uncoiling of the chains in this phase. From Figures \ref{fig:EndToEnd_rate} and \ref{fig:RadGyr_rate} we observe that the amount of uncoiling is fairly low at the strain rate $\dot\epsilon=2.5\times10^9$/sec. This keeps the number of new entanglements formed in the system very small. Hence, contribution to the strain is primarily due to deformation in the internal structure of the chain. This is also seen in Figure \ref{fig:MeanBondLenStrain_rate}, in which a relaxation in the bond length is observed between $\epsilon=0.2-0.5$ at low strain rates. In contrast, this is absent at higher strain rates. Balance between the contribution to the stress due to bond deformations and excluded volume effects results in a very small increment in stress at faster strain rates. At slow strain rates though, this is not completely balanced and as a result we observe that stiffness is little higher. Below, we discuss how the strain rate affects the stress and the tangent modulus in terms of the variations in the structure of the elastomer.

The evolution of the structure of the elastomer is determined as it is strained for different values of the strain rate parameter. The objective is to establish the major modes of deformations contributing to the stress. In this regard, we first study the evolution of the mean-square bond length. This is shown in Figure~\ref{fig:MeanBondLenStrain_rate}. We observe that initially the evolution of the mean-square bond length is similar, for all strain rates, up to a strain of $\epsilon = 0.2$. Thereafter, we observe that the bond length relaxes a little at the slower strain rates in the range $\dot{\epsilon}=2.5\times 10^8 \mbox{ to } 5.0\times 10^8/ \mbox{sec}$. In the faster strain rate cases of $\dot{\epsilon} = 1.0\times 10^9 \mbox{ and } 2.5\times 10^9/ \mbox{sec}$, in the region of strain between $\epsilon=0.2-0.5$, there is no relaxation similar to that observed at lower strain rates. However, beyond $\epsilon=0.5$ the rate of increase in the bond length is almost same irrespective of the strain rate. At lower strain rates there is significant difference between the time scale of external deformation and the time scale of the internal motion of the united atoms. This enables chains to relax the structure and acquire a lower energy state. In high strain rate cases, the time scale of external deformation reaches very close to the time scale of the internal motion of the united atoms. As a result the molecules do not get enough time to respond to the external deformation and hence very small relaxation. This is the reason we see reduction in the bond length in the middle of the strain loading at low strain rates where as this is not found at higher strain rates. 

The effect of loading rate on the evolution of the bond angle is depicted in Figure~\ref{fig:BondAngle_rate}. We do not find any significant difference in the evolution of the bond angles at different rates. 

The evolution of mean-square end-to-end length and mean-square radius of gyration is shown in Figure~\ref{fig:EndToEnd_rate} and Figure~\ref{fig:RadGyr_rate}, respectively. Here we observe a peculiar behavior in that the variation with strain of both of these parameters is completely different at $\dot\epsilon=2.5\times 10^9/\mbox{sec}$ than at all other values of strain rates. In fact, at slower strain rates there is no significant difference in the variation of these two parameters. This indicates that uncoiling of the chains is suppressed at faster strain rates, and hence, the major deformation in the system is due to the change in the mean-square bond length. Also, the overall arrangement of the chains in the system, and in turn, excluded volume interaction plays a significant role in the behavior of the system at different rates. 

\begin{figure}
 \centering
 \subfloat[$g_2/g_1$]{\label{fig:MassRatiog2_rate}\includegraphics[width=0.45\textwidth]{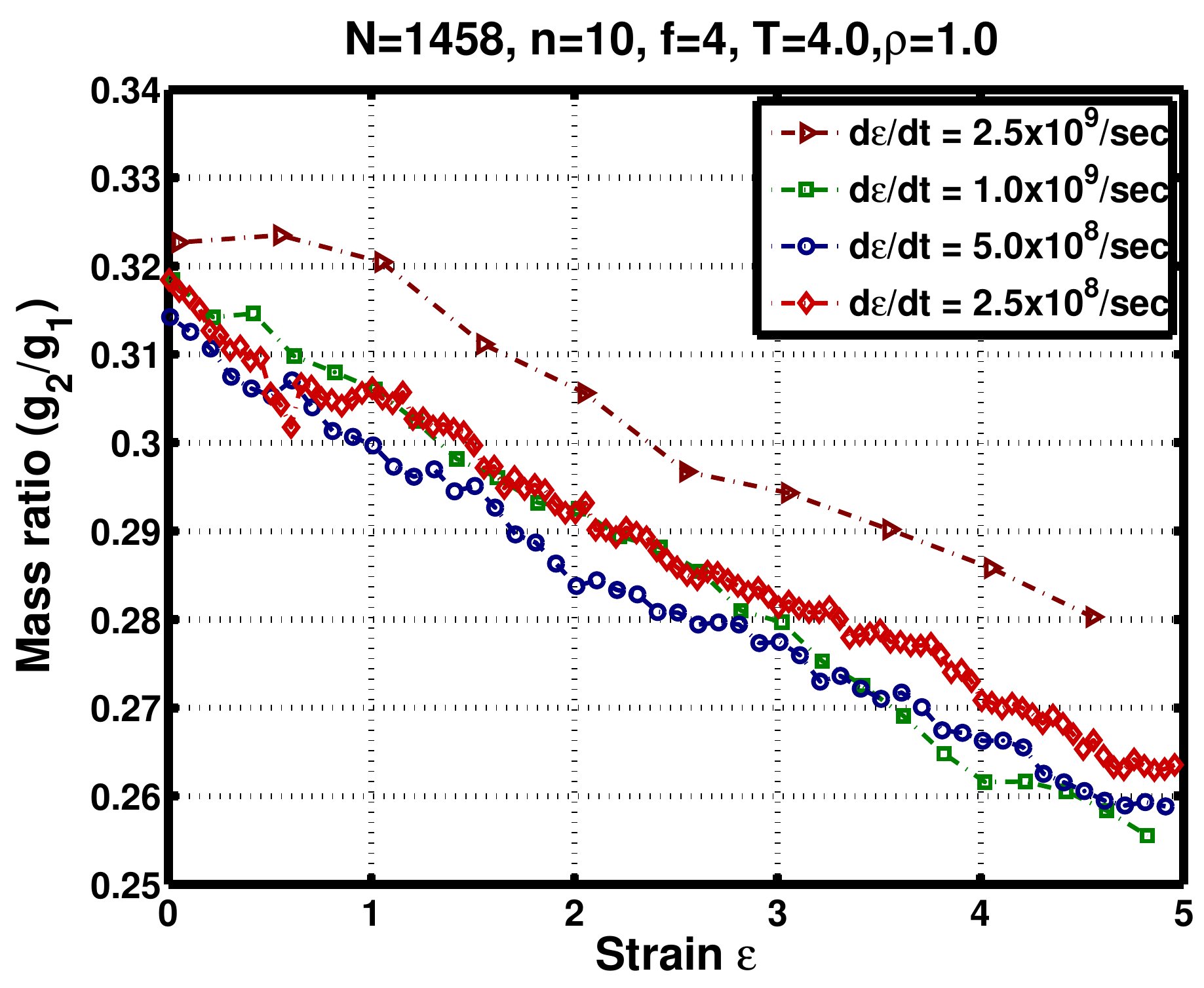}} 
 \subfloat[$g_3/g_1$]{\label{fig:MassRatiog3_rate}\includegraphics[width=0.45\textwidth]{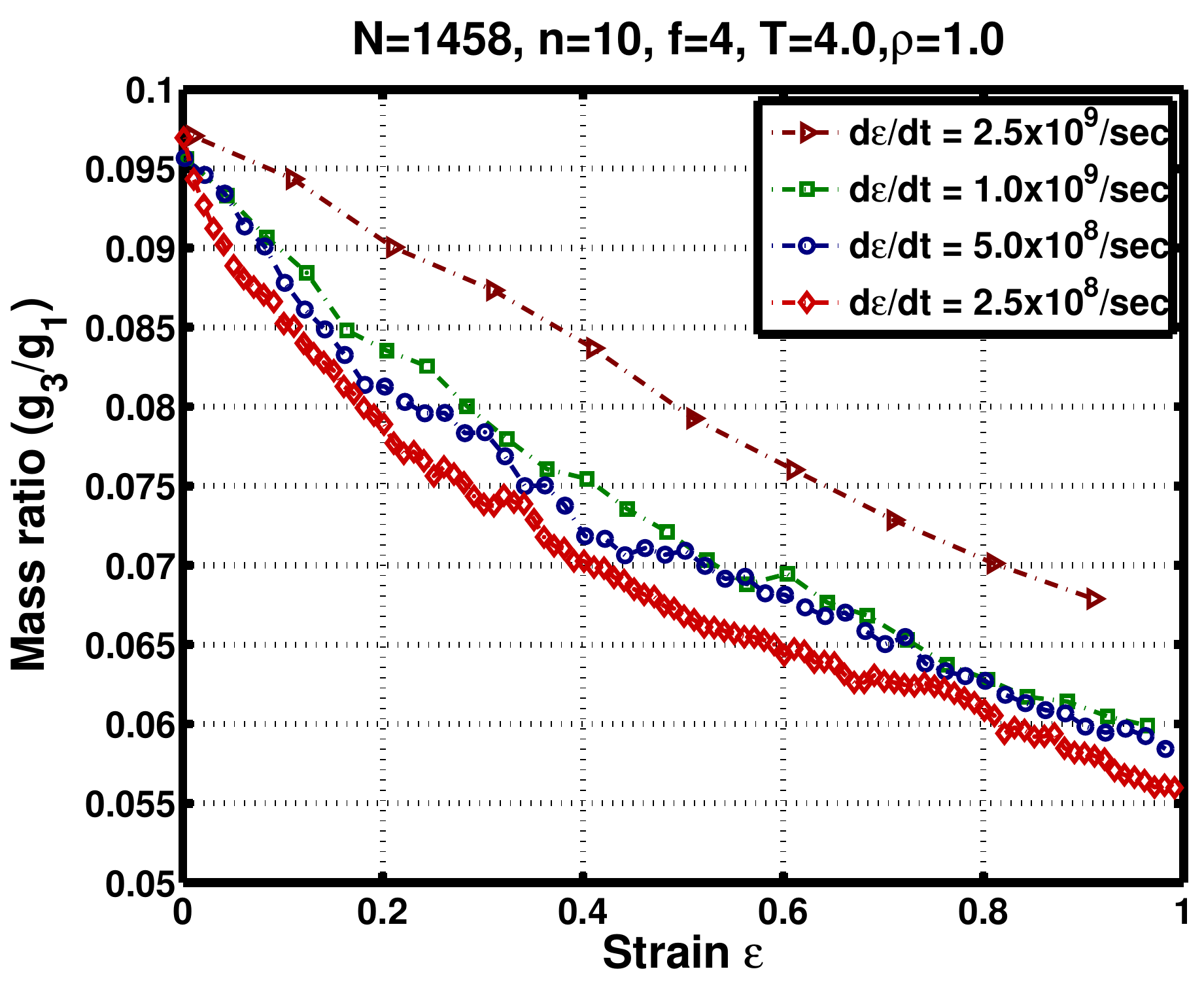}}
 \caption{Effect of strain rate on mass ratios}
 \label{fig:MassRatios_rate}
\end{figure}

The variation of the mass ratios is shown in Figure~\ref{fig:MassRatios_rate}. At $\dot\epsilon=2.5\times 10^9/\mbox{sec}$ the chains are more spherical during the stretch as compared to all other values of the strain rates. We observe that there is a small reduction in the mass ratios at the above mentioned value of the strain rate which indicates that the shape of the chains do not change significantly. At strain rates other than $\dot\epsilon=2.5\times10^9$/sec, the variation in the mass ratios with strain is similar and there is a larger reduction in the mass ratios as compared to that at $\dot\epsilon=2.5\times 10^9/\mbox{sec}$. Hence, at all these strain rates there is a significant change in the shape of the chains and they take an almost planar configuration whereas at $\dot\epsilon=2.5\times10^9$/sec there very small change in shape of the chains.

\begin{figure}
 \centering
 \includegraphics[width=0.45\textwidth]{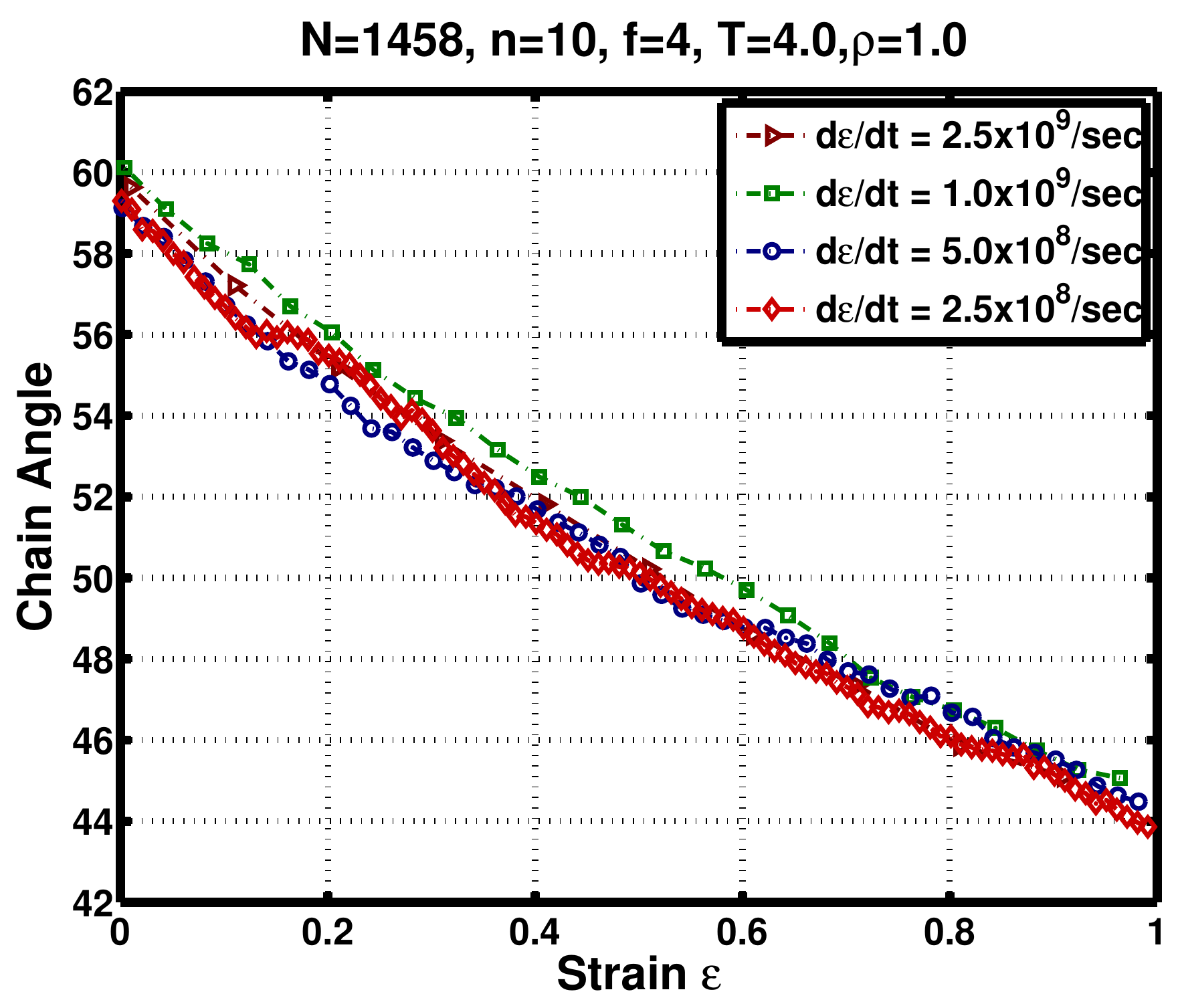}
 \caption{Effect of strain rate on chain angle}
 \label{fig:ChainAng_rate}
\end{figure}

The evolution of the chain angle is not affected much with change in strain rate as shown in Figure~\ref{fig:ChainAng_rate}. This indicates that the overall alignment of the chain is not affected by the rate of the loading. 

In summary, only bond length is affected significantly by the rate of the loading. End-to-end length, radius of gyration and mass ratios are affected significantly only at a high value of strain rates. Other parameters do not show any significant variation at different strain rates. 

\begin{figure}
 \centering
 \subfloat[Bond length]{\label{fig:LenBondLen_rate}\includegraphics[width=0.45\textwidth]{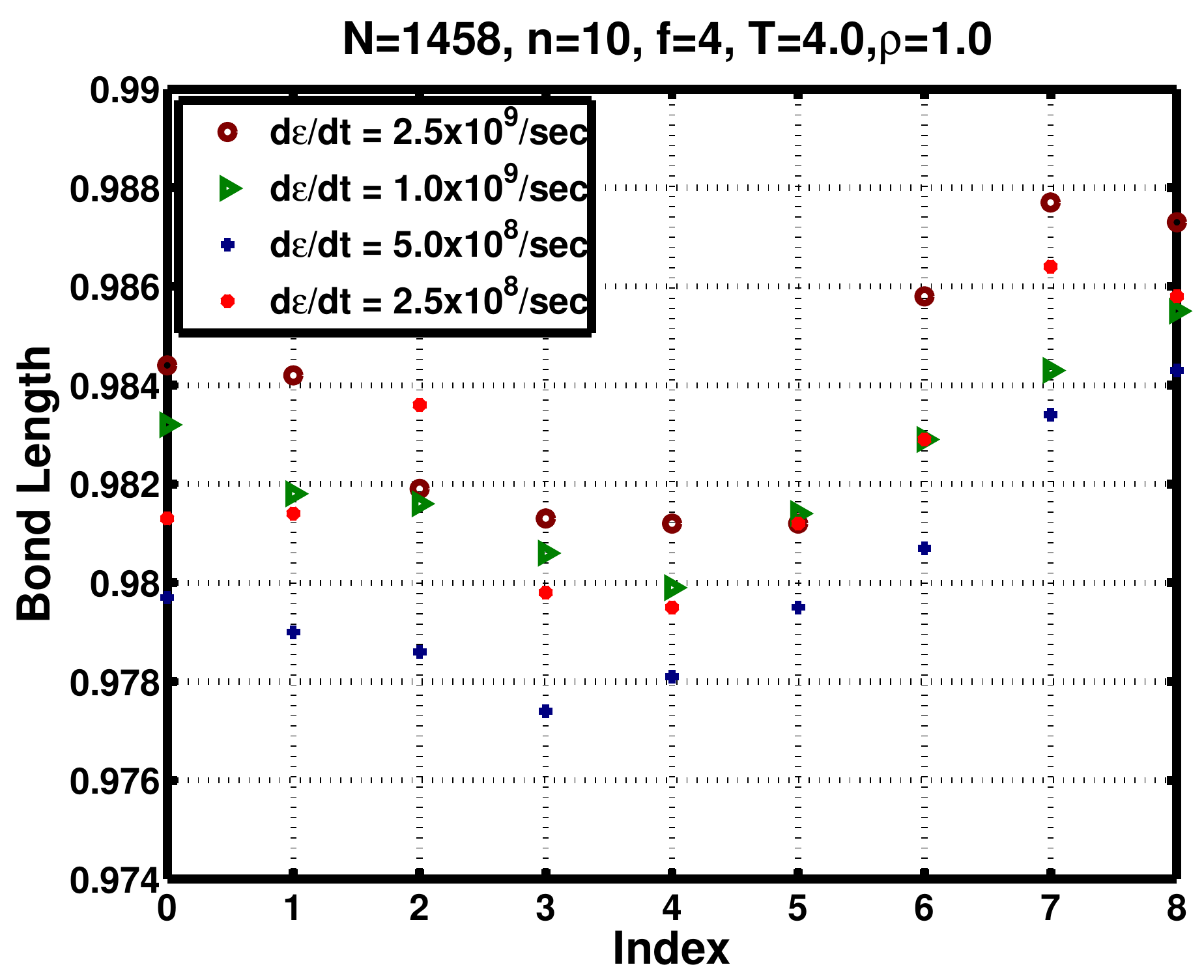}}
 \subfloat[Bond angle]{\label{fig:LenBondAng_rate}\includegraphics[width=0.45\textwidth]{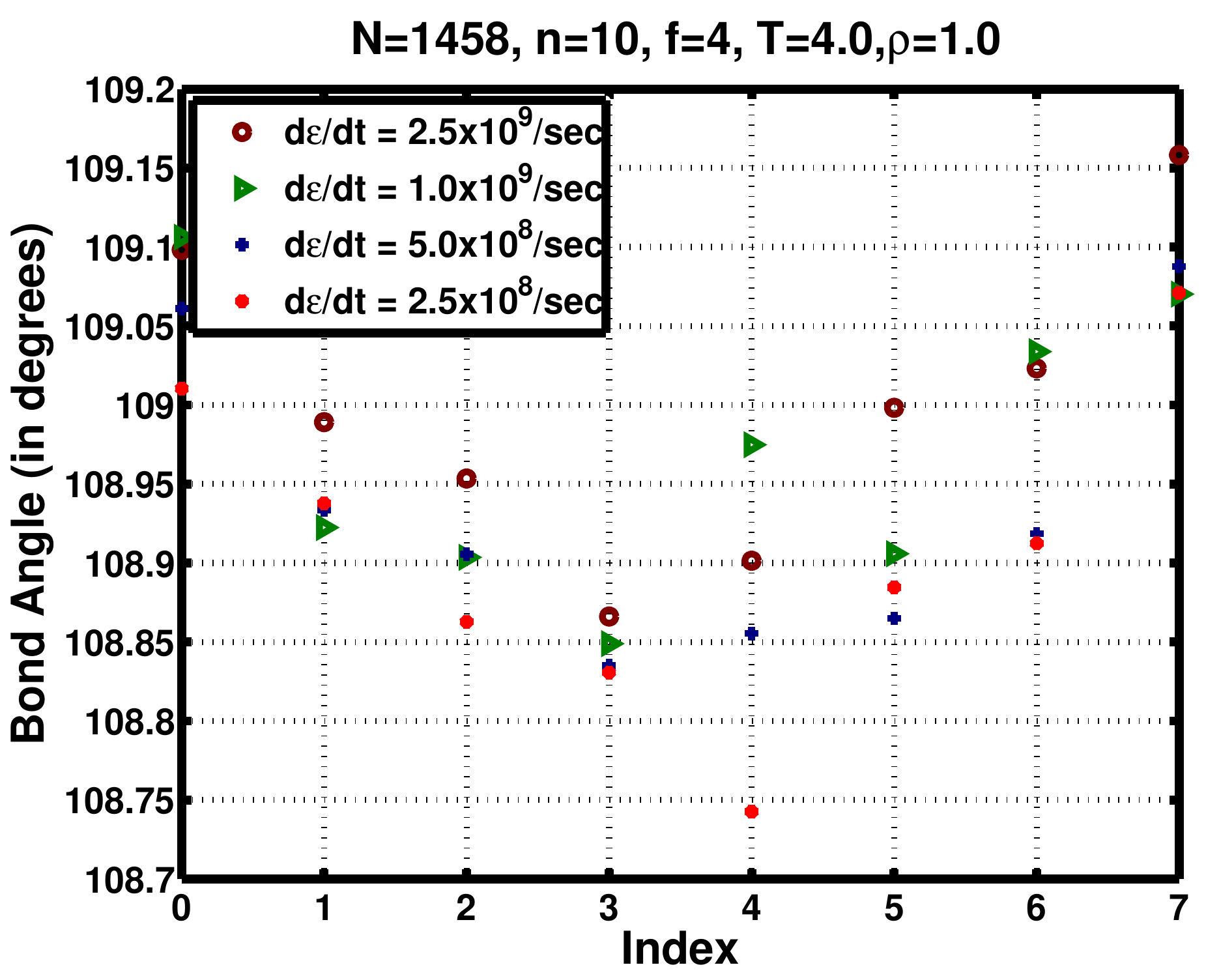}}\\
 \caption{Effect of strain rate: Distribution of structural properties along the length of the chain for constant strain rate loading}
 \label{fig:LenDist_rate}
\end{figure}

The variation of the mean bond length and mean bond angle along the length of the chain is shown in Figure~\ref{fig:LenDist_rate}. We observe that at faster strain rates, the values of the mean bond angle are higher along the chain length  with some fluctuations. At slower strain rates, however, the bond lengths show a behavior wherein the bond lengths are higher at the chain ends and decrease uniformly towards the middle from both ends. 

A similar trend is also noted in the variation of the mean bond angle.  At $\dot\epsilon=2.5\times 10^9/\mbox{sec}$, the bond angles fluctuate along the length of the chain about the mean value. At slower strain rates we observe that at the chain ends, bond angles are higher. Lower the strain rate, lower is the dip in chain angle in the middle of the chain indicating the relaxation of the bonds. But higher angles at the chain ends makes the mean bond angles more or less uniform at all strain rates. This is the reason the evolution of the mean bond angle is similar at all strain rates as observed in Figure~\ref{fig:BondAngle_rate}.

\subsection{Effect of chain length}
\label{subsec:control_para_chain_length}

\begin{figure}
  \centering
  \subfloat[Stress-strain curve]{\label{fig:fit_stress_strain_ChainLen}\includegraphics[width=0.45\textwidth]{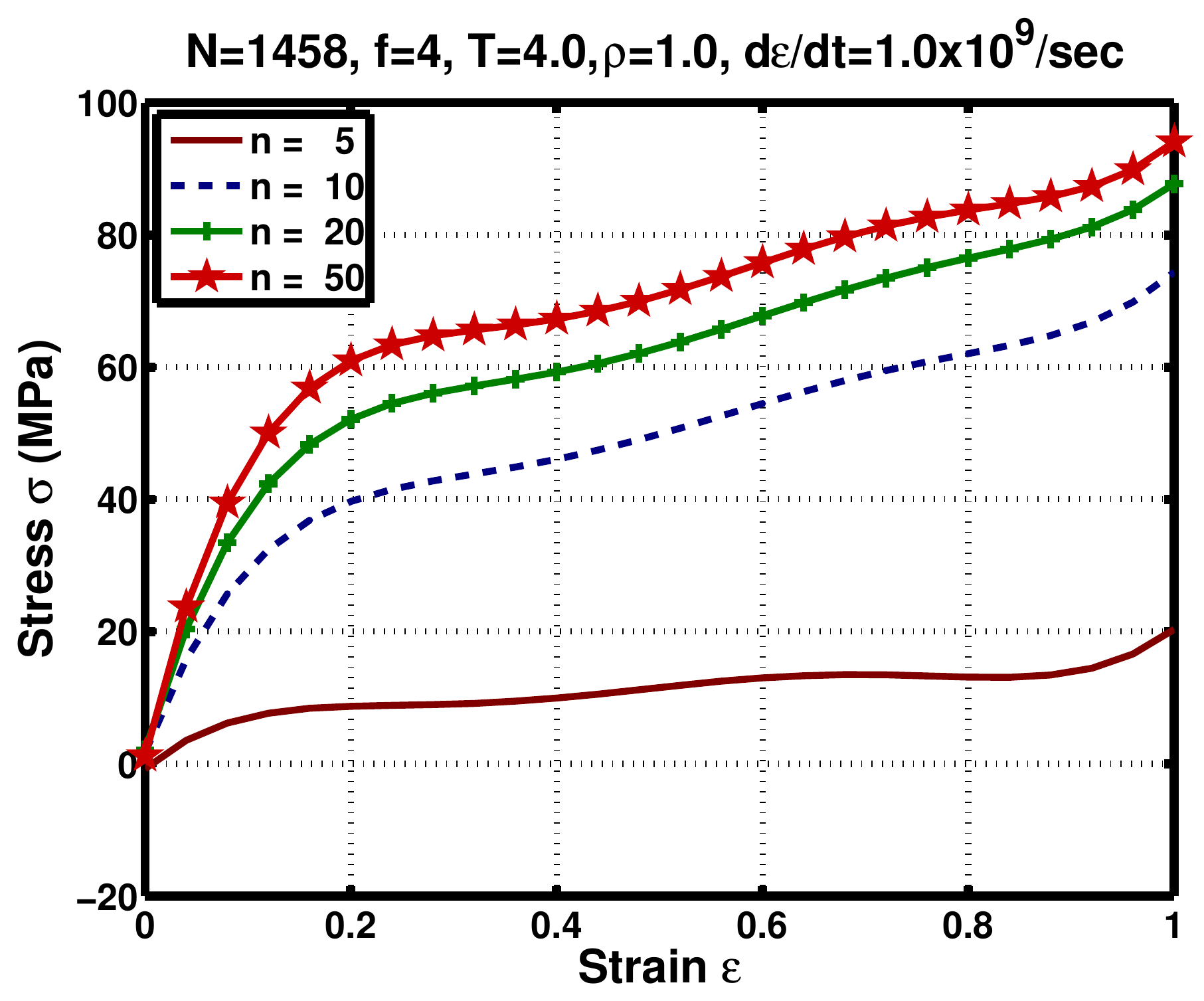}}
  \subfloat[Modulus vs. strain]{\label{fig:modulus_ChainLen}\includegraphics[width=0.45\textwidth]{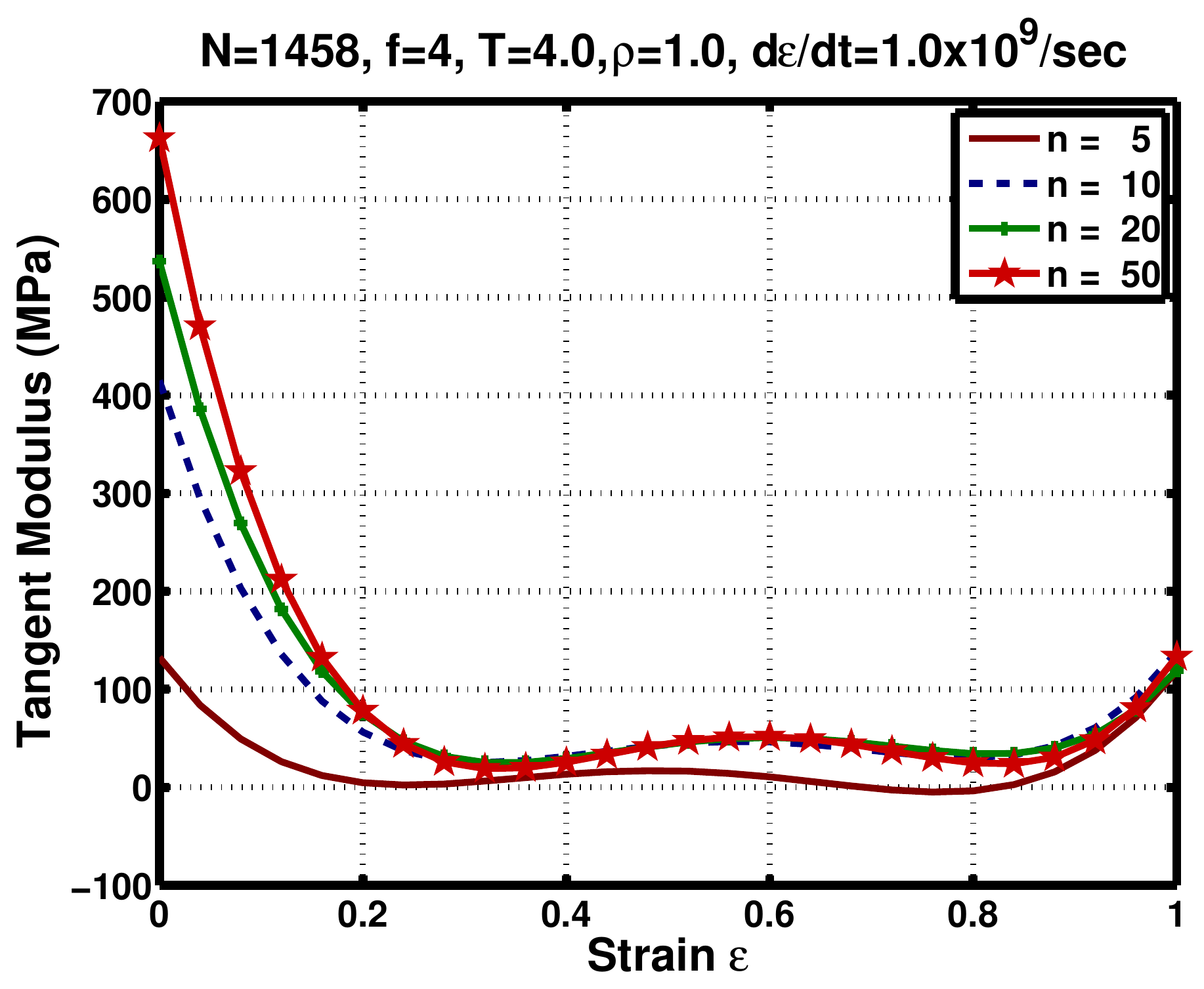}}
  \caption{Effect of chain length on stress response}
  \label{fig:StressStrainMod_ChainLen}
\end{figure}

We now perform simulations on systems of various chain lengths. The stress response of the system at various chain lengths is shown in Figure~\ref{fig:fit_stress_strain_ChainLen}. We observe that the stress response is significantly affected by the size of the chains. Long chain polymers have higher stress induced. In fact, a large jump in the stress is found from $n=5$ to $n=10$. Thereafter this jump is smaller and decreases as the chain length is increased. In long chain polymers, as they are strained, chains get entangled very easily due to large number of constraints from the neighboring molecules. Short chains move very easily in the system and in case they get entangled, they can easily slip over one another. This is the primary mode through which short chains contribute to the overall deformation. As a result, stress induced in very short chain systems such as $n=5$ is very low. In long chain polymers, slipping is very difficult as there could be the possibility of more than one entanglement in a single chain as chain size increases. As a result we observe higher stress induced in a long chain system. If the length of the chains is less than the entanglement length of the system, entanglement effects are not significant \citep{Grest1992} and hence the stress is lower in short chain systems. Dynamics of the chain also completely changes across the entanglement length. Below the entanglement length, $n<n_e$, chains show Rouse behavior, where-as above the entanglement length, the behavior is better described by a reptation model \citep{Kremer1990}. The modulus curves at different chain lengths are also shown in Figure~\ref{fig:modulus_ChainLen}. Here, we note that a system with longer chain length is stiffer as compared to a short chain system. 

\begin{figure}
 \centering
 \subfloat[Mean-square bond length]{\label{fig:MeanBondLenStrain_ChainLen}\includegraphics[width=0.45\textwidth]{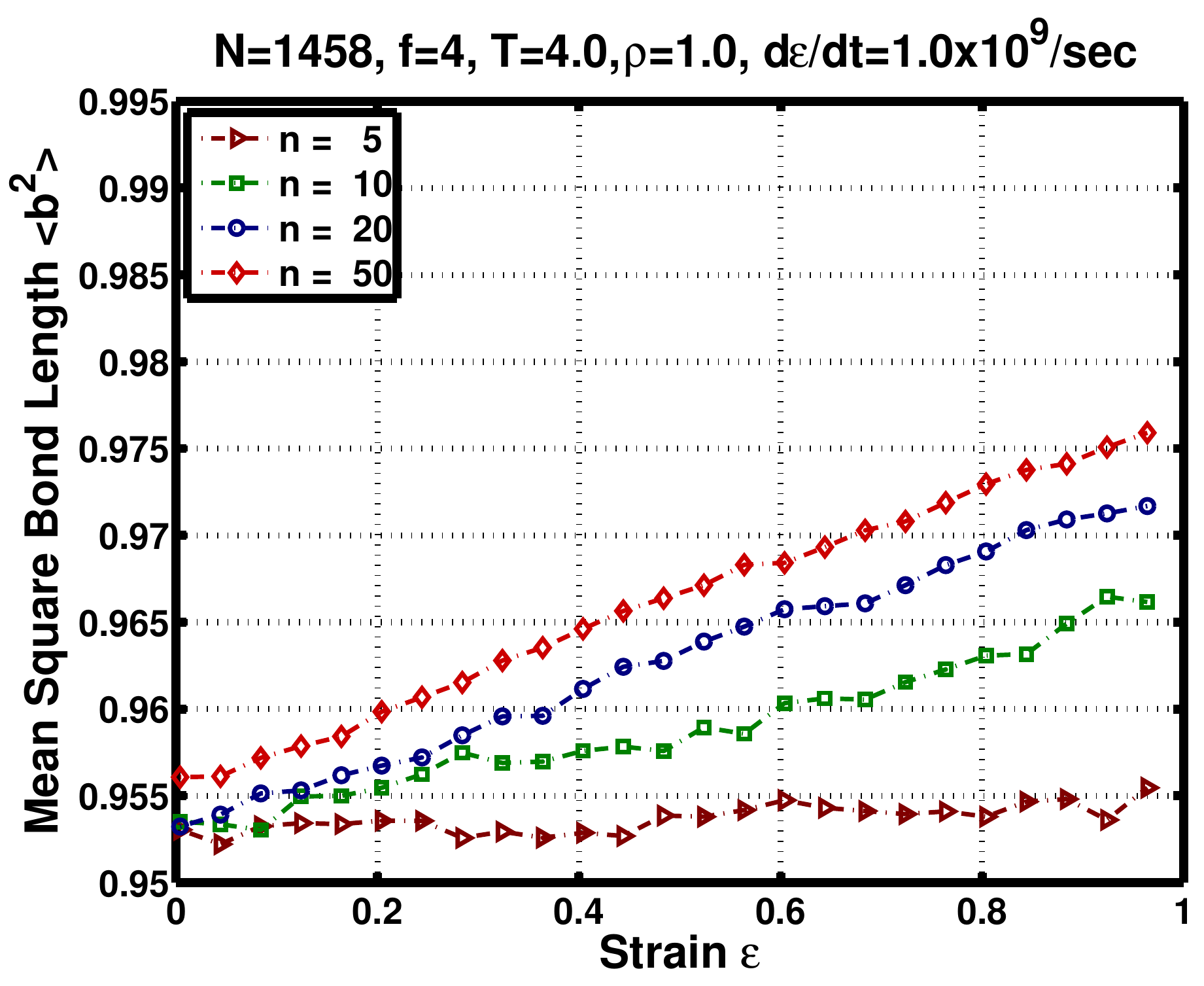}}
 \subfloat[Mean bond angle]{\label{fig:BondAngle_ChainLen}\includegraphics[width=0.45\textwidth]{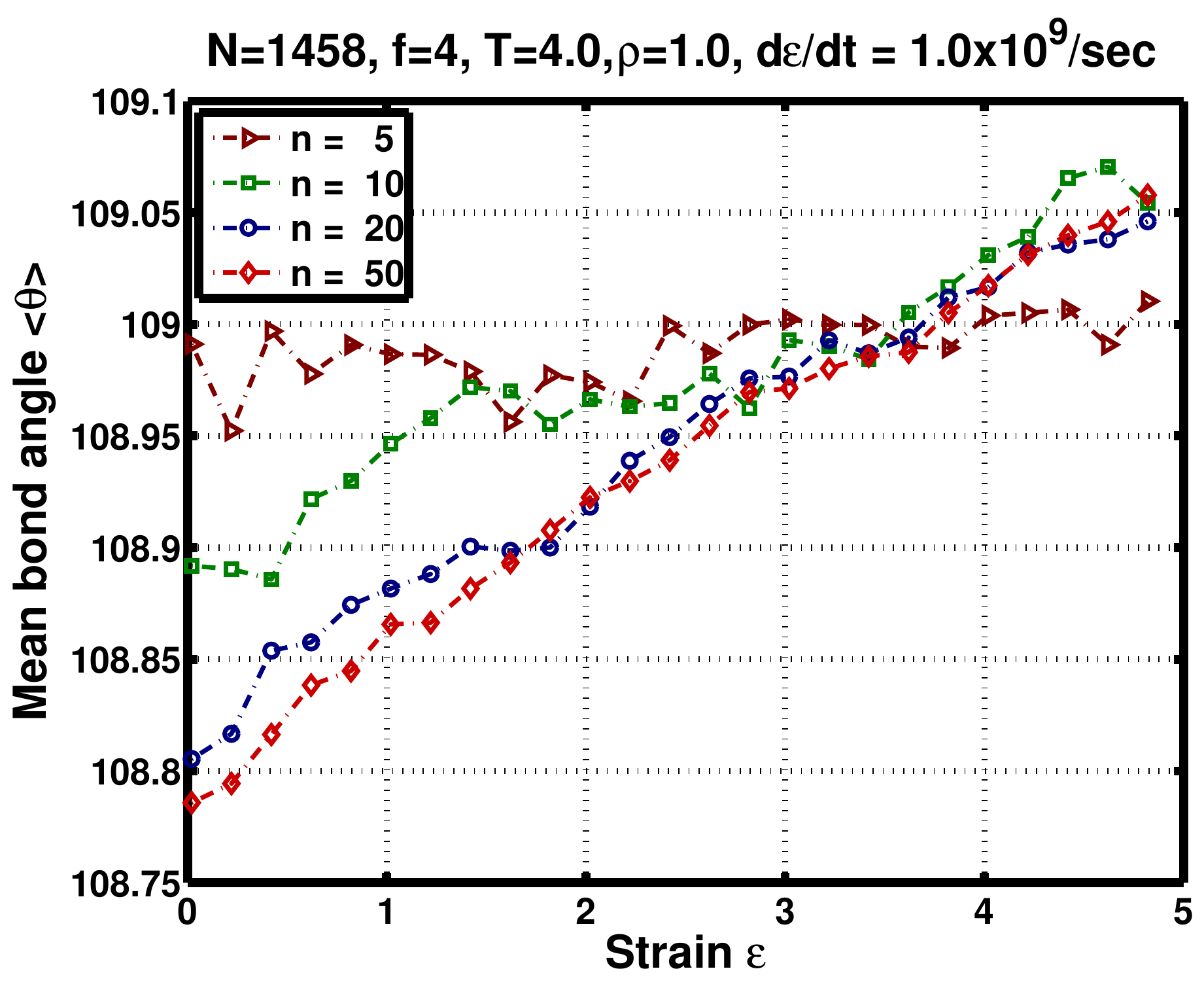}}\\
 \subfloat[End-to-end length]{\label{fig:EndToEnd_ChainLen}\includegraphics[width=0.45\textwidth]{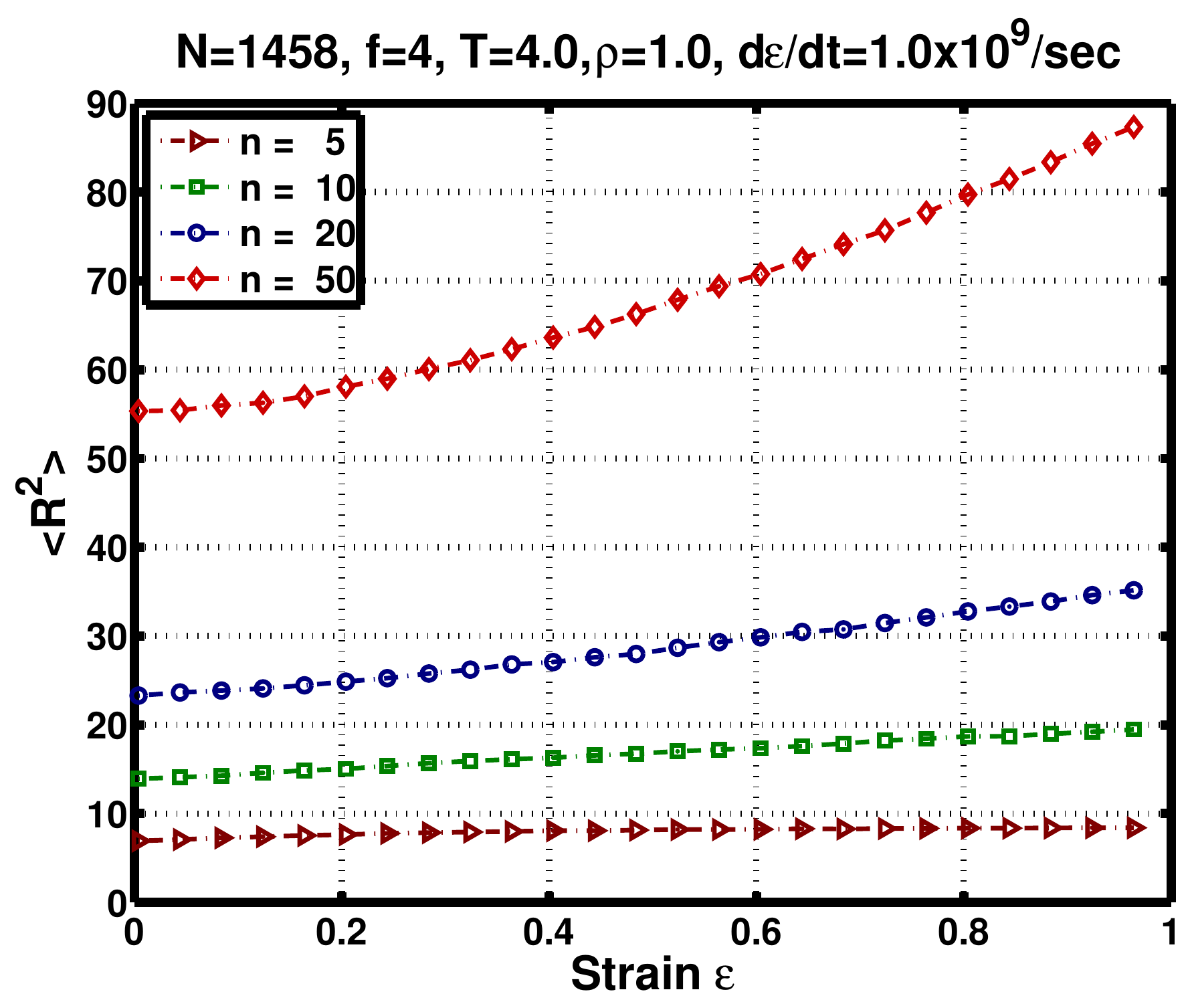}}
 \subfloat[Radius of gyration]{\label{fig:RadGyr_ChainLen}\includegraphics[width=0.45\textwidth]{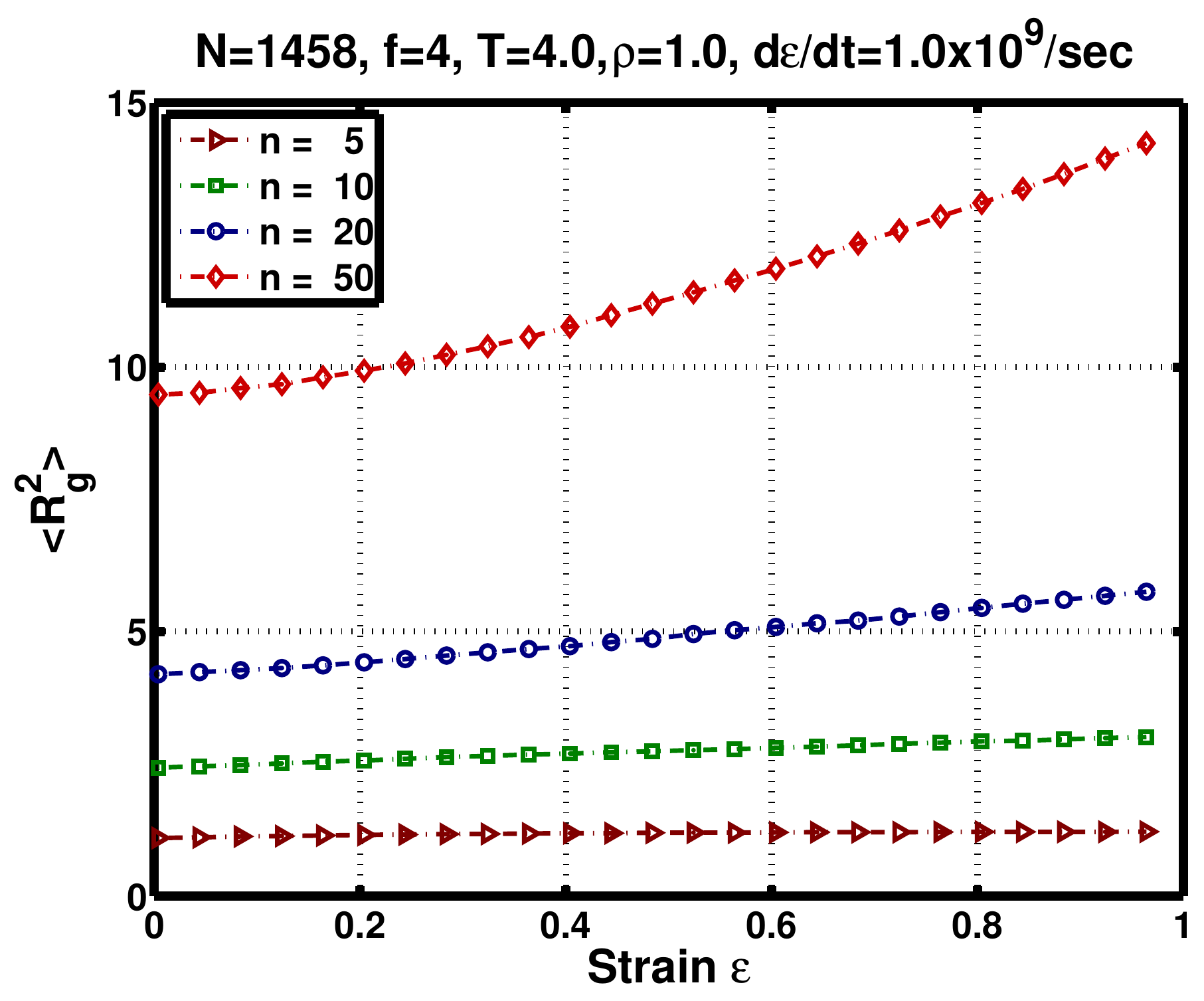}}
 \caption{Variation of structural properties for uniaxial constant strain rate loading: Effect of chain length}
 \label{fig:Properties_ChainLen}
\end{figure}

A detailed understanding of the above behavior can be obtained by examining the evolution of various structure parameters with strain. Towards this end, we study the variation of the mean-square bond length with strain shown in Figure~\ref{fig:MeanBondLenStrain_ChainLen}. We find a large change in the bond length with strain in long chain systems. This change is significant between $n=5$ to $n=10$. This is primarily because small chains can slip over one another very easily and as a result there is no bond deformation. Slipping gets increasingly difficult as the chain length grows and hence ways to contribute to deformation are through bond deformation and uncoiling of the chains. The variation of mean bond angle reveals that at higher chain lengths, the bond angles in the equilibrium configuration are smaller. This is shown in Figure~\ref{fig:BondAngle_ChainLen}, and is due to the increased constraints from the neighboring molecules. But as the system is strained, it is observed that there is a large change in the bond angle in a long chain system as compared to a short chain system. In fact, near $\epsilon=1.0$, the bond angles at all chain lengths come close to each other even though the simulation starts at lower bond angles in long chain polymers. 
 
The variation of the mean-square end-to-end length and radius of gyration are shown in Figure~\ref{fig:EndToEnd_ChainLen} and Figure~\ref{fig:RadGyr_ChainLen}. As expected, change in both the parameters increases with the length of the chain as the system is strained. In long chain systems, there is better possibility for uncoiling of the chain due to the very fact that in the initial configuration, chains will be coiled. In very short chain systems chains are not coiled to the extent found in long chain systems in their initial configuration. As a result we observe a very small variation in both of the above parameters for very short chains. 

\begin{figure}
 \centering
 \subfloat[$g_2/g_1$]{\label{fig:MassRatiog2_ChainLen}\includegraphics[width=0.4\textwidth]{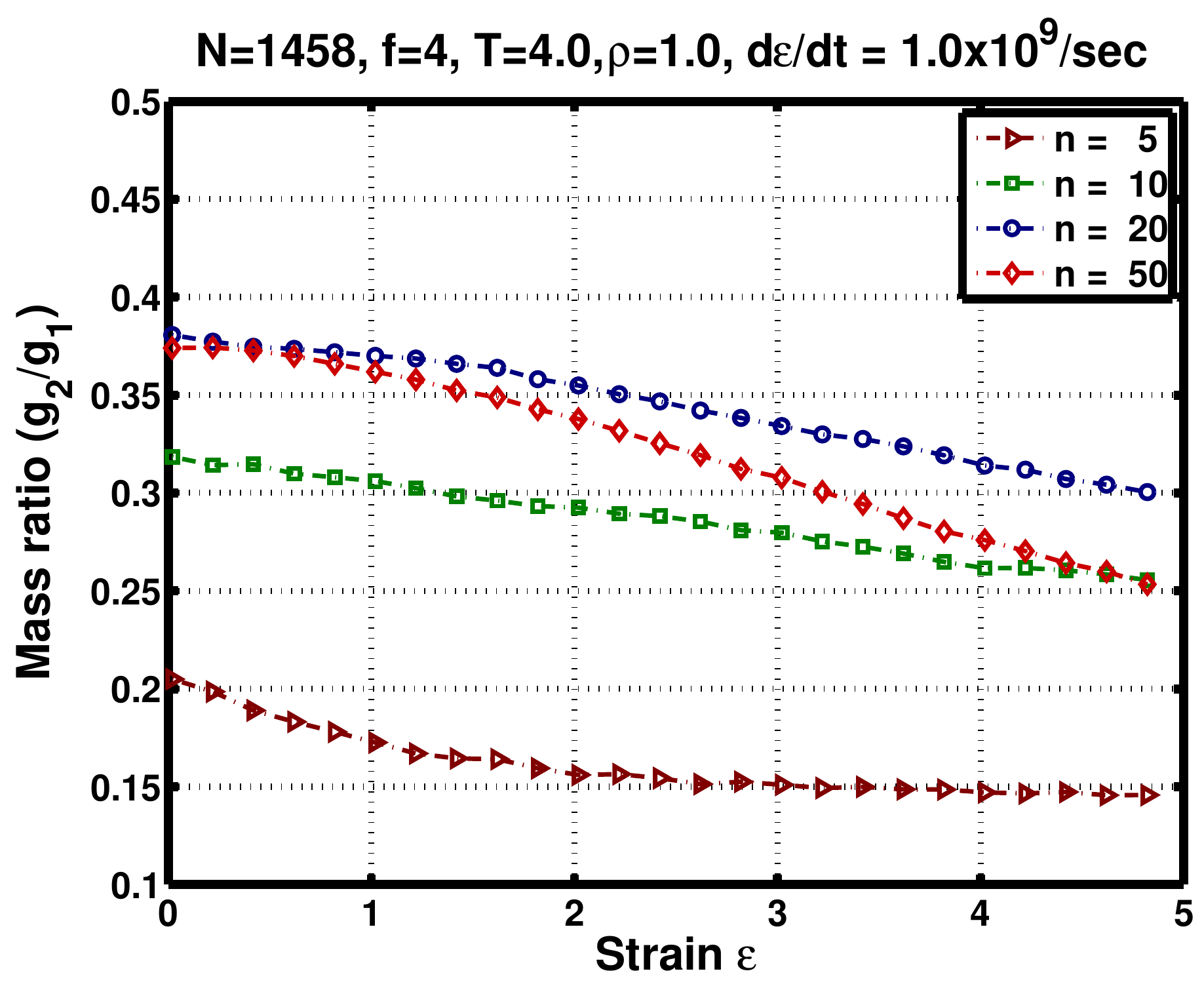}} 
 \subfloat[$g_3/g_1$]{\label{fig:MassRatiog3_ChainLen}\includegraphics[width=0.4\textwidth]{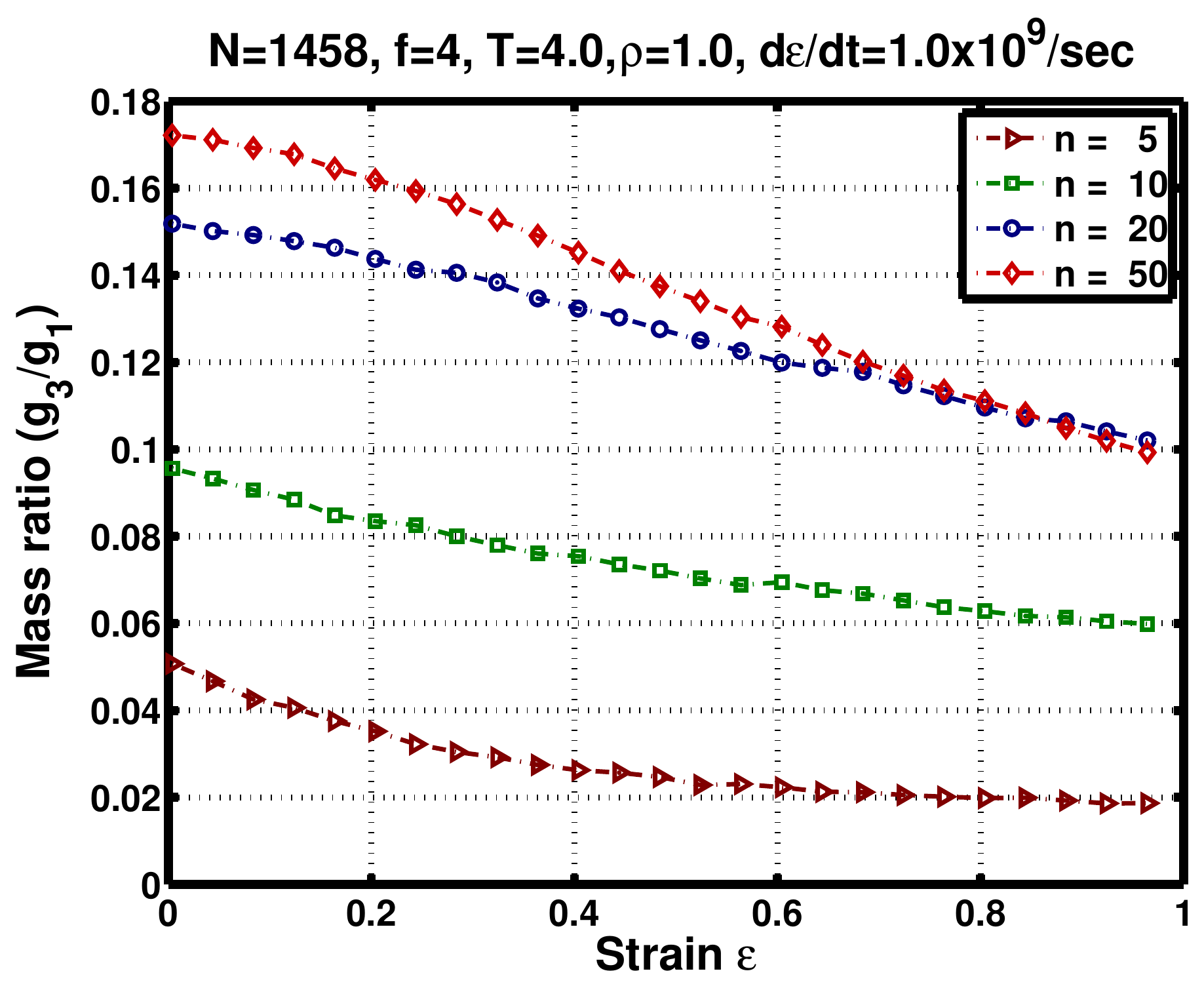}} 
 \caption{Effect of chain length on mass ratios}
 \label{fig:MassRatios_ChainLen}
\end{figure}

The variation of the mass ratios with strain at different values of the chain lengths are shown in Figure~\ref{fig:MassRatios_ChainLen}. Chains are more and more spherical longer the length of the chain. This also indirectly indicates that the degree of coiling of the longer chains is more as compared to that for shorter chains.  Also, we note that a very short chain system takes almost a planar configuration as the mass distribution $g_3$ is very small. The change in the mass ratio with strain is greater in a long chain system indicative of more uncoiling as well as change in shape of the chains in such polymers. Furthermore, it can also be observed that from $n=20$ to $n=50$, there is not enough change in the mass ratios which indicates that both are coiled approximately to the same degree. Some more studies at longer chain lengths are needed to comment on the behavior of mass ratio with chain length at chain lengths beyond $n=50$. Even though there is stretching of the bonds, the spherical shape of the chains does not give rise to stress anisotropy and hence we do not observe significant change in stress contribution due to this in systems at, and above, $n=10$.

\begin{figure}
 \centering
 \includegraphics[width=0.4\textwidth]{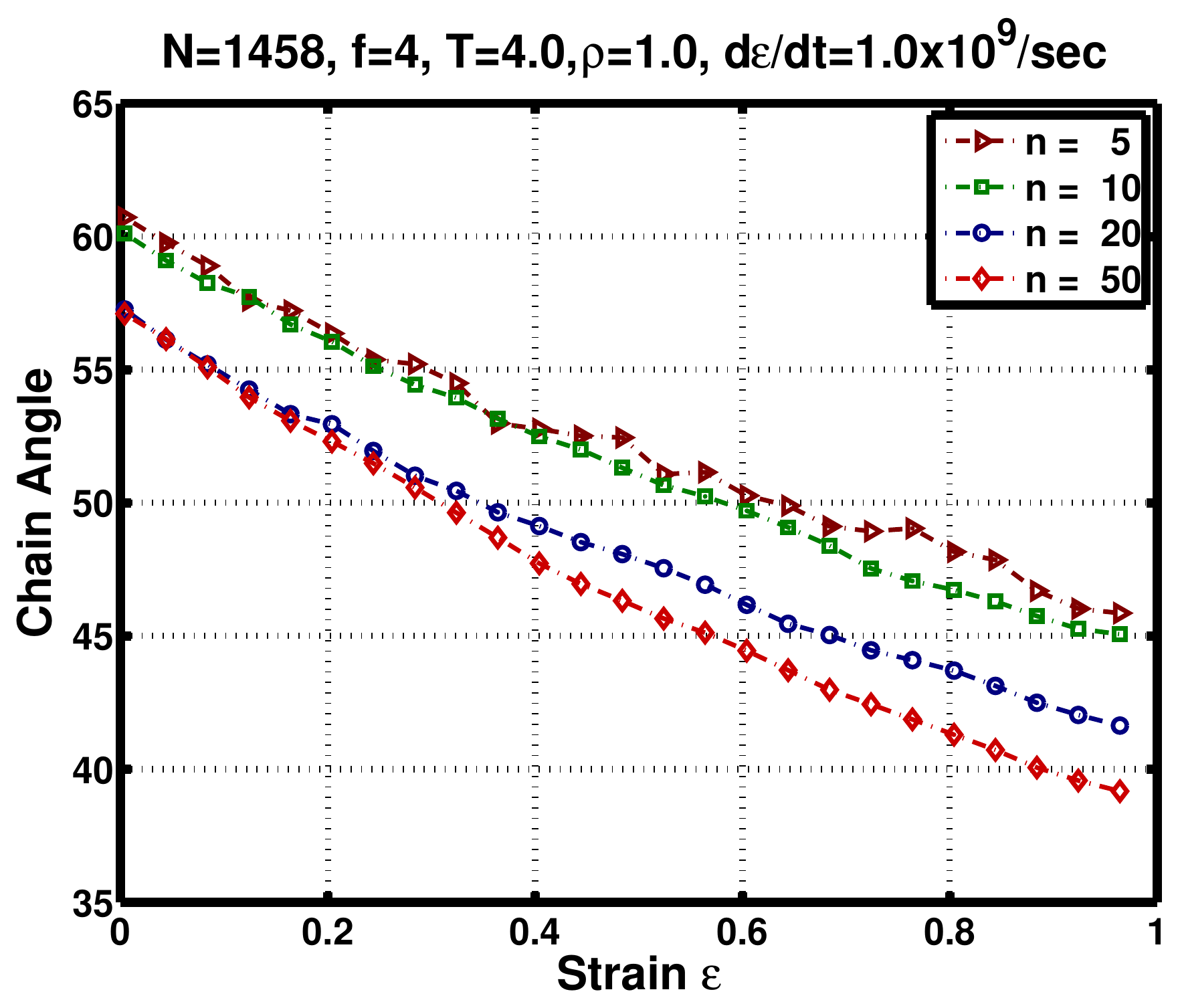}
 \caption{Effect of chain length on chain angle}
 \label{fig:ChainAng_ChainLen}
\end{figure}

Chains are oriented more in the loading direction as the chain length is increased as shown in Figure \ref{fig:ChainAng_ChainLen}. Furthermore, alignment of the chains in the loading direction also increases with the length of the chain when the system is strained.
 
\section{Conclusion}\label{sec:conclusion}

We have presented a study of the stress response of a polymer network subject to uniaxial constant strain-rate loading. The tool used for this study is a united atom molecular dynamics simulation. The variation of micro-structure parameters with strain are computed that give a detailed insight into the deformation behavior of the system. These are also used to explain the observed response in terms of the different kinds of chain deformation. The study on the effect of cross-linking reveals that stress levels increase in a cross-linked polymer network as compared to a linear polymeric system \cite{Srivastava2010}. We observe that temperature reduces the stress anisotropy in the system even though the bond stretch in greater at higher temperatures. This is because of randomness in the bond orientation at higher temperatures. Chain uncoiling is also more at higher temperatures. Density has a very significant effect on stress. There is a large change in the stress level with increasing density. This is because of large changes in the mean-square bond length and mean-square bond angle. Also, at higher densities, the uncoiling of the chains is suppressed and the major contribution to the deformation is by internal deformation of the chains. This is the reason for the large jump in the stress at higher densities. Rate of the loading affects the response of the system. At faster rates of loading, stress anisotropy increases. At a very high strain-rate, $\dot\epsilon=2.5\times 10^9/\mbox{sec}$, we observe that uncoiling of the chain is suppressed by a significant amount. The deformation mechanism is mostly internal deformation rather that overall shape and size. We notice that stress levels increase with increasing chain length of the polymer. This jump in the stress level exponentially decreases with increasing chain length. At very small chain lengths, the internal deformation in the chains ceases to exist and contribution to the overall deformation is due to slipping of the chains over one another or by uncoiling of the same. Chain opening is not very significant at very low chain lengths because in the equilibrium configuration itself, chains are not significantly coiled. As result a very small amount of stress is induced in short polymer chain networks. As the chain length increases, contribution to the stress increases. For longer chain lengths, we note that there are internal deformations in the chains together with their uncoiling. In an overall sense, excluded volume interactions play a significant role in governing the behavior of the system. Internal deformation such as bond stretching and bond bending are dominant and affect the stress response of the system.  Apart from these, stacking of the chains in the system also plays a dominant role in the behavior in terms of excluded volume interactions. Low density, high temperature and small chain length facilitate chain uncoiling and chain slipping in cross-linked polymers.

\bibliographystyle{jfm_ase}	
\bibliography{references_22Apr2011}	
\addcontentsline{toc}{section}{References}

 \end{document}